\documentclass[prd,aps,floatfix,twocolumn]{revtex4-1}
\usepackage{amsfonts,amsmath,amssymb,amsthm}
\usepackage{bm}
\usepackage{booktabs}
\usepackage{braket}
\usepackage{cases}
\usepackage{color}
\usepackage{dcolumn}
\usepackage{epsfig}
\usepackage{epstopdf}   
\usepackage{graphicx}
\usepackage[colorlinks,citecolor=blue,anchorcolor=red,menucolor=red,linkcolor=red,filecolor=red,runcolor=red,urlcolor=blue,frenchlinks=red]{hyperref}
\usepackage{indentfirst}
\usepackage{latexsym}
\usepackage{longtable}
\usepackage{multirow}
\usepackage{rotating}
\usepackage{slashed}  
\usepackage{subfigure}

\newcommand{\rmMeV}{\mathrm{MeV}}

\allowdisplaybreaks[4]
\begin{document}
\title{Hidden-charm pentaquarks and $P_c$ states}
\author{Xin-Zhen Weng$^1$}
\email{xzhweng@pku.edu.cn}
\author{Xiao-Lin Chen$^1$}
\email{chenxl@pku.edu.cn}
\author{Wei-Zhen Deng$^1$}
\email{dwz@pku.edu.cn}
\author{Shi-Lin Zhu$^{1,2,3}$}
\email{zhusl@pku.edu.cn}
\affiliation{
$^1$School of Physics and State Key Laboratory of Nuclear Physics and Technology, Peking University, Beijing 100871, China \\
$^2$Center of High Energy Physics, Peking University, Beijing 100871, China \\
$^3$Collaborative Innovation Center of Quantum Matter, Beijing
100871, China }

\begin{abstract}

Recently, the LHCb Collaboration reported three $P_c$ states in the
${J/\psi}p$ channel.
We systematically study the mass spectrum of the hidden charm
pentaquark in the framework of an extended chromomagnetic model.
For the $nnnc\bar{c}$ pentaquark with $I=1/2$, we find that
(i) the lowest state is $P_{c}(4327.0,1/2,1/2^{-})$ [We use
$P_{c}(m,I,J^{P})$ to denote the $nnnc\bar{c}$ pentaquark], which
corresponds to the $P_{c}(4312)$.
Its dominant decay mode is $\Lambda_{c}\bar{D}^{*}$.
(ii) We find two states in the vicinity of $P_{c}(4380)$.
The first one is $P_{c}(4367.4,1/2,3/2^{-})$ and decays dominantly
to $N{J/\psi}$ and $\Lambda_{c}\bar{D}^{*}$.
The other one is $P_{c}(4372.4,1/2,1/2^{-})$.
Its dominant decay mode is $\Lambda_{c}\bar{D}$, and its partial
decay width of $N\eta_{c}$ channel is comparable to that of
$N{J/\psi}$.
(iii) In higher mass region, we find $P_{c}(4476.3,1/2,3/2^{-})$ and
$P_{c}(4480.9,1/2,1/2^{-})$, which correspond to $P_{c}(4440)$ and
$P_{c}(4457)$.
In the open charm channels, both of them decay dominantly to the
$\Lambda_{c}\bar{D}^{*}$.
(iv) We predict two states above $4.5~\text{GeV}$, namely
$P_{c}(4524.5,1/2,3/2^{-})$ and $P_{c}(4546.0,1/2,5/2^{-})$.
The masses of the $nnnc\bar{c}$ state with $I=3/2$ are all over
$4.6~\text{GeV}$.
Moreover, we use the model to explore the $nnsc\bar{c}$,
$ssnc\bar{c}$, and $sssc\bar{c}$ pentaquark states.
%

\end{abstract}

\maketitle
\thispagestyle{empty} 

\section{Introduction}
\label{Sec:Introduction}

Before the birth of quantum chromodynamics (QCD), the possible
existence of tetraquark ($qq\bar{q}\bar{q}$) and pentaquark
($qqqq\bar{q}$) had been anticipated when
Gell-Mann~\cite{GellMann:1964nj} and Zweig~\cite{Zweig:1964jf} first
proposed the quark model.
In 1976, Jaffe studied the light tetraquark in the framework of the
MIT bag model~\cite{Jaffe:1976ig,Jaffe:1976ih}.
Chan and H{\"o}gaasen also studied this topic in the color-magnetic
spin-spin interaction from the one-gluon
exchange~\cite{Chan:1977st}.
Chao further considered the
hidden-charm~\cite{Chao:1979mm,Chao:1979tg} and
full-charm~\cite{Chao:1980dv} tetraquarks.
Meanwhile, the pentaquark was also studied in many models, such as
the color-magnetic hyperfine
interaction~\cite{Fukugita:1978sn,Hogaasen:1978jw} and the MIT bag
model~\cite{Strottman:1979qu}.

In despite of the theoretical investigations, the first experimental
evidence of the exotic states did not appear until 2003, when the
Belle Collaboration observed the $X(3872)$ state in the exclusive
$B^{\pm}{\to}K^{\pm}\pi^{+}\pi^{-}J/\psi$ decays~\cite{Choi:2003ue}.
Later, the CDF~\cite{Acosta:2003zx}, D0~\cite{Abazov:2004kp},
$BABAR$~\cite{Aubert:2004ns}, LHCb~\cite{Aaij:2011sn},
CMS~\cite{Chatrchyan:2013cld}, and BESIII~\cite{Ablikim:2013dyn}
Collaborations confirmed this state, and the LHCb Collaboration
further determined its quantum number to be
$I^{G}J^{PC}=0^{+}1^{++}$~\cite{Aaij:2011sn}.
For over a decade, lots of charmoniumlike $XYZ$ states have been
observed, such as $Y(3940)$~\cite{Abe:2004zs},
$Y(4140)$~\cite{Aaltonen:2009tz}, $Y(4260)$~\cite{Aubert:2005rm},
$Y(4360)$~\cite{Aubert:2007zz}, $Y(4660)$~\cite{Wang:2007ea}, and so
on.
Many of the $XYZ$ states do not fit into the conventional $q\bar{q}$
meson spectrum in the quark model.
To explain their nature, theorists have interpreted some of them to
be the molecular state~\cite{Swanson:2003tb,Guo:2017jvc}, hybrid
meson~\cite{Zhu:2005hp,Esposito:2016itg},
tetraquark~\cite{Cui:2006mp,Park:2013fda}, etc.
More detailed reviews can be found in
Refs.~\cite{Lebed:2016hpi,Esposito:2016noz,Chen:2016qju,Ali:2017jda,Guo:2017jvc,Liu:2019zoy}
and references therein.

Compared to the tetraquark candidates, the experimental observation
of the pentaquark states is more difficult.
In 2015, the LHCb Collaboration measured the
$\Lambda_{b}{\to}{J/\psi}K^{-}p$ decays, and observed two
resonances, $P_{c}(4380)$ and $P_{c}(4450)$, in the ${J/\psi}p$
channel, which indicates that they have a minimal quark content of
$uudc\bar{c}$~\cite{Aaij:2015tga}.
Very recently, the LHCb Collaboration reported the observation of
three narrow peaks in the ${J/\psi}p$ invariant mass spectrum of the
$\Lambda_{b}{\to}{J/\psi}Kp$ decays~\cite{Aaij:2019vzc}.
They found that the $P_{c}(4450)^{+}$ is actually composed of two
narrow resonances, $P_{c}(4440)^{+}$ and $P_{c}(4457)^{+}$.
Moreover, they also reported a new state below the
$\Sigma_{c}\bar{D}$ threshold, namely the $P_{c}(4312)^{+}$.
Their masses and widths are as follows:
\begin{eqnarray*}
P_{c}(4312)^{+}:M&=&4311.9\pm0.7_{-0.6}^{+6.8}~\text{MeV},\\
\Gamma&=&9.8\pm2.7_{-4.5}^{+3.7}~\text{MeV},\\
P_{c}(4440)^{+}:M&=&4440.3\pm1.3_{-4.7}^{+4.1}~\text{MeV},\\
\Gamma&=&20.6\pm4.9_{-10.1}^{+8.7}~\text{MeV},\\
P_{c}(4457)^{+}:M&=&4457.3\pm0.6_{-1.7}^{+4.1}~\text{MeV},\\
\Gamma&=&6.4\pm2.0_{-1.9}^{+5.7}~\text{MeV}.
\end{eqnarray*}
Since their masses are slightly below the $\Sigma_{c}\bar{D}$,
$\Sigma_{c}^{*}\bar{D}$, and $\Sigma_{c}\bar{D}^{*}$ thresholds
respectively, they can be interpreted as molecules composed of a
charm baryon and an anticharm
meson~\cite{Chen:2015loa,Chen:2015moa,
    Huang:2015uda,Meissner:2015mza,Roca:2015dva,
    Azizi:2016dhy,Chen:2016heh,Chen:2016otp,Chen:2019asm,Chen:2019bip,Guo:2019fdo,He:2019ify,Liu:2019tjn}.
For example, Chen~\cite{Chen:2019bip} interpreted them as bound
states of $\Sigma_{c}\bar{D}$ with $J^{P}=1/2^{-}$,
$\Sigma_{c}^{*}\bar{D}$ with $J^{P}=3/2^{-}$, and
$\Sigma_{c}\bar{D}^{*}$ with $J^{P}=3/2^{-}$, while Chen {\it et
al.}~\cite{Chen:2019asm}, He~\cite{He:2019ify}, and Liu {\it et
al.}~\cite{Liu:2019tjn} interpreted the $P_{c}(4312)$, $P_{c}(4440)$,
and $P_{c}(4457)$ as loosely bound $\Sigma_{c}\bar{D}$ with
($I=1/2$, $J^{P}=1/2^{-}$), $\Sigma_{c}\bar{D}^{*}$ with ($I=1/2$,
$J^{P}=1/2^{-}$), and $\Sigma_{c}\bar{D}^{*}$ with ($I=1/2$,
$J^{P}=3/2^{-}$).

Another interesting possibility is that some of the $P_c$ states
might be tightly bound pentaquark states.
The light $q^4\bar{q}$ pentaquark states was first studied with the
color-magnetic interaction among the
quarks~\cite{Fukugita:1978sn,Hogaasen:1978jw}.
Later, Strottman used the MIT bag model to discuss this system,
where the mass spectra mostly depend on the chromomagnetic
interaction between the quarks (or
antiquark)~\cite{Strottman:1979qu}.
The hidden-charm pentaquarks were also studied in constituent quark
model~\cite{Lebed:2015tna,Maiani:2015vwa,Mironov:2015ica,Wang:2015epa,Zhu:2015bba,Santopinto:2016pkp,Richard:2017una,Ali:2019npk}.

The quark model is widely used to investigate the mass spectra of
hadrons
\cite{Neeman:1961jhl,GellMann:1962xb,GellMann:1964nj,Zweig:1964jf,
Eichten:1978tg,
DeRujula:1975qlm,Isgur:1977ef,
Basdevant:1984rk,
Godfrey:1985xj,Capstick:1986bm}.
In the quark model, each quark (antiquark) carries the kinetic
energy $\sqrt{p^2+m^2}$.
In the nonrelativistic limit, the kinetic energy reduces to
$m+p^2/2m$, and the interquark potential contains the lattice
QCD-inspired linear confinement interaction and the short-range
one-gluon-exchange (OGE) interaction.
Usually the OGE interaction consists of the spin-independent color
Coulomb-type terms, the spin-spin chromomagnetic interaction, the
tensor interaction, and the spin-orbit interactions etc.

We can use the chromomagnetic  model to study the ground state
hadrons~\cite{Sakharov:1966tua,Sakharov:1967,DeRujula:1975qlm,DeGrand:1975cf,Jaffe:1976ig,Jaffe:1976ih,
Cui:2005az,
Buccella:2006fn,
Wu:2017weo}.
In the chromomagnetic model, the mass of the ground state hadrons
consists of the effective quark masses and the chromomagnetic
hyperfine interaction.
This simple model reproduced the hyperfine splitting of hadrons
quite well.
Compared to the quark model, the chromoelectric interaction has been
absorbed by the effective quark masses.
However, the one-body effective quark masses are not enough to
account for the two-body chromoelectric effects.
In Ref.~\cite{Karliner:2016zzc}, Karliner {\it et al.} found that
the color-related binding terms are needed when they considered the
interactions between a heavy (anti-)quark and a strange (or heavy)
quark.
Similarly, H\o{}gaasen {\it et al.} generalized the chromomagnetic
model and included a chromoelectric term
$H_{\text{CE}}=-\sum_{i,j}A_{ij}\tilde{\bm\lambda}_i\cdot\tilde{\bm\lambda}_j$
to study the hidden-beauty partners of the
$X(3872)$~\cite{Hogaasen:2013nca}.
Note that in 1978, Fukugita {\it et al.} had already used the color
and chromomagnetic interactions to investigate the
pseudobaryons~\cite{Fukugita:1978sn}.
Chan {\it et al.} also used these interactions to study the
properties of di/triquarks, which are constituents of
multiquarks~\cite{Chan:1978nk}.

In Ref.~\cite{Weng:2018mmf}, we extended the chromomagnetic model
and included the effect of color interaction.
According to color algebra, we further introduced the quark pair
mass parameters ($m_{qq}$ and $m_{q\bar{q}}$) to account for both
the effective quark masses ($m_{q}$) and the color interaction
($A_{qq}$ and $A_{q\bar{q}}$) between the two quarks.
Then we used this model to calculate the masses of multiheavy
baryons.
Our calculated mass of $\Xi_{cc}$, $3633.3\pm9.3~\text{MeV}$ is very
close to the LHCb's experiment
[$3621.40\pm0.72(\text{stat.}\pm0.27(\text{syst.})\pm0.14(\Lambda_{c})~\text{MeV}$]~\cite{Aaij:2017ueg}.

In this paper, we systematically study the mass spectrum of the
$qqqc\bar{c}$ ($q=n,s$, and $n=u,d$) pentaquarks in the extended
chromomagnetic model.
In Sec.~\ref{Sec:Model} we introduce the extended chromomagnetic
model.
In Sec.~\ref{Sec:Parameter} we present the model parameters.
Then we calculate and discuss the numerical results in
Sec.~\ref{sec:c.c}.
We conclude in Sec.~\ref{Sec:Conclusion}.
%

\section{The Extended Chromomagnetic Model}
\label{Sec:Model}

In the chromomagnetic (CM) model, the mass of hadron is governed by
the
Hamiltonian~\cite{DeRujula:1975qlm,Chan:1977st,Cui:2005az,Buccella:2006fn,Wu:2017weo}
\begin{equation}
H= \sum_{i}m_{i} -\sum_{i<j}v_{ij} \bm{S}_{i}\cdot\bm{S}_{j}
\bm{F}_{i}\cdot\bm{F}_{j},
\end{equation}
where $m_i$ is the $i$th constituent quark's (or antiquark's)
effective mass, which includes the constituent quark mass, the
kinetic energy, and so on, and $\bm{S}_{i}=\bm{\sigma}_i/2$ and
$\bm{F}_{i}=\tilde{\bm{\lambda}}_i/2$ are the quark spin and color
operators, respectively.
For the antiquark, $\bm{S}_{\bar{q}}=-\bm{S}_{q}^{*}$ and
$\bm{F}_{\bar{q}}=-\bm{F}_{q}^{*}$.
The coefficient $v_{ij}$ depends on the spatial wave function and
the quark masses
\begin{equation}
  v_{ij}= \frac{8\pi}{3m_i m_j}
  \Braket{\alpha_s(r)\delta^3(\bm{r})}.
\end{equation}

As pointed out in
Refs.~\cite{Hogaasen:2013nca,Karliner:2016zzc,Weng:2018mmf}, the
effective quark masses are not enough to absorb  all the two-body
chromoelectric effects.
To solve this problem, H{\o}gaasen {\it et al.} generalized the
chromomagnetic model by including a chromoelectric
term~\cite{Hogaasen:2013nca}
\begin{equation}
  H_{\text{CE}} = -\sum_{i,j} A_{ij}
  \tilde{\bm\lambda}_i\cdot\tilde{\bm\lambda}_j.
\end{equation}
Since
\begin{eqnarray}\label{eqn:m+color=color}
&&\sum_{i<j} \left( m_i + m_j \right) \bm{F}_{i} \cdot \bm{F}_{j} \notag \\
&&= \left(\sum_i m_i\bm{F}_i\right) \cdot \left(\sum_i
\bm{F}_i\right) - \frac43 \sum_i m_i \,,
\end{eqnarray}
and the total color operator $\sum_i\bm{F}_i$ nullifies any
colorless physical state,
we introduced a new quark pair mass parameter
\begin{equation}\label{eqn:para:color+m}
  m_{ij} = \left( m_i + m_j\right) + \frac{16}3 A_{ij} \,,
\end{equation}
and rewrite the model Hamiltonian as~\cite{Weng:2018mmf}
\begin{equation}\label{eqn:hamiltonian:final}
  H_{\text{CM}} = -\frac34\sum_{i<j} m_{ij} V^{\text{C}}_{ij}
  - \sum_{i<j} v_{ij} V^{\text{CM}}_{ij},
\end{equation}
where
\begin{align}
  V^{\text{C}}_{ij} =& \bm{F}_{i} \cdot \bm{F}_{j} \,, \\
  V^{\text{CM}}_{ij} =& \bm{S}_{i}\cdot \bm{S}_{j}
  \bm{F}_{i}^{a}\cdot\bm{F}_{j}^{a} \,,
\end{align}
are the color and CM interactions between quarks.

To investigate the mass spectra of the pentaquark states, we need to construct the wave functions.
A detailed construction of the pentaquark wave functions in the $(q_1q_2{\otimes}q_{3})\otimes(q_{4}\bar{q}_{5})$ configuration can be found in Appendix~\ref{App:wavefunc}.
Diagonalizing the Hamiltonian in these bases, we can obtain the mass spectrum and eigenvector of the hidden charm pentaquark states.
%

\section{Numerical results}
\label{Sec:Result}

\subsection{Parameters}
\label{Sec:Parameter}

\begin{table*}
    \centering
    \caption{Parameters of the $q\bar{q}$ and $qq$ pairs (in units of $\rmMeV$).}
    \label{table:parameter}
    \begin{tabular}{cccccccccccc}
        \toprule[1pt]
        \toprule[1pt]
        $m_{n\bar{n}}$  & $m_{n\bar{s}}$ & $m_{s\bar{s}}$ & $m_{n\bar{c}}$ & $m_{s\bar{c}}$ & $m_{c\bar{c}}$ & $m_{n\bar{b}}$ & $m_{s\bar{b}}$ & $m_{c\bar{b}}$ & $m_{b\bar{b}}$ \\
        $615.95$ & $794.22$ & $936.40$ & $1973.22$ & $2076.14$ & $3068.53$ & $5313.35$ & $5403.25$ & $6322.27$ & $9444.97$ \\
        \midrule[1pt]
        $v_{n\bar{n}}$  & $v_{n\bar{s}}$ & $v_{s\bar{s}}$ & $v_{n\bar{c}}$ & $v_{s\bar{c}}$ & $v_{c\bar{c}}$ & $v_{n\bar{b}}$ & $v_{s\bar{b}}$ & $v_{c\bar{b}}$ & $v_{b\bar{b}}$ \\
        $477.92$ & $298.57$ & $249.18$ & $106.01$ & $107.87$ & $85.12$ & $33.89$ & $36.43$ & $47.18$ & $45.98$ \\
        \midrule[1pt]
        $m_{nn}$ & $m_{ns}$ &  $m_{ss}$ & $m_{nc}$ & $m_{sc}$ & $m_{cc}$ & $m_{nb}$ & $m_{sb}$ & $m_{cb}$ & $m_{b{b}}$ \\
        $724.85$ & $906.65$ & $1049.36$ & $2079.96$ & $2183.68$ & $3171.51$ & $5412.25$ & $5494.80$ & $6416.07$ & $9529.57$ \\
        \midrule[1pt]
        $v_{n{n}}$  & $v_{n{s}}$ & $v_{ss}$ & $v_{n{c}}$ & $v_{s{c}}$ & $v_{c{c}}$ & $v_{n{b}}$ & $v_{s{b}}$ & $v_{c{b}}$ & $v_{b{b}}$  \\
        $305.34$ & $212.75$ & $195.30$ & $62.81$ & $70.63$ & $56.75$ & $19.92$ & $8.47$ & $31.45$ & $30.65$ \\
        \bottomrule[1pt]
        \bottomrule[1pt]
    \end{tabular}
\end{table*}

In Ref.~\cite{Weng:2018mmf}, we have carefully extracted the
parameters of the extended chromomagnetic model from the ground
state mesons and baryons.
Specifically, the parameters $m_{q\bar{q}}$ and $v_{q\bar{q}}$ are
extracted from the mesons.
The $m_{qq}$ and $v_{qq}$ with at most one heavy quark are extracted
from the light and singly heavy baryons, and those with two heavy
quarks are estimated from a quark model consideration.
With these parameters, we calculated the mass of $\Xi_{cc}$ to be
$3633.3\pm9.3~\text{MeV}$, which is very close to the LHCb's result,
$M_{\Xi_{cc}}=3621.40\pm0.72~\text{MeV}$~\cite{Aaij:2017ueg}.
All parameters are listed in Table~\ref{table:parameter}.
In this work, we use the same parameters to study the mass spectrum
of the $S$-wave $qqqc\bar{c}$ pentaquark states.
%

\subsection{The hidden-charm pentaquarks}
\label{sec:c.c}

\subsubsection{The $nnnc\bar{c}$ system}
\label{sec:nnnc.c}

\begin{table*}
\centering \caption{Pentaquark masses and eigenvectors of the
$nnnc\bar{c}$ system. The masses are all in units of MeV.}
\label{table:mass:nnnc.c}
\begin{tabular}{ccccccc}
\toprule[1pt] \toprule[1pt]
System & $J^{P}$ & Mass & Eigenvector & Scattering state \\
\midrule[1pt]
$(nnnc\bar{c})^{I=3/2}$ & $\frac{1}{2}^{-}$ & $4320.8$ & $\{0.070,-0.217,0.974\}$ & $\Delta{J/\psi}(4329)$ \\
&& $4601.9$ & $\{0.733,-0.651,-0.197\}$ \\
&& $4717.1$ & $\{0.677,0.727,0.114\}$ \\
& $\frac{3}{2}^{-}$ & $4217.5$ & $\{-0.119,-0.016,-0.993\}$ & $\Delta\eta_{c}(4216)$ \\
&& $4336.0$ & $\{-0.052,-0.998,0.022\}$ & $\Delta{J/\psi}(4329)$ \\
&& $4633.0$ & $\{0.992,-0.054,-0.118\}$ \\
& $\frac{5}{2}^{-}$ & $4336.8$ & $\{1\}$ & $\Delta{J/\psi}(4329)$ \\
\midrule[1pt]
$(nnnc\bar{c})^{I=1/2}$ & $\frac{1}{2}^{-}$ & $3907.0$ & $\{0.111,-0.112,0.013,0.001,0.987\}$ & $N\eta_{c}(3923)$ \\
&& $4024.2$ & $\{0.042,0.042,0.130,-0.990,-0.000\}$ & $N{J/\psi}(4036)$ \\
&& $4327.0$ & $\{0.773,-0.356,0.501,0.084,-0.134\}$ \\
&& $4372.4$ & $\{0.189,0.907,0.358,0.093,0.077\}$ \\
&& $4480.9$ & $\{0.594,0.193,-0.777,-0.069,-0.035\}$ \\
& $\frac{3}{2}^{-}$ & $4028.2$ & $\{0.035,0.078,-0.032,-0.996\}$ & $N{J/\psi}(4036)$ \\
&& $4367.4$ & $\{-0.570,-0.356,0.737,-0.072\}$ \\
&& $4476.3$ & $\{0.819,-0.312,0.482,-0.011\}$ \\
&& $4524.5$ & $\{0.057,0.877,0.473,0.056\}$ \\
& $\frac{5}{2}^{-}$ & $4546.0$ & $\{1\}$ \\
\bottomrule[1pt] \bottomrule[1pt]
\end{tabular}
\end{table*}
%

\begin{figure*}
	\includegraphics[width=400pt]{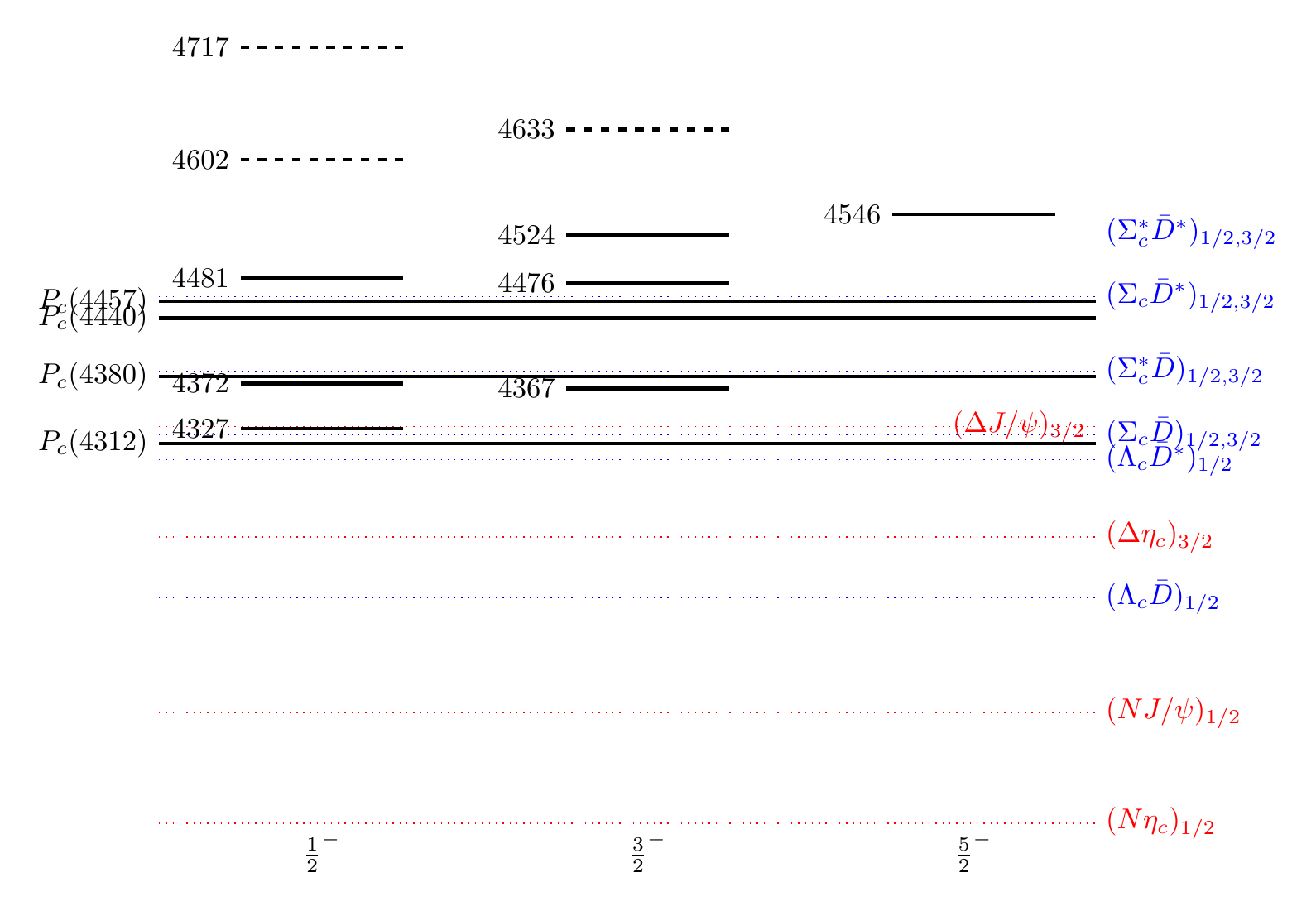}
	\caption{Mass spectra of the $I=\frac12$ (solid) and $I=\frac32$ (dashed) $nnnc\bar{c}$ pentaquark states.
		The dotted lines indicate various meson-baryon thresholds and the long solid lines indicate the observed $P_c$ states.
		The masses are all in units of MeV.}
	\label{fig:nnnccbar}
\end{figure*}
%

\begin{table}
\centering \caption{The partial width ratios for the hidden-charm
decays of the $nnnc\bar{c}$ pentaquark states. For each state, we
chose one mode as the reference channel, and the partial width
ratios of the other channels are calculated relative to this
channel. The masses are all in units of MeV.}
\label{table:R:hidden:nnnc.c}
\begin{tabular}{ccccccccccccc}
\toprule[1pt] \toprule[1pt] $I$ & $J^P$ & Mass & ${\Delta{J/\psi}}$
& ${\Delta\eta_{c}}$ & ${N{J/\psi}}$
& ${N\eta_{c}}$ \\
\midrule[1pt] $\frac{3}{2}$ & $\frac{1}{2}^{-}$
& $4601.9$ & $1$ &&& \\
&& $4717.1$ & $1$ &&& \\
& $\frac{3}{2}^{-}$
& $4633.0$ & $1$ & $5.5$ && \\
\midrule[1pt] $\frac{1}{2}$ & $\frac{1}{2}^{-}$
& $4327.0$ &&& $1$ & $3.0$ \\
&& $4372.4$ &&& $1$ & $0.8$ \\
&& $4480.9$ &&& $1$ & $0.3$ \\
& $\frac{3}{2}^{-}$
& $4367.4$ &&& $1$ & \\
&& $4476.3$ &&& $1$ & \\
&& $4524.5$ &&& $1$ & \\
& $\frac{5}{2}^{-}$ & $4546.0$ &&&& \\
\bottomrule[1pt] \bottomrule[1pt]
\end{tabular}
\end{table}
%
\begin{table}
    \centering
\caption{The partial width ratios for the open charm decays of the
$nnnc\bar{c}$ pentaquark states. For each state, we chose one
mode as the reference channel, and the partial width ratios of
other channels are calculated relative to this channel.
        The masses are all in units of MeV.}
    \label{table:R:open:nnnc.c}
    \begin{tabular}{ccccccccccccc}
        \toprule[1pt]
        \toprule[1pt]
        $I$ & $J^P$ & Mass
        & ${\Sigma_{c}^{*}\bar{D}^{*}}$
        & ${\Sigma_{c}^{*}\bar{D}}$
        & ${\Sigma_{c}\bar{D}^{*}}$
        & ${\Sigma_{c}\bar{D}}$
        & ${\Lambda_{c}\bar{D}^{*}}$
        & ${\Lambda_{c}\bar{D}}$ \\
        \midrule[1pt]
        $\frac{3}{2}$ & $\frac{1}{2}^{-}$
        & $4601.9$ & $0.05$ && $1$ & $0.4$ \\
        && $4717.1$ & $7.0$ && $1$ & $0.2$ \\
        & $\frac{3}{2}^{-}$
        & $4633.0$ & $5.3$ & $3.1$ & $1$ & \\
        \midrule[1pt]
        $\frac{1}{2}$ & $\frac{1}{2}^{-}$
        & $4327.0$ & $0$ && $0$ & $1.3$ & $1$ & $0.02$ \\
        && $4372.4$ & $0$ && $0$ & $0.7$ & $1$ & $57.6$ \\
        && $4480.9$ & $0$ && $0.3$ & $0.09$ & $1$ & $0.07$ \\
        & $\frac{3}{2}^{-}$
        & $4367.4$ & $0$ & $0$ & $0$ && $1$ & \\
        && $4476.3$ & $0$ & $0.2$ & $1.9$ && $1$ & \\
        && $4524.5$ & $0$ & $0.58$ & $0.63$ && $1$ & \\
        & $\frac{5}{2}^{-}$ & $4546.0$ & $1$ &&&&& \\
        \bottomrule[1pt]
        \bottomrule[1pt]
    \end{tabular}
\end{table}

The calculated eigenvalues and eigenvectors of the $nnnc\bar{c}$
system are listed in Table~\ref{table:mass:nnnc.c}.
First we consider the $nnnc\bar{c}$ state with isospin $I=1/2$.
The lowest state has mass of $3097.0~\text{MeV}$ with
$J^{P}=1/2^{-}$.
This state, $\sum_{i}b_{i}\Psi_{Di}^{1/2}$ with
\begin{equation}
\{b_i\} = \{0.111,-0.112,0.013,0.001,0.987\},
\end{equation}
has a dominant component of $\Psi_{D5}^{1/2}$.
Notice that in the $nnn{\otimes}c\bar{c}$ configuration,
$\Psi_{D5}^{1/2}$ can be written as a direct product of a baryon and
a meson,
\begin{equation}
\Psi_{D5}^{1/2}=N\otimes\eta_{c}.
\end{equation}
In other words, this state couples almost completely to the
$N\eta_{c}$ scattering state.
Therefore it has probably a very broad width and is just a part of
the continuum.
It is worth stressing that this kind of state also exists in the
calculation of the $q\bar{q}c\bar{c}$ tetraquark, where the lowest
state couples strongly to a heavy charmonium and a light
meson~\cite{Hogaasen:2005jv,Cui:2006mp}.
Moreover, the states of $4024.2~\text{MeV}$ (with $J^{P}=1/2^{-}$)
and $4028.2~\text{MeV}$ (with $J^{P}=3/2^{-}$) couple strongly to
$N$ and $J/\psi$ channel.
The above states are also scattering states.
We label these scattering states in the fifth column of
Table~\ref{table:mass:nnnc.c}.
The situation of the $nnnc\bar{c}$ states with $I=3/2$ is similar.
There are four low mass states.
The lowest one, $4217.5~\text{MeV}$ with $J^{P}=3/2^{-}$, is a
scattering state of $\Delta$ and $\eta_{c}$, and the other three
states, $4320.8~\text{MeV}$ with $J^{P}=1/2^{-}$,
$4336.0~\text{MeV}$ with $J^{P}=3/2^{-}$ and $4336.8~\text{MeV}$
with $J^{P}=3/2^{-}$, couple very strongly to $\Delta$ and $J/\psi$.

After identifying the scattering states, the other states are
genuine pentaquarks.
We plot their relative position in Fig.~\ref{fig:nnnccbar}.
For simplicity, we use $P_{c}(m,I,J^{P})$ to denote the
$nnnc\bar{c}$ pentaquark states.
From Table~\ref{table:mass:nnnc.c}, we see that the lightest state
is $P_{c}(4327.0,1/2,1/2^{-})$.
This state is very close to the recently observed $P_{c}(4312)$.
If the future experiment does confirm the quantum number of
$P_{c}(4312)$ to be $1/2^{-}$, it is likely a tightly bound
pentaquark state.
We find two states in the vicinity of $P_{c}(4380)$, namely the $P_{c}(4367.4,1/2,3/2^{-})$ and $P_{c}(4372.4,1/2,1/2^{-})$.
If $P_{c}(4380)$ truly corresponds to one of the two states,
the other state should also exist, which can be searched for 
in future experiment.
In higher energy region, we find the $P_{c}(4476.3,1/2,3/2^{-})$ and
$P_{c}(4480.9,1/2,1/2^{-})$, which can be identified with
$P_{c}(4440)$ and $P_{c}(4457)$~\cite{Aaij:2019vzc}.
Above $4.5~\text{GeV}$, there are $P_{c}(4524.5,1/2,3/2^{-})$ and
$P_{c}(4546.0,1/2,5/2^{-})$.
The $nnnc\bar{c}$ pentaquark with isospin $I=3/2$ are all above
$4.6~\text{GeV}$.

Besides the mass spectrum, the eigenvectors also provide important
information about the decay
properties~\cite{Jaffe:1976ig,Strottman:1979qu,Zhao:2014qva,Wang:2015epa}.
We can calculate the overlap between the pentaquark and a particular
baryon~$\times$~meson state.
Then we can determine the decay amplitude of the pentaquark into
that particular baryon~$\times$~meson channel.
To calculate the overlap, we transform the eigenvectors of the
pentaquark states into the $nnc{\otimes}n\bar{c}$ configuration (see Table~\ref{table:wavefunc:nnnc.c} of Appendix~\ref{App:eigenvec}).
Normally, the $nnc$ and $n\bar{c}$ components inside the pentaquark
can be either of color-singlet or of color-octet.
The former one can easily dissociate into a $S$-wave meson and a
$S$-wave baryon (the so-called ``Okubo-Zweig-Iizuka (OZI)-superallowed''
decays~\cite{Jaffe:1976ig}).
The latter one cannot fall apart without the gluon exchange.
For simplicity, we follow Refs.~\cite{Jaffe:1976ig,Strottman:1979qu}
and focus on the ``OZI-superallowed'' decays in this work.
For the color-singlet part, we can rewrite the base states as a
direct product of a baryon and a meson.
For each decay mode, the branching fraction is proportional to the
square of the coefficient of the corresponding component in the
eigenvectors, and also depends on the phase space.
For the two body $L$-wave decay, its partial width
reads~\cite{Gao-1992-Group}
\begin{equation}
\label{eqn:width}
\Gamma_{i}=\gamma_{i}\alpha\frac{k^{2L+1}}{m^{2L}}{\cdot}|c_i|^2,
\end{equation}
where $\alpha$ is an effective coupling constant, $\gamma_{i}$ is a
quantity determined by the decay dynamics, $m$ is the mass of the
parent particle, $k$ is the momentum of the daughter particles in
the rest frame of the parent particle, and $c_i$ is the coefficient
of the corresponding component.
For the decay processes which we are interested in, $(k/m)^2$ is of
$\mathcal{O}(10^{-2})$ or even smaller.
Thus we only consider the $S$-wave decays since the higher wave
decays are all suppressed.
Next we have to estimate the $\gamma_{i}$.
Generally, $\gamma_{i}$ depends on the spatial wave functions of the
initial pentaquark and final meson and baryon, which are different
for each decay process.
In the quark model, the spatial wave functions of the ground state
scalar and vector meson are the same.
And in the heavy quark limit, $\Sigma_{c}$ and $\Sigma_{c}^{*}$ have
the same spatial wave function.
Furthermore, the spatial wave function of $\Lambda_{c}$ does not
differ much from that of $\Sigma_{c}$.
Then for each pentaquark,
\begin{equation}
\gamma_{\Delta{J/\psi}}=\gamma_{\Delta\eta_{c}},
\end{equation}
\begin{equation}
\gamma_{N{J/\psi}}=\gamma_{N\eta_{c}},
\end{equation}
and
\begin{equation}
\gamma_{\Sigma_{c}^{*}\bar{D}^{*}}= \gamma_{\Sigma_{c}^{*}\bar{D}}=
\gamma_{\Sigma_{c}\bar{D}^{*}}= \gamma_{\Sigma_{c}\bar{D}}\approx
\gamma_{\Lambda_{c}\bar{D}^{*}}= \gamma_{\Lambda_{c}\bar{D}}.
\end{equation}
The values of the relative widths of different decay modes are
listed in Tables~\ref{table:R:hidden:nnnc.c}
and~\ref{table:R:open:nnnc.c}.

First we consider the $I=1/2$ case.
The lowest state, $P_{c}(4327.0,1/2,1/2^{-})$, has two hidden charm
decay modes, namely $N{J/\psi}$ and $N\eta_{c}$.
Their partial decay width ratio is
\begin{equation}
\frac{\Gamma[P_{c}(4327.0,1/2,1/2^{-}){\to}{N\eta_{c}}]}{\Gamma[P_{c}(4327.0,1/2,1/2^{-}){\to}{N{J/\psi}}]}=3.0,
\end{equation}
which indicates that the partial decay width of the $N\eta_{c}$
channel is larger than that of the $N{J/\psi}$.
On the other hand, $P_{c}(4327.0,1/2,1/2^{-})$ also has open charm
decay modes.
From Table~\ref{table:R:open:nnnc.c}, we see that
$\Sigma_{c}\bar{D}$ and $\Lambda_{c}\bar{D}^{*}$ are its dominant
decay modes.
It is worth stressing that the calculated mass of this state is just
several MeV higher than the threshold of $\Sigma_{c}\bar{D}$
($4321~\text{MeV}$); considering the error of the model (Taking
$\Xi_{cc}$ for example, our calculation differs from the experiment
by $12~\text{MeV}$~\cite{Weng:2018mmf}), this state may probably lie
below the $\Sigma_{c}\bar{D}$ threshold and thus cannot decay into
this channel.
If we assume that the $P_{c}(4327.0,1/2,1/2^{-})$ corresponds to the
observed $P_{c}(4312)$ state, we have
\begin{equation}
\frac{\Gamma[P_{c}(4312){\to}N\eta_{c}]}{\Gamma[P_{c}(4312){\to}N{J/\psi}]}
=3.1.
\end{equation}
If the $P_{c}(4312)$ is observed in the $N\eta_{c}$ channel, and its
partial decay width is larger than that of the $N{J/\psi}$ channel,
then the $P_{c}(4312)$ is very likely a tightly bound pentaquark
which corresponds the $P_{c}(4327.0,1/2,1/2^{-})$.
If the $P_{c}(4312)$ does not appear in the $N\eta_{c}$ channel, or
its partial decay width is much smaller than that of the $N{J/\psi}$
channel, the $P_{c}(4312)$ may not be a tightly bound pentaquark.
Moreover,
\begin{equation}
\frac{\Gamma[P_{c}(4312){\to}\Lambda_{c}\bar{D}]}{\Gamma[P_{c}(4312){\to}\Lambda_{c}\bar{D}^{*}]}
= 0.02.
\end{equation}
We hope the future experiments can search for the $P_{c}(4312)$ in
the $N\eta_{c}$ and $\Lambda_{c}\bar{D}^{*}$ channels.

Next we consider the two states in the vicinity of $P_{c}(4380)$.
The $P_{c}(4367.4,1/2,3/2^{-})$ only has one hidden charm decay mode
$N{J/\psi}$, while the $P_{c}(4372.4,1/2,1/2^{-})$ can decay to both
$N{J/\psi}$ and $N\eta_{c}$.
Moreover,
\begin{equation}
\frac{\Gamma[P_{c}(4372.4,1/2,1/2^{-}){\to}N\eta_{c}]}{\Gamma[P_{c}(4372.4,1/2,1/2^{-}){\to}N{J/\psi}]}
=0.8.
\end{equation}
Thus this state can also be found in the $N\eta_{c}$ channel.
On the other hand, $P_{c}(4367.4,1/2,3/2^{-})$ can only decay to
$\Lambda_{c}\bar{D}^{*}$, and $P_{c}(4372.4,1/2,1/2^{-})$ decays
dominantly to $\Lambda_{c}\bar{D}$.

Then we consider the $P_{c}(4476.3,1/2,3/2^{-})$ and
$P_{c}(4480.9,1/2,1/2^{-})$.
Both of them couple weakly to the hidden charm channel(s).
Note that the former state can only decay to $N{J/\psi}$ while the
latter state can also decay to $N\eta_{c}$, which can be used to
distinguish the two states.
In the open charm channels, both of the two states decay dominantly
to the $\Lambda_{c}\bar{D}^{*}$ channel.
The $\Sigma_{c}\bar{D}^{*}$ mode is also important for
$P_{c}(4476.3,1/2,3/2^{-})$.
The mass difference between the $\Sigma_{c}\bar{D}^{*}$ threshold
($4462~\text{MeV}$) and the two states is only $\sim10~\text{MeV}$,
which is within the error of the CM model.
The two states probably lie below the $\Sigma_{c}\bar{D}^{*}$
threshold and cannot decay through this mode.
$P_{c}(4476.3,1/2,3/2^{-})$ can also decay to
$\Sigma_{c}^{*}\bar{D}$ with a not-so-small fraction.
If $P_{c}(4476.3,1/2,3/2^{-})$ and $P_{c}(4480.9,1/2,1/2^{-})$ truly
correspond to the $P_{c}(4440)$ and $P_{c}(4457)$ respectively, we
have
\begin{equation}
\frac{\Gamma[P_{c}(4440){\to}\Sigma_{c}^{*}\bar{D}]}{\Gamma[P_{c}(4440){\to}\Lambda_{c}\bar{D}^{*}]}=0.16,
\end{equation}
\begin{equation}
\frac{\Gamma[P_{c}(4457){\to}N\eta_{c}]}{\Gamma[P_{c}(4457){\to}N{J/\psi}]}=0.29,
\end{equation}
and
\begin{align}
&\Gamma[P_{c}(4457){\to}\Sigma_{c}\bar{D}]:
\Gamma[P_{c}(4457){\to}\Lambda_{c}\bar{D}^{*}]
\notag\\
&:\Gamma[P_{c}(4457){\to}\Lambda_{c}\bar{D}]= 0.09:1:0.07.
\end{align}

Finally, we consider the two states over $4.5~\text{GeV}$.
We see that $P_{c}(4524.5,1/2,3/2^{-})$ may also be observed in the
$N{J/\psi}$ channel, while $P_{c}(4546.0,1/2,5/2^{-})$ can only
decay to this mode through higher partial waves, which is
suppressed.
The dominant decay modes of $P_{c}(4524.5,1/2,3/2^{-})$ are
$\Sigma_{c}^{*}\bar{D}$, $\Sigma_{c}\bar{D}^{*}$ and
$\Lambda_{c}\bar{D}^{*}$.
Note that the $\Sigma_{c}^{*}\bar{D}^{*}$ mode has the largest
coefficient in the eigenvector, but this mode is suppressed by phase
space.
And the $P_{c}(4546.0,1/2,5/2^{-})$ can only decay to
$\Sigma_{c}^{*}\bar{D}^{*}$.

There are three $nnnc\bar{c}$ pentaquark states with $I=3/2$.
Their masses are all above $4.6~\text{GeV}$.
Their couplings to $\Delta{J/\psi}$ are not very small (see Table~\ref{table:mass:nnnc.c} or see Table~\ref{table:wavefunc:nnnc.c} of Appendix~\ref{App:eigenvec}), thus they can be observed in the
$\Delta{J/\psi}$ channel in the future experiments.
We also calculate the partial decay width ratio of each mode.
For $P_{c}(4601.9,3/2,1/2^{-})$ and $P_{c}(4717.1,3/2,1/2^{-})$
states, we have
\begin{equation}
\Gamma_{\Sigma_{c}^{*}\bar{D}^{*}}: \Gamma_{\Sigma_{c}\bar{D}^{*}}:
\Gamma_{\Sigma_{c}\bar{D}}=
0.05:1:0.4
\end{equation}
and
\begin{equation}
\Gamma_{\Sigma_{c}^{*}\bar{D}^{*}}: \Gamma_{\Sigma_{c}\bar{D}^{*}}:
\Gamma_{\Sigma_{c}\bar{D}}=
7.0:1:0.2
\end{equation}
respectively.
And for $P_{c}(4633.0,3/2,3/2^{-})$, we have
\begin{equation}
\Gamma_{\Delta{J/\psi}}: \Gamma_{\Delta\eta_{c}}= 1:5.5
\end{equation}
and
\begin{align}
\Gamma_{\Sigma_{c}^{*}\bar{D}^{*}}: \Gamma_{\Sigma_{c}^{*}\bar{D}}:
\Gamma_{\Sigma_{c}\bar{D}^{*}}= 5.3:3.1:1.
\end{align}

In both cases, the $P_{c}$ states have a large decay fraction to the
open charm channels.
Since all $P_{c}$ states are observed in the $NJ/\psi$ channel, it
is very helpful if the future experiments can search for the open
charm channels.
%

\subsubsection{The $nnsc\bar{c}$ system}
\label{sec:nnsc.c}

\begin{table*}
    \centering
    \caption{Pentaquark masses and eigenvectors of the $nnsc\bar{c}$ systems.
        The masses are all in units of MeV.}
    \label{table:mass:nnsc.c}
    \begin{tabular}{ccccccc}
        \toprule[1pt]
        \toprule[1pt]
        System & $J^{P}$ & Mass & Eigenvectors & Scattering state \\
        \midrule[1pt]
        $(nnsc\bar{c})^{I=1}$ & $\frac{1}{2}^{-}$ & $4145.5$ & $\{0.095,-0.017,-0.170,0.108,-0.020,-0.0002,0.003,-0.975\}$ & $\Sigma\eta_{c}(4177)$ \\
        && $4264.9$ & $\{0.037,0.115,0.079,0.038,0.135,0.005,-0.979,-0.014\}$ & $\Sigma{J/\psi}(4290)$ \\
        && $4442.8$ & $\{-0.122,0.224,0.837,-0.190,0.351,0.144,0.134,-0.190\}$ \\
        && $4466.7$ & $\{-0.086,0.199,0.173,0.043,-0.180,-0.942,0.006,-0.033\}$ & $\Sigma^{*}{J/\psi}(4481)$ \\
        && $4522.2$ & $\{-0.565,-0.169,-0.141,-0.760,-0.178,-0.008,-0.105,-0.108\}$ \\
        && $4612.6$ & $\{0.019,-0.437,0.454,0.188,-0.732,0.138,-0.106,-0.034\}$ \\
        && $4696.3$ & $\{-0.621,0.566,-0.084,0.408,-0.260,0.229,0.017,-0.005\}$ \\
        && $4808.1$ & $\{-0.512,-0.598,0.019,0.412,0.437,-0.140,-0.012,-0.007\}$ \\
        & $\frac{3}{2}^{-}$ & $4269.7$ & $\{0.033,-0.057,-0.121,0.040,-0.001,-0.004,0.990\}$ & $\Sigma{J/\psi}(4290)$ \\
        && $4366.8$ & $\{-0.080,-0.020,-0.009,0.128,-0.019,-0.988,-0.009\}$ & $\Sigma^{*}\eta_{c}(4368)$ \\
        && $4485.9$ & $\{-0.183,0.421,0.232,-0.323,0.789,-0.053,0.072\}$ & \\
        && $4488.4$ & $\{0.306,-0.570,-0.288,0.343,0.610,0.023,-0.091\}$ & \\
        && $4584.9$ & $\{-0.235,-0.680,0.581,-0.375,-0.022,-0.021,0.054\}$ \\
        && $4636.2$ & $\{-0.268,-0.171,-0.709,-0.625,-0.016,-0.048,-0.062\}$ \\
        && $4728.8$ & $\{0.859,0.053,0.094,-0.478,-0.062,-0.132,0.004\}$ \\
        & $\frac{5}{2}^{-}$ & $4487.8$ & $\{0.006,0.99998\}$ & $\Sigma^{*}{J/\psi}(4481)$ \\
        && $4644.3$ & $\{0.99998,-0.006\}$ \\
        \midrule[1pt]
        $(nnsc\bar{c})^{I=0}$ & $\frac{1}{2}^{-}$ & $4086.1$ & $\{-0.126,-0.059,0.022,0.146,0.001,0.002,0.979\}$ & $\Lambda\eta_{c}(4100)$ \\
        && $4197.4$ & $\{0.045,0.350,0.130,0.547,0.361,0.652,-0.059\}$ & \\
        && $4208.6$ & $\{-0.038,0.381,0.250,0.479,0.136,-0.735,-0.057\}$ & \\
        && $4386.6$ & $\{-0.208,0.102,-0.327,0.435,-0.797,0.095,-0.078\}$ \\
        && $4465.0$ & $\{0.735,0.323,-0.572,-0.036,0.040,-0.091,0.132\}$ \\
        && $4489.6$ & $\{0.152,-0.763,-0.255,0.502,0.238,-0.112,-0.096\}$ \\
        && $4607.0$ & $\{0.612,-0.179,0.649,0.085,-0.397,0.076,0.041\}$ \\
        & $\frac{3}{2}^{-}$ & $4209.5$ & $\{0.041,0.088,0.024,-0.033,0.994\}$ & $\Lambda{J/\psi}(4212)$ \\
        && $4387.3$ & $\{0.101,0.074,-0.402,-0.906,-0.031\}$ \\
        && $4501.5$ & $\{0.521,0.335,0.743,-0.242,-0.077\}$ \\
        && $4603.6$ & $\{-0.845,0.258,0.396,-0.249,-0.006\}$ \\
        && $4656.0$ & $\{0.037,0.899,-0.360,0.239,-0.065\}$ \\
        & $\frac{5}{2}^{-}$ & $4680.6$ & $\{1\}$ \\
        \bottomrule[1pt]
        \bottomrule[1pt]
    \end{tabular}
\end{table*}
%

\begin{figure*}
	\includegraphics[width=400pt]{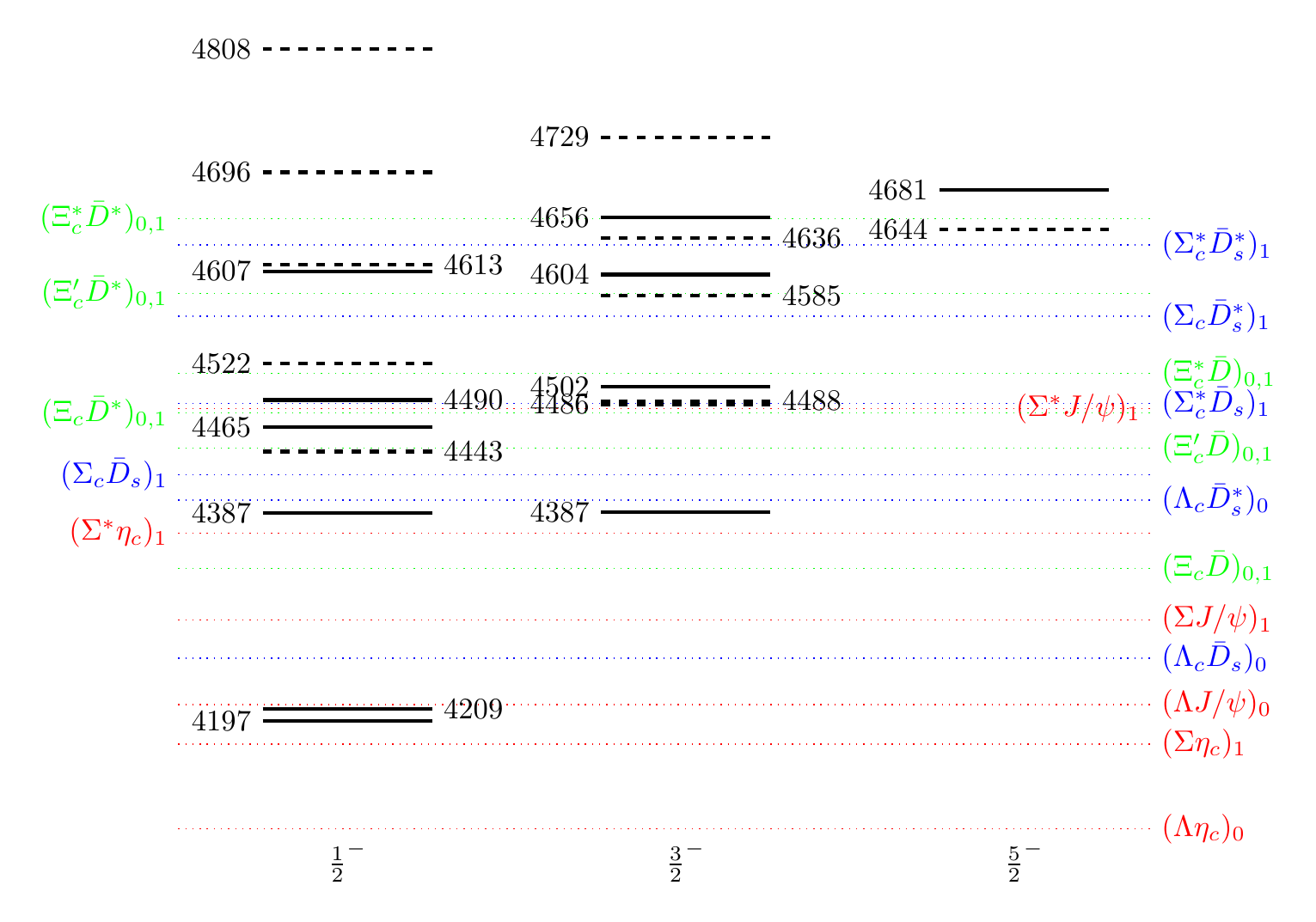}
	\caption{Mass spectra of the $I=0$ (solid) and $I=1$ (dashed) $nnsc\bar{c}$ pentaquark states.
		The dotted lines indicate various meson-baryon thresholds.
		The masses are all in units of MeV.}
	\label{fig:nnsccbar}
\end{figure*}
%

\begin{table}
    \centering
    \caption{The partial width ratios for the hidden-charm decays of
    the $nnsc\bar{c}$ pentaquark states. The masses are all in units of MeV.}
    \label{table:R:hidden:nnsc.c}
    \begin{tabular}{ccccccccccccccccccccc}
        \toprule[1pt]
        \toprule[1pt]
        $I$ & $J^{P}$ & Mass
        & ${\Sigma^{*}{J/\psi}}$
        & ${\Sigma^{*}\eta_{c}}$
        & ${\Sigma{J/\psi}}$
        & ${\Sigma\eta_{c}}$
        & ${\Lambda{J/\psi}}$
        & ${\Lambda\eta_{c}}$ \\
        \midrule[1pt]
        $1$ & $\frac{1}{2}^{-}$
        & $4442.8$ &0&&1&2.7&& \\
        && $4522.2$ &0.002&&1&1.3&& \\
        && $4612.6$ &1.1&&1&0.12&& \\
        && $4696.3$ &1&&0.008&0.0008&& \\
        && $4808.1$ &1&&0.009&0.003&& \\
        & $\frac{3}{2}^{-}$
        & $4485.9$ &18.5&0.4&1&&& \\
        && $4488.4$ &8.6&0.05&1&&& \\
        && $4584.9$ &0.10&0.14&1&&& \\
        && $4636.2$ &0.05&0.54&1&&& \\
        && $4728.8$ &1&5.5&0.007&&& \\
        & $\frac{5}{2}^{-}$
        & $4644.3$ &1&&&&& \\
        \midrule[1pt]
        $0$ & $\frac{1}{2}^{-}$
        & $4197.4$ &&&&&0&1 \\
        && $4208.6$ &&&&&0&1 \\
        && $4386.6$ &&&&&1&0.87 \\
        && $4465.0$ &&&&&1&2.6 \\
        && $4489.6$ &&&&&1&0.87 \\
        && $4607.0$ &&&&&1&0.33 \\
        & $\frac{3}{2}^{-}$
        & $4387.3$ &&&&&1& \\
        && $4501.5$ &&&&&1& \\
        && $4603.6$ &&&&&1& \\
        && $4656.0$ &&&&&1& \\
        & $\frac{5}{2}^{-}$
        & $4680.6$ &&&&&& \\
        \bottomrule[1pt]
        \bottomrule[1pt]
    \end{tabular}
\end{table}
%
\begin{table}
    \centering
    \caption{The partial width ratios for the $nnc{\otimes}s\bar{c}$ open
    charm decays of the $nnsc\bar{c}$ pentaquark states. The masses are all in units of MeV.}
    \label{table:R:open:nnc:nnsc.c}
    \begin{tabular}{ccccccccccccccccccccc}
        \toprule[1pt]
        \toprule[1pt]
        $I$ & $J^{P}$ & Mass
        & ${\Sigma_{c}^{*}\bar{D}_{s}^{*}}$
        & ${\Sigma_{c}^{*}\bar{D}_{s}}$
        & ${\Sigma_{c}\bar{D}_{s}^{*}}$
        & ${\Sigma_{c}\bar{D}_{s}}$
        & ${\Lambda_{c}\bar{D}_{s}^{*}}$
        & ${\Lambda_{c}\bar{D}_{s}}$ \\
        \midrule[1pt]
        $1$ & $\frac{1}{2}^{-}$
        & $4442.8$ &0&&0&1&& \\
        && $4522.2$ &0&&0&1&& \\
        && $4612.6$ &0&&1&0.0001&& \\
        && $4696.3$ &0.00002&&1.9&1&& \\
        && $4808.1$ &37.7&&5.6&1&& \\
        & $\frac{3}{2}^{-}$
        & $4485.9$ &0&0&0&&& \\
        && $4488.4$ &0&1&0&&& \\
        && $4584.9$ &0&1&176.1&&& \\
        && $4636.2$ &17.9&1&0.33&&& \\
        && $4728.8$ &0.65&1&0.21&&& \\
        & $\frac{5}{2}^{-}$
        & $4644.3$ &1&&&&& \\
        \midrule[1pt]
        $0$ & $\frac{1}{2}^{-}$
        & $4197.4$ &&&&&0&0 \\
        && $4208.6$ &&&&&0&0 \\
        && $4386.6$ &&&&&0&1 \\
        && $4465.0$ &&&&&1&0.03 \\
        && $4489.6$ &&&&&1&134.1 \\
        && $4607.0$ &&&&&13.1&1 \\
        & $\frac{3}{2}^{-}$
        & $4387.3$ &&&&&0& \\
        && $4501.5$ &&&&&1& \\
        && $4603.6$ &&&&&1& \\
        && $4656.0$ &&&&&1& \\
        & $\frac{5}{2}^{-}$
        & $4680.6$ &&&&&& \\
        \bottomrule[1pt]
        \bottomrule[1pt]
    \end{tabular}
\end{table}
%
\begin{table}
    \centering
    \caption{The partial width ratios for the $nsc{\otimes}n\bar{c}$ open
    charm decays of the $nnsc\bar{c}$ pentaquark states. The masses are all in units of MeV.}
    \label{table:R:open:nsc:nnsc.c}
    \begin{tabular}{ccccccccccccccccccccc}
        \toprule[1pt]
        \toprule[1pt]
        $I$ & $J^{P}$ & Mass
        & ${\Xi_{c}^{*}\bar{D}^{*}}$
        & ${\Xi_{c}^{*}\bar{D}}$
        & ${\Xi_{c}'\bar{D}^{*}}$
        & ${\Xi_{c}'\bar{D}}$
        & ${\Xi_{c}\bar{D}^{*}}$
        & ${\Xi_{c}\bar{D}}$ \\
        \midrule[1pt]
        $1$ & $\frac{1}{2}^{-}$
        & $4442.8$ &0&&0&0&0&1 \\
        && $4522.2$ &0&&0&2.6&0.006&1 \\
        && $4612.6$ &0&&2.2&0.24&3.4&1 \\
        && $4696.3$ &0.14&&0.27&0.21&4.6&1 \\
        && $4808.1$ &151.4&&7.4&5.0&12.8&1 \\
        & $\frac{3}{2}^{-}$
        & $4485.9$ &0&0&0&&1& \\
        && $4488.4$ &0&0&0&&1& \\
        && $4584.9$ &0&0.007&0&&1& \\
        && $4636.2$ &0&0.38&0.49&&1& \\
        && $4728.8$ &14.6&8.0&3.4&&1& \\
        & $\frac{5}{2}^{-}$
        & $4644.3$ &0&&&&& \\
        \midrule[1pt]
        $0$ & $\frac{1}{2}^{-}$
        & $4197.4$ &0&&0&0&0&0 \\
        && $4208.6$ &0&&0&0&0&0 \\
        && $4386.6$ &0&&0&0&0&1 \\
        && $4465.0$ &0&&0&1&0&7.9 \\
        && $4489.6$ &0&&0&0.88&0.50&1 \\
        && $4607.0$ &0&&7.4&1.6&2.4&1 \\
        & $\frac{3}{2}^{-}$
        & $4387.3$ &0&0&0&&0& \\
        && $4501.5$ &0&0&0&&1& \\
        && $4603.6$ &0&1&0.35&&18.0& \\
        && $4656.0$ &0.87&1&1.3&&0.65& \\
        & $\frac{5}{2}^{-}$
        & $4680.6$ &1&&&&& \\
        \bottomrule[1pt]
        \bottomrule[1pt]
    \end{tabular}
\end{table}

Now we turn to the $nnsc\bar{c}$ systems.
The mass spectrum of the $nnsc\bar{c}$ system is listed in
Table~\ref{table:mass:nnsc.c}.
Similar to the $nnnc\bar{c}$ case, we first identify the scattering
states composed of a $nns$ baryon and a charmonium.
For the $I=0$ case, the $\Lambda{\otimes}\eta_{c}$ scattering state
corresponds to the spin-$1/2$ state around $4086.1~\text{MeV}$.
The $\Lambda{\otimes}J/\psi$ scattering states can be of spin-$1/2$
and -$3/2$.
The latter one has a mass $4209.5~\text{MeV}$, while the former one
is more complex.
Actually, there are two states correspond to the spin-$1/2$
$\Lambda{\otimes}J/\psi$ scattering state.
Their masses are $4197.4~\text{MeV}$ and $4208.6~\text{MeV}$
respectively.
Since they all have large fractions of color-octet components
($57\%$ and $46\%$),
we still consider them as pentaquarks.
We also reproduce most of the scattering states with $I=1$.
The scattering state of $\Sigma^{(*)}$ and $\eta_{c}$ has
$J^{P}=1/2^{-}$($3/2^{-}$) and mass $4145.5~\text{MeV}$
($4366.8~\text{MeV}$).
And the $\Sigma{\otimes}J/\psi$ scattering states can be of
$J^{P}=1/2^{-}$ ($4264.9~\text{MeV}$) and $J^{P}=3/2^{-}$
($4269.7~\text{MeV}$).
We only reproduce two $\Sigma^{*}{\otimes}J/\psi$ scattering states,
namely the $J^{P}=1/2^{-}$ one with mass $4466.7~\text{MeV}$ and the
$J^{P}=5/2^{-}$ one with mass $4487.8~\text{MeV}$.
For the spin-$3/2$ $\Lambda{\otimes}J/\psi$ case, there are two
$J^{P}=3/2^{-}$ states couple strongly to
$\Sigma^{*}{\otimes}J/\psi$.
Their masses are $4485.9~\text{MeV}$ and $4488.4~\text{MeV}$,
respectively.
They also have large fractions of color-octet components ($37\%$ and
$62\%$). Thus we consider them as pentaquarks.
For clarity, we add a fifth column in Table~\ref{table:mass:nnsc.c}
to label these scattering states.
In the following, we will use $P_{c,s}(m,I,J^{P})$ to denote the
$nnsc\bar{c}$ pentaquark states.

In Fig.~\ref{fig:nnsccbar}, we show the relative position of the
$nnsc\bar{c}$ pentaquark states.
We also plot all the meson-baryon thresholds which they can decay to
through quark rearrangement.
Compared to the $nnnc\bar{c}$ case, the $nnsc\bar{c}$ system has
larger numbers of states and decay patterns.
There are 18 channels that the $nnsc\bar{c}$ pentaquarks may decay
to.
From the figure, we can easily identify the decay constrains of the
isospin conservation and kinetics.
Next we study their decay properties.
Similar to the $nnnc\bar{c}$ case, we need to consider the $\gamma_{i}$.
In the quark model, the spatial wave functions of the ground state
scalar and vector meson are the same.
And the same spatial wave function of $\Sigma^{*}$ does not differ
much from that of $\Sigma$.
In the heavy quark limit, $\Sigma_{c}$ and $\Sigma_{c}^{*}$ have the
same spatial wave function.
Similarly, the $\Xi_{c}^{*}$ and $\Xi_{c}'$ have the same spatial
wave function, and their spatial wave functions do not differ much
from that of $\Xi_{c}$.
Thus for each $nnsc\bar{c}$ pentaquark
\begin{equation}
\gamma_{\Sigma^*{J/\psi}}= \gamma_{\Sigma^*\eta_{c}}\approx
\gamma_{\Sigma{J/\psi}}= \gamma_{\Sigma\eta_{c}},
\end{equation}
\begin{equation}
\gamma_{\Lambda{J/\psi}}= \gamma_{\Lambda\eta_{c}},
\end{equation}
\begin{equation}
\gamma_{\Sigma_{c}^{*}\bar{D}_{s}^{*}}=
\gamma_{\Sigma_{c}^{*}\bar{D}_{s}}=
\gamma_{\Sigma_{c}\bar{D}_{s}^{*}}= \gamma_{\Sigma_{c}\bar{D}_{s}},
\end{equation}
\begin{equation}
\gamma_{\Lambda_{c}\bar{D}_{s}^{*}}=
\gamma_{\Lambda_{c}\bar{D}_{s}},
\end{equation}
and
\begin{equation}
\gamma_{\Xi_{c}^{*}\bar{D}^{*}}= \gamma_{\Xi_{c}^{*}\bar{D}}=
\gamma_{\Xi'_{c}\bar{D}^{*}}= \gamma_{\Xi'_{c}\bar{D}}\approx
\gamma_{\Xi_{c}\bar{D}^{*}}= \gamma_{\Xi_{c}\bar{D}}.
\end{equation}
Combining the eigenvectors in the $nnc{\otimes}s\bar{c}$ and 
$nsc{\otimes}n\bar{c}$ configurations (see Tables~\ref{table:wavefunc:nnsc.c:I=1}--\ref{table:wavefunc:nnsc.c:I=0} of Appendix~\ref{App:eigenvec}), we can calculate the relative 
partial widths of different decay modes, as listed in
Tables~\ref{table:R:hidden:nnsc.c}--\ref{table:R:open:nsc:nnsc.c}.

From the eigenvectors, we find a new type of scattering state, which
consists of a charm baryon plus an anticharm meson.
The $P_{c,s}(4584.9,1,3/2^{-})$ has $82\%$ of the
$\Sigma_{c}\bar{D}_{s}^{*}$ component, while both the
$P_{c,s}(4636.2,1,3/2^{-})$ and $P_{c,s}(4644.3,1,5/2^{-})$ have
more than $85\%$ of the $\Sigma_{c}^{*}\bar{D}_{s}^{*}$ component.
Some other states, namely the $P_{c,s}(4386.6,0,1/2^{-})$,
$P_{c,s}(4387.3,0,3/2^{-})$, $P_{c,s}(4680.6,0,5/2^{-})$ and
$P_{c,s}(4442.8,1,1/2^{-})$ states, also have quite large fractions
of the color-singlet open charm components.
They are expected to be broad. But we still cannot rule out the
possibility that they are pentaquark states.
To obtain a more definite conclusion, one needs to consider the
dynamics inside the pentaquark, which is beyond the present work.

Two of the lowest $nnsc\bar{c}$ pentaquark states are the
$P_{c,s}(4197.4,0,1/2^{-})$ and $P_{c,s}(4208.6,0,1/2^{-})$.
From Fig.~\ref{fig:nnsccbar}, we see that they can only decay to
$\Lambda\eta_{c}$, thus they should have narrow widths.
However, their wave functions have large overlaps with the
$\Lambda{J/\psi}$, and their predicted masses are just below the
$\Lambda{J/\psi}$ threshold.
Considering the error of the present model, their masses can
probably be larger than the $\Lambda{J/\psi}$ threshold.
In that case, they will decay easily to $\Lambda{J/\psi}$ and be
broader.
Both $P_{c,s}(4386.6,0,1/2^{-})$ and $P_{c,s}(4387.3,0,3/2^{-})$
decay dominantly to $\Lambda{J/\psi}$.
But $P_{c,s}(4386.6,0,1/2^{-})$ can also decay to $\Lambda\eta_{c}$,
with
\begin{equation}
\frac{\Gamma[P_{c,s}(4386.6,0,1/2^{-}){\to}\Lambda\eta_{c}]}{\Gamma[P_{c,s}(4386.6,0,1/2^{-}){\to}\Lambda{J/\psi}]}
=0.87.
\end{equation}
The $P_{c,s}(4465.0,0,1/2^{-})$ and $P_{c,s}(4489.6,0,1/2^{-})$ have
the same quantum numbers and decay channels, but we can still use
their relative size of partial decay widths to distinguish them.
For $P_{c,s}(4465.0,0,1/2^{-})$, we have
\begin{equation}
\Gamma_{\Lambda{J/\psi}}: \Gamma_{\Lambda\eta_{c}}= 1:2.6,
\end{equation}
\begin{equation}
\Gamma_{\Lambda_{c}\bar{D}_{s}^{*}}:
\Gamma_{\Lambda_{c}\bar{D}_{s}}= 1:0.03,
\end{equation}
and
\begin{equation}
\Gamma_{\Xi_{c}'\bar{D}}: \Gamma_{\Xi_{c}\bar{D}^{*}}:
\Gamma_{\Xi_{c}\bar{D}}= 1:0:7.9.
\end{equation}
And the $P_{c,s}(4489.6,0,1/2^{-})$ has
\begin{equation}
\Gamma_{\Lambda{J/\psi}}: \Gamma_{\Lambda\eta_{c}}= 1:0.87,
\end{equation}
\begin{equation}
\Gamma_{\Lambda_{c}\bar{D}_{s}^{*}}:
\Gamma_{\Lambda_{c}\bar{D}_{s}}= 1:134.1,
\end{equation}
and
\begin{equation}
\Gamma_{\Xi_{c}'\bar{D}}: \Gamma_{\Xi_{c}\bar{D}^{*}}:
\Gamma_{\Xi_{c}\bar{D}}= 1:0.57:1.1.
\end{equation}
Its dominant decay mode is $\Lambda_{c}\bar{D}_{s}$.
For the $P_{c,s}(4501.5,0,3/2^{-})$,
its dominant decay modes are $\Lambda_{c}\bar{D}_{s}^{*}$ and
$\Xi_{c}\bar{D}^{*}$.
It can also decay to $\Lambda{J/\psi}$.
We also obtain three states above $4.6~\text{GeV}$.
We further study the $I=1$ $nnsc\bar{c}$ pentaquark states.
Their partial decay width ratios are also listed in
Tables~\ref{table:R:hidden:nnsc.c}--\ref{table:R:open:nsc:nnsc.c}.
There are three states above all meson-baryon thresholds
[$P_{c,s}(4680.6,0,5/2^{-})$ is not included since it is a
scattering state; see Fig.~\ref{fig:nnsccbar}].
They may be broad since they can decay freely to many open charm
channels.

Experimentally, three $P_c$ states have been observed in the
$N{J/\psi}$ channel. It is quite possible that the $P_{c,s}$ states
can be found in the $\Lambda{J/\psi}$ and $\Sigma^{(*)}{J/\psi}$
channels.
Moreover, we can also use open charm channels to search for these
states.
%

\subsubsection{The $ssnc\bar{c}$ and $sssc\bar{c}$ systems}
\label{sec:ssnc.c+sssc.c}

\begin{table*}
    \centering
    \caption{Pentaquark masses and eigenvectors of the $ssnc\bar{c}$ systems.
        The masses are all in units of MeV.}
    \label{table:mass:ssnc.c}
    \begin{tabular}{ccccccc}
        \toprule[1pt]
        \toprule[1pt]
        System & $J^{P}$ & Mass & Eigenvector & Scattering state \\
        \midrule[1pt]
        $ssnc\bar{c}$ & $\frac{1}{2}^{-}$ & $4288.0$ & $\{0.123,-0.021,-0.169,0.115,-0.019,-0.0001,0.003,-0.971\}$ & $\Xi\eta_{c}(4302)$ \\
        && $4406.0$ & $\{-0.043,-0.160,-0.095,-0.043,-0.146,0.004,0.969,0.015\}$ & $\Xi{J/\psi}(4415)$ \\
        && $4573.4$ & $\{0.222,-0.379,-0.781,0.179,-0.303,0.068,-0.171,0.199\}$ \\
        && $4604.7$ & $\{-0.050,0.239,-0.087,0.130,-0.231,-0.928,0.003,0.024\}$ & $\Xi^{*}{J/\psi}(4630)$ \\
        && $4621.7$ & $\{-0.700,-0.198,-0.167,-0.605,-0.205,-0.035,-0.136,-0.123\}$ \\
        && $4728.5$ & $\{0.157,-0.625,0.561,0.075,-0.492,-0.090,-0.111,-0.046\}$ \\
        && $4787.6$ & $\{0.479,-0.330,-0.051,-0.578,0.480,-0.306,0.010,-0.001\}$ \\
        && $4902.2$ & $\{-0.434,-0.484,-0.013,0.479,0.564,-0.174,0.007,0.004\}$ \\
        & $\frac{3}{2}^{-}$ & $4413.7$ & $\{-0.042,0.058,0.119,-0.038,-0.001,-0.003,-0.990\}$ & $\Xi{J/\psi}(4415)$ \\
        && $4509.4$ & $\{-0.141,0.018,0.008,0.100,-0.020,-0.985,0.007\}$ & $\Xi^{*}\eta_{c}(4517)$ \\
        && $4614.5$ & $\{0.548,-0.582,-0.350,0.469,0.041,-0.047,-0.118\}$ \\
        && $4630.6$ & $\{-0.027,-0.034,-0.007,0.073,-0.996,0.031,-0.004\}$ & $\Xi^{*}{J/\psi}(4630)$ \\
        && $4715.2$ & $\{0.374,0.804,-0.363,0.284,-0.015,-0.013,-0.023\}$ \\
        && $4769.1$ & $\{-0.460,-0.091,-0.849,-0.226,0.006,0.034,-0.079\}$ \\
        && $4819.0$ & $\{-0.570,0.033,0.0998,0.795,0.077,0.162,0.007\}$ \\
        & $\frac{5}{2}^{-}$ & $4631.7$ & $\{-0.006,0.99998\}$ & $\Xi^{*}{J/\psi}(4630)$ \\
        && $4790.0$ & $\{0.99998,0.006\}$ \\
        \bottomrule[1pt]
        \bottomrule[1pt]
    \end{tabular}
\end{table*}
%

\begin{table}
    \centering
    \caption{The partial width ratios for the hidden charm decays of
        the $ssnc\bar{c}$ pentaquark states. The masses are all in units of MeV.}
    \label{table:R:hidden:ssnc.c}
    \begin{tabular}{ccccccccccccccccccccc}
        \toprule[1pt]
        \toprule[1pt]
        $J^{P}$ & Mass
        & ${\Xi^{*}{J/\psi}}$
        & ${\Xi^{*}\eta_{c}}$
        & ${\Xi{J/\psi}}$
        & ${\Xi\eta_{c}}$ \\
        \midrule[1pt]
        $\frac{1}{2}^{-}$
        & $4573.4$ &0&&1&1.8 \\
        & $4621.7$ &0&&1&1.02 \\
        & $4728.5$ &0.38&&1&0.20 \\
        & $4787.6$ &1&&0.001&0.00003 \\
        & $4902.2$ &1&&0.002&0.0006 \\
        $\frac{3}{2}^{-}$
        & $4614.5$ &0&0.11&1& \\
        & $4715.2$ &0.23&0.26&1& \\
        & $4769.1$ &0.004&0.16&1& \\
        & $4819.0$ &1&5.7&0.01& \\
        $\frac{5}{2}^{-}$
        & $4790.0$ &1&&& \\
        \bottomrule[1pt]
        \bottomrule[1pt]
    \end{tabular}
\end{table}
%
\begin{table}
    \centering
    \caption{The partial width ratios for the $ssc{\otimes}n\bar{c}$ open
    charm decays of the $ssnc\bar{c}$ pentaquark states. The masses are all in units of MeV.}
    \label{table:R:open:ssc:ssnc.c}
    \begin{tabular}{ccccccccccccccccccccc}
        \toprule[1pt]
        \toprule[1pt]
        $J^{P}$ & Mass
        & ${\Omega_{c}^{*}\bar{D}^{*}}$
        & ${\Omega_{c}^{*}\bar{D}}$
        & ${\Omega_{c}\bar{D}^{*}}$
        & ${\Omega_{c}\bar{D}}$ \\
        \midrule[1pt]
        $\frac{1}{2}^{-}$
        & $4573.4$ &0&&0&1 \\
        & $4621.7$ &0&&0&1 \\
        & $4728.5$ &0&&2.5&1 \\
        & $4787.6$ &0.10&&3.9&1 \\
        & $4902.2$ &46.0&&4.6&1 \\
        $\frac{3}{2}^{-}$
        & $4614.5$ &0&0&0& \\
        & $4715.2$ &0&1&10.3& \\
        & $4769.1$ &0&1&0.72& \\
        & $4819.0$ &2.1&1&0.33& \\
        $\frac{5}{2}^{-}$
        & $4790.0$ &1&&& \\
        \bottomrule[1pt]
        \bottomrule[1pt]
    \end{tabular}
\end{table}
%
\begin{table}
    \centering
    \caption{The partial width ratios for the $nsc{\otimes}s\bar{c}$ open
    charm decays of the $ssnc\bar{c}$ pentaquark states. The masses are all in units of MeV.}
    \label{table:R:open:nsc:ssnc.c}
    \begin{tabular}{ccccccccccccccccccccc}
        \toprule[1pt]
        \toprule[1pt]
        $J^{P}$ & Mass
        & ${\Xi_{c}^{*}\bar{D}_{s}^{*}}$
        & ${\Xi_{c}^{*}\bar{D}_{s}}$
        & ${\Xi_{c}'\bar{D}_{s}^{*}}$
        & ${\Xi_{c}'\bar{D}_{s}}$
        & ${\Xi_{c}\bar{D}_{s}^{*}}$
        & ${\Xi_{c}\bar{D}_{s}}$ \\
        \midrule[1pt]
        $\frac{1}{2}^{-}$
        & $4573.4$ &0&&0&1&0&0.003 \\
        & $4621.7$ &0&&0&0.0005&1&28.1 \\
        & $4728.5$ &0&&15.3&1&67.4&5.9 \\
        & $4787.6$ &0.04&&3.0&1&0.09&0.0009 \\
        & $4902.2$ &42.4&&5.0&1&0.06&0.19 \\
        $\frac{3}{2}^{-}$
        & $4614.5$ &0&1&0&&14.5& \\
        & $4715.2$ &0&1&35.9&&31.5& \\
        & $4769.1$ &12.9&1&1.7&&14.9& \\
        & $4819.0$ &1.0&1&0.29&&2.2& \\
        $\frac{5}{2}^{-}$
        & $4790.0$ &1&&&&& \\
        \bottomrule[1pt]
        \bottomrule[1pt]
    \end{tabular}
\end{table}
%

\begin{table}
    \centering
    \caption{Pentaquark masses and eigenvectors of the $sssc\bar{c}$ systems.
        The masses are all in units of MeV.}
    \label{table:mass:sssc.c}
    \begin{tabular}{ccccccc}
        \toprule[1pt]
        \toprule[1pt]
        System & $J^{P}$ & Mass & Eigenvector & Scattering state \\
        \midrule[1pt]
        $sssc\bar{c}$ & $\frac{1}{2}^{-}$ & $4736.0$ & $\{0.164,-0.386,0.908\}$ & $\Omega{J/\psi}(4769)$ \\
        && $4894.4$ & $\{0.756,-0.542,-0.367\}$ \\
        && $5009.4$ & $\{0.633,0.747,0.203\}$ \\
        & $\frac{3}{2}^{-}$ & $4645.1$ & $\{-0.190,-0.021,-0.982\}$ & $\Omega\eta_{c}(4656)$ \\
        && $4767.5$ & $\{-0.082,-0.996,0.037\}$ & $\Omega{J/\psi}(4769)$ \\
        && $4924.1$ & $\{0.978,-0.087,-0.187\}$ \\
        & $\frac{5}{2}^{-}$ & $4768.6$ & $\{1\}$ & $\Omega{J/\psi}(4769)$ \\
        \bottomrule[1pt]
        \bottomrule[1pt]
    \end{tabular}
\end{table}
%

\begin{table}
    \centering
    \caption{The partial width ratios for the hidden charm decays of
    the $sssc\bar{c}$ pentaquark states. The masses are all in units of MeV.}
    \label{table:R:hidden:sssc.c}
    \begin{tabular}{ccccccccccccc}
        \toprule[1pt]
        \toprule[1pt]
        $J^P$ & Mass
        & ${\Omega{J/\psi}}$
        & ${\Omega\eta_{c}}$ \\
        \midrule[1pt]
        $\frac{1}{2}^{-}$
        & $4894.4$ & $1$ & \\
        & $5009.4$ & $1$ & \\
        $\frac{3}{2}^{-}$
        & $4924.1$ & $1$ & $6.1$ \\
        \bottomrule[1pt]
        \bottomrule[1pt]
    \end{tabular}
\end{table}
%
\begin{table}
    \centering
    \caption{The partial width ratios for the open charm decays of
    the $sssc\bar{c}$ pentaquark states. The masses are all in units of MeV.}
    \label{table:R:open:sssc.c}
    \begin{tabular}{ccccccccccccc}
        \toprule[1pt]
        \toprule[1pt]
        $J^P$ & Mass
        & ${\Omega_{c}^{*}\bar{D}_{s}^{*}}$
        & ${\Omega_{c}^{*}\bar{D}_{s}}$
        & ${\Omega_{c}\bar{D}_{s}^{*}}$
        & ${\Omega_{c}\bar{D}_{s}}$ \\
        \midrule[1pt]
        $\frac{1}{2}^{-}$
        & $4894.4$ & $0.01$ && $1$ & $0.2$ \\
        & $5009.4$ & $10.2$ && $1$ & $0.2$ \\
        $\frac{3}{2}^{-}$
        & $4924.1$ & $4.3$ & $3.3$ & $1$ & \\
        \bottomrule[1pt]
        \bottomrule[1pt]
    \end{tabular}
\end{table}
%

\begin{figure*}
	\includegraphics[width=400pt]{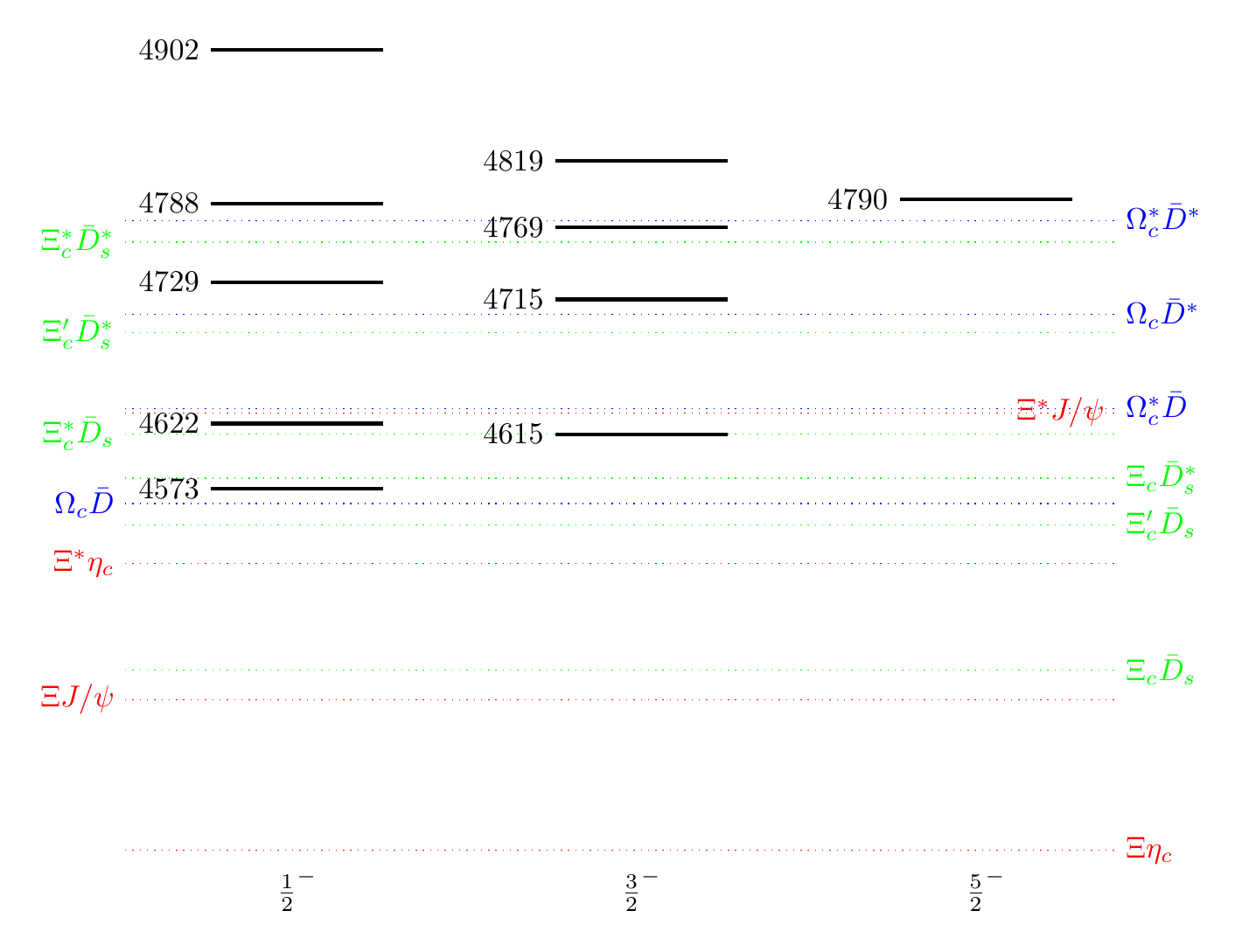}
	\caption{Mass spectra of the $ssnc\bar{c}$ pentaquark states.
		The dotted lines indicate various meson-baryon thresholds.
		The masses are all in units of MeV.}
	\label{fig:ssnccbar}
\end{figure*}
%
\begin{figure*}
	\includegraphics[width=400pt]{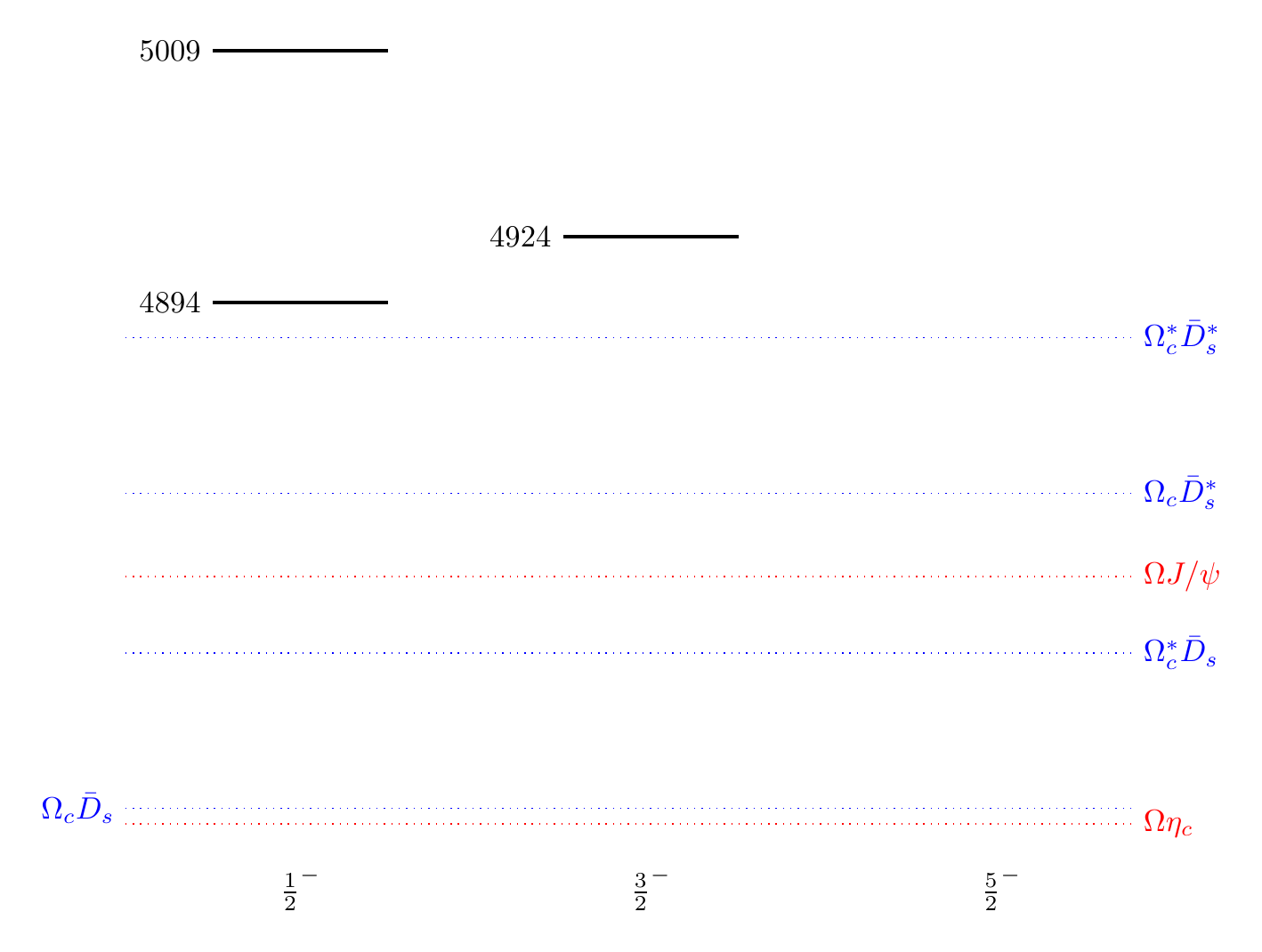}
	\caption{Mass spectra of the $sssc\bar{c}$ pentaquark states.
		The dotted lines indicate various meson-baryon thresholds.
		The masses are all in units of MeV.}
	\label{fig:sssccbar}
\end{figure*}

The $ssnc\bar{c}$ system is similar to the $I=1$ $nnsc\bar{c}$
system.
We present their mass spectra in Table~\ref{table:mass:ssnc.c}.
As indicated in the last column, we reproduce the scattering states
of $\Xi\eta_{c}$ ($4288.0~\text{MeV}$ with $J^{P}=1/2^{-}$),
$\Xi{J/\psi}$ ($4406.0~\text{MeV}$ with $J^{P}=1/2^{-}$ and
$4413.7~\text{MeV}$ with $J^{P}=3/2^{-}$), $\Xi^{*}\eta_{c}$
($4509.4~\text{MeV}$ with $J^{P}=3/2^{-}$) and $\Xi^{*}{J/\psi}$
($4604.7~\text{MeV}$ with $J^{P}=1/2^{-}$, $4630.6~\text{MeV}$ with
$J^{P}=3/2^{-}$ and $4631.7~\text{MeV}$ with $J^{P}=5/2^{-}$).
In the following, we will use $P_{c,ss}(m,J^{P})$ to denote the
$ssnc\bar{c}$ pentaquark.

We plot the relative position of the $ssnc\bar{c}$ pentaquark states
and all the relevant meson-baryon thresholds in
Fig.~\ref{fig:ssnccbar}.
We also transform the eigenvectors to the $ssc{\otimes}n\bar{c}$ and
$nsc{\otimes}n\bar{c}$ configurations (see Table~\ref{table:wavefunc:ssnc.c} of Appendix~\ref{App:eigenvec}).
The only state with $J^{P}=5/2^{-}$, $P_{c,ss}(4790.0,5/2^{-})$,
lies over all thresholds and
\begin{equation}
P_{c,ss}(4790.0,5/2^{-})
=0.94487\Omega_{c}^{*}{\otimes}\bar{D}^{*}+\cdots.
\end{equation}
It is a scattering state of $\Omega_{c}^{*}\bar{D}^{*}$.
Its dominant decay mode is $\Omega_{c}^{*}\bar{D}^{*}$ and it should
be broad.
Similar to the $nnnc\bar{c}$ and $nnsc\bar{c}$, for each
$ssnc\bar{c}$ pentaquark state,
\begin{equation}
\gamma_{\Xi^*{J/\psi}}= \gamma_{\Xi^*\eta_{c}}=
\gamma_{\Xi{J/\psi}}= \gamma_{\Xi\eta_{c}},
\end{equation}
\begin{equation}
\gamma_{\Omega_{c}^{*}\bar{D}^{*}}= \gamma_{\Omega_{c}^{*}\bar{D}}=
\gamma_{\Omega_{c}\bar{D}^{*}}= \gamma_{\Omega_{c}\bar{D}},
\end{equation}
\begin{equation}
\gamma_{\Xi_{c}^{*}\bar{D}_{s}^{*}}=
\gamma_{\Xi_{c}^{*}\bar{D}_{s}}= \gamma_{\Xi'_{c}\bar{D}_{s}^{*}}=
\gamma_{\Xi'_{c}\bar{D}_{s}}\approx \gamma_{\Xi_{c}\bar{D}_{s}^{*}}=
\gamma_{\Xi_{c}\bar{D}_{s}}.
\end{equation}
The calculated partial decay width ratios are listed in
Tables~\ref{table:R:hidden:ssnc.c}--\ref{table:R:open:nsc:ssnc.c}.

The last class of the hidden-charm pentaquark is the $sssc\bar{c}$
system.
They are similar to the $nnnc\bar{c}$ states with isospin $I=3/2$.
We present their mass spectra in Table~\ref{table:mass:sssc.c}.
We find three scattering states ($4736.0~\text{MeV}$ with
$J^{P}=1/2^{-}$, $4767.5~\text{MeV}$ with $J^{P}=3/2^{-}$ and
$4768.6~\text{MeV}$ with $J^{P}=5/2^{-}$) which couple very strongly
to the $\Omega{J/\psi}$ and a scattering state ($4645.1~\text{MeV}$
with $J^{P}=3/2^{-}$) which couples strongly to the
$\Omega\eta_{c}$.
We will focus on the other $sssc\bar{c}$ pentaquark states.
To study their decay properties, we transform their wave functions
to the $ssc{\otimes}s\bar{c}$ configuration (see
Table~\ref{table:wavefunc:sssc.c} of Appendix~\ref{App:eigenvec}).
And we also plot their relative position in
Fig.~\ref{fig:sssccbar}, along with all possible decay channels.
We find that they are all above the open charm thresholds and have
large overlap with the $\Omega_{c}^{(*)}\otimes\bar{D}_{s}^{(*)}$
component.
Thus they should all be very broad.
The partial decay width ratios can be found in
Tables~\ref{table:R:hidden:sssc.c}--\ref{table:R:open:sssc.c}.
%

\section{Conclusions}
\label{Sec:Conclusion}

In this work, we have systematically studied the mass spectrum of
the hidden charm pentaquark in the framework of an extended
chromomagnetic model.
In addition to the chromomagnetic interaction, the effect of color
interaction is also considered in this model.
With the eigenvectors obtained, we have further investigated the
decay properties of the pentaquarks.

For the $nnnc\bar{c}$ pentaquark with $I=1/2$, we find that the
masses of the experimentally observed $P_{c}$ states are compatible
with such pentaquark states.
The lowest state $P_{c}(4327.0,1/2,1/2^{-})$ corresponds to the
$P_{c}(4312)$.
This state has two hidden charm channels, namely the $N{J/\psi}$ and
$N\eta_{c}$ channels.
And its partial decay width of the $N\eta_{c}$ mode is larger than
that of the $N{J/\psi}$ mode.
In the open charm decay channel, $P_{c}(4327.0,1/2,1/2^{-})$ decays
dominantly to the $\Lambda_{c}\bar{D}^{*}$ mode.
We hope the future experiments can search for the $P_{c}(4312)$ in
the $N\eta_{c}$ and $\Lambda_{c}\bar{D}^{*}$ channels.

There are two states, $P_{c}(4367.4,1/2,3/2^{-})$ and
$P_{c}(4372.4,1/2,1/2^{-})$, in the vicinity of the $P_{c}(4380)$.
$P_{c}(4367.4,1/2,3/2^{-})$ decays into the $N{J/\psi}$ and
$\Lambda_{c}\bar{D}^{*}$ modes, while the other hidden-charm (like
$N\eta_{c}$) or open charm decay modes are all suppressed.
Its partner state, $P_{c}(4372.4,1/2,1/2^{-})$ can decay into both
$N{J/\psi}$ and $N\eta_{c}$ modes.
And their partial decay widths are comparable.
In the open charm channel, $P_{c}(4372.4,1/2,1/2^{-})$ decays
dominantly to the $\Lambda_{c}\bar{D}$ mode.
If $P_{c}(4380)$ truly corresponds to the
$P_{c}(4367.4,1/2,3/2^{-})$, this partner state should also exit,
which can be searched for in future experiments.

In the higher mass region, we find $P_{c}(4476.3,1/2,3/2^{-})$ and
$P_{c}(4480.9,1/2,1/2^{-})$.
They may correspond to the $P_{c}(4440)$ and $P_{c}(4457)$,
respectively.
Both of them couple weakly to the hidden charm channel(s).
Note that the former state can only decay to $N{J/\psi}$ while the
latter state can also decay to $N\eta_{c}$, which can be used to
distinguish the two states.
In the open charm channels, both of them decay dominantly to the
$\Lambda_{c}\bar{D}^{*}$.
And the $P_{c}(4476.3,1/2,3/2^{-})$ can also decay to
$\Sigma_{c}^{*}\bar{D}$ with a not-so-small fraction.

Moreover, we predict two states above $4.5~\text{GeV}$, namely
$P_{c}(4524.5,1/2,3/2^{-})$ and $P_{c}(4546.0,1/2,5/2^{-})$.
Like the observed $P_{c}$ states, $P_{c}(4524.5,1/2,3/2^{-})$ can
also be observed in the $N{J/\psi}$ channel.
In the open charm channel, it decays dominantly into
$\Lambda_{c}\bar{D}^{*}$, while the $\Sigma_{c}^{*}\bar{D}$,
$\Sigma_{c}\bar{D}^{*}$ modes are also important.
On the other hand, $P_{c}(4546.0,1/2,5/2^{-})$ can only decay to
$\Sigma_{c}^{*}\bar{D}^{*}$, all other decay modes are suppressed.

There are three $nnnc\bar{c}$ pentaquark states with $I=3/2$, their
masses are all over $4.6~\text{GeV}$.
They can decay into the $\Delta{J/\psi}$ channel, while
$P_{c}(4633.0,3/2,3/2^{-})$ can also decays to ${\Delta\eta_{c}}$.
In the open charm channel,
$P_{c}(4601.9,3/2,1/2^{-})$ decays dominantly to the
${\Sigma_{c}\bar{D}^{*}}$ and ${\Sigma_{c}\bar{D}}$ modes,
$P_{c}(4717.1,3/2,1/2^{-})$ decays dominantly to the
${\Sigma_{c}^{*}\bar{D}^{*}}$ and ${\Sigma_{c}\bar{D}^{*}}$ modes,
and $P_{c}(4633.0,3/2,3/2^{-})$ can decay to the
${\Sigma_{c}^{*}\bar{D}^{*}}$, ${\Sigma_{c}^{*}\bar{D}}$, and
${\Sigma_{c}\bar{D}^{*}}$ modes.

We have also used this model to explore the $nnsc\bar{c}$,
$ssnc\bar{c}$, and $sssc\bar{c}$ pentaquark states.
With the obtained eigenvectors, we further explore the hidden and
open charm decays of these pentaquark states.
We hope that future experiments in LHCb and other collaborations can
search for these states.
%

\section*{Acknowledgments}

X.~Z.~W. is grateful to G.~J.~Wang and L.~Meng for helpful comments and discussions.
This project is supported by the National Natural Science Foundation
of China under Grants No.~11575008, No.~11621131001 and National Key Basic
Research Program of China (2015CB856700).
%

\begin{appendix}

\section{The pentaquark wave functions}
\label{App:wavefunc}

In this section, we construct the pentaquark wave functions in the 
$(q_1q_2{\otimes}q_{3})\otimes(q_{4}\bar{q}_{5})$ configuration.
In principle, the total wave function is a direct product of the
orbital, color, spin and flavor bases.
Since we only consider the ground states, the orbital wave function
is symmetric and irrelevant for the effective Hamiltonian [see
Eq.~(\ref{eqn:hamiltonian:final})].
Moreover, the Hamiltonian does not contain a flavor operator
explicitly.
Thus we first construct the color-spin wave function, and then
incorporate the flavor wave function to account for the Pauli
principle.

The spins of the pentaquark states can be $1/2$, $3/2$, and $5/2$.
In the $(qq{\otimes}q){\otimes}q\bar{q}$ configuration, the possible
color-spin wave functions are listed as follows,
\begin{enumerate}
	\item $J^{P}=1/2^{-}$:
	\begin{align}\label{eqn:wavefunc:colorspin:1/2}
		\beta^{1/2}_{1} &=
		\ket{[(q_1q_2)_{1}^{6}q_3]_{3/2}^{8}(q_4\bar{q}_5)_1^{8}}_{1/2},
		\notag\\
		\beta^{1/2}_{2} &=
		\ket{[(q_1q_2)_{1}^{6}q_3]_{1/2}^{8}(q_4\bar{q}_5)_1^{8}}_{1/2},
		\notag\\
		\beta^{1/2}_{3} &=
		\ket{[(q_1q_2)_{0}^{6}q_3]_{1/2}^{8}(q_4\bar{q}_5)_1^{8}}_{1/2},
		\notag\\
		\beta^{1/2}_{4} &=
		\ket{[(q_1q_2)^{6}_{1}q_3]_{1/2}^{8}(q_4\bar{q}_5)_0^{8}}_{1/2},
		\notag\\
		\beta^{1/2}_{5} &=
		\ket{[(q_1q_2)_{0}^{6}q_3]_{1/2}^{8}(q_4\bar{q}_5)_0^{8}}_{1/2},
		\notag\\
		\beta^{1/2}_{6} &=
		\ket{[(q_1q_2)_{1}^{\bar{3}}q_3]_{3/2}^{8}(q_4\bar{q}_5)_1^{8}}_{1/2},
		\notag\\
		\beta^{1/2}_{7} &=
		\ket{[(q_1q_2)_{1}^{\bar{3}}q_3]_{1/2}^{8}(q_4\bar{q}_5)_1^{8}}_{1/2},
		\notag\\
		\beta^{1/2}_{8} &=
		\ket{[(q_1q_2)_{0}^{\bar{3}}q_3]_{1/2}^{8}(q_4\bar{q}_5)_1^{8}}_{1/2},
		\notag\\
		\beta^{1/2}_{9} &=
		\ket{[(q_1q_2)_{1}^{\bar{3}}q_3]_{1/2}^{8}(q_4\bar{q}_5)_0^{8}}_{1/2},
		\notag\\
		\beta^{1/2}_{10} &=
		\ket{[(q_1q_2)_{0}^{\bar{3}}q_3]_{1/2}^{8}(q_4\bar{q}_5)_0^{8}}_{1/2},
		\notag\\
		\beta^{1/2}_{11} &=
		\ket{[(q_1q_2)_{1}^{\bar{3}}q_3]_{3/2}^{1}(q_4\bar{q}_5)_1^{1}}_{1/2},
		\notag\\
		\beta^{1/2}_{12} &=
		\ket{[(q_1q_2)_{1}^{\bar{3}}q_3]_{1/2}^{1}(q_4\bar{q}_5)_1^{1}}_{1/2},
		\notag\\
		\beta^{1/2}_{13} &=
		\ket{[(q_1q_2)_{0}^{\bar{3}}q_3]_{1/2}^{1}(q_4\bar{q}_5)_1^{1}}_{1/2},
		\notag\\
		\beta^{1/2}_{14} &=
		\ket{[(q_1q_2)_{1}^{\bar{3}}q_3]_{1/2}^{1}(q_4\bar{q}_5)_0^{1}}_{1/2},
		\notag\\
		\beta^{1/2}_{15} &=
		\ket{[(q_1q_2)_{0}^{\bar{3}}q_3]_{1/2}^{1}(q_4\bar{q}_5)_0^{1}}_{1/2},
	\end{align}
	\item $J^{P}=3/2^{-}$:
	\begin{align}\label{eqn:wavefunc:colorspin:3/2}
		\beta^{3/2}_{1} &=
		\ket{[(q_1q_2)_{1}^{6}q_3]_{3/2}^{8}(q_4\bar{q}_5)_1^{8}}_{3/2},
		\notag\\
		\beta^{3/2}_{2} &=
		\ket{[(q_1q_2)_{1}^{6}q_3]_{3/2}^{8}(q_4\bar{q}_5)_0^{8}}_{3/2},
		\notag\\
		\beta^{3/2}_{3} &=
		\ket{[(q_1q_2)_{1}^{6}q_3]_{1/2}^{8}(q_4\bar{q}_5)_1^{8}}_{3/2},
		\notag\\
		\beta^{3/2}_{4} &=
		\ket{[(q_1q_2)_{0}^{6}q_3]_{1/2}^{8}(q_4\bar{q}_5)_1^{8}}_{3/2},
		\notag\\
		\beta^{3/2}_{5} &=
		\ket{[(q_1q_2)_{1}^{\bar{3}}q_3]_{3/2}^{8}(q_4\bar{q}_5)_1^{8}}_{3/2},
		\notag\\
		\beta^{3/2}_{6} &=
		\ket{[(q_1q_2)_{1}^{\bar{3}}q_3]_{3/2}^{8}(q_4\bar{q}_5)_0^{8}}_{3/2},
		\notag\\
		\beta^{3/2}_{7} &=
		\ket{[(q_1q_2)_{1}^{\bar{3}}q_3]_{1/2}^{8}(q_4\bar{q}_5)_1^{8}}_{3/2},
		\notag\\
		\beta^{3/2}_{8} &=
		\ket{[(q_1q_2)_{0}^{\bar{3}}q_3]_{1/2}^{8}(q_4\bar{q}_5)_1^{8}}_{3/2},
		\notag\\
		\beta^{3/2}_{9} &=
		\ket{[(q_1q_2)_{1}^{\bar{3}}q_3]_{3/2}^{1}(q_4\bar{q}_5)_1^{1}}_{3/2},
		\notag\\
		\beta^{3/2}_{10} &=
		\ket{[(q_1q_2)_{1}^{\bar{3}}q_3]_{3/2}^{1}(q_4\bar{q}_5)_0^{1}}_{3/2},
		\notag\\
		\beta^{3/2}_{11} &=
		\ket{[(q_1q_2)_{1}^{\bar{3}}q_3]_{1/2}^{1}(q_4\bar{q}_5)_1^{1}}_{3/2},
		\notag\\
		\beta^{3/2}_{12} &=
		\ket{[(q_1q_2)_{0}^{\bar{3}}q_3]_{1/2}^{1}(q_4\bar{q}_5)_1^{1}}_{3/2},
	\end{align}
	\item $J^{P}=5/2^{-}$:
	\begin{align}\label{eqn:wavefunc:colorspin:5/2}
		\beta^{5/2}_{1} =&
		\ket{[(q_1q_2)_{1}^{6}q_3]_{3/2}^{8}(q_4\bar{q}_5)_1^{8}}_{5/2},
		\notag\\
		\beta^{5/2}_{2} =&
		\ket{[(q_1q_2)_{1}^{\bar{3}}q_3]_{3/2}^{8}(q_4\bar{q}_5)_1^{8}}_{5/2},
		\notag\\
		\beta^{5/2}_{3} =&
		\ket{[(q_1q_2)_{1}^{\bar{3}}q_3]_{3/2}^{1}(q_4\bar{q}_5)_1^{1}}_{5/2},
	\end{align}
\end{enumerate}
where the superscript $1$, $\bar{3}$, $6$, or $8$ denotes the color,
and the subscript denotes the spin $0$, $1$, $1/2$, $3/2$, or $5/2$.
These wave functions have definite symmetry under the exchange of
the first two quarks.
$(q_1q_2)_{1}^{6}$ and $(q_1q_2)_{0}^{\bar{3}}$ are symmetric, while
$(q_1q_2)_{1}^{\bar{3}}$ and $(q_1q_2)_{0}^{6}$ are antisymmetric.

Next we consider the flavor wave function.
Taking the Pauli principle into account, we can obtain four types of
total wave functions.
\begin{enumerate}
	\item Type A [$\text{Flavor}=\{(nnsQ\bar{Q})^{I=1},ssnQ\bar{Q}\}$]:
	\begin{enumerate}
		\item $J^{P}=1/2^{-}$:
		\begin{align}\label{eqn:wavefunc:total:A1/2}
			\Psi_{A1}^{1/2} = q_1q_2q_3'Q_4\bar{Q}_5\otimes\beta^{1/2}_{3},
			\notag\\
			\Psi_{A2}^{1/2} = q_1q_2q_3'Q_4\bar{Q}_5\otimes\beta^{1/2}_{5},
			\notag\\
			\Psi_{A3}^{1/2} = q_1q_2q_3'Q_4\bar{Q}_5\otimes\beta^{1/2}_{6},
			\notag\\
			\Psi_{A4}^{1/2} = q_1q_2q_3'Q_4\bar{Q}_5\otimes\beta^{1/2}_{7},
			\notag\\
			\Psi_{A5}^{1/2} = q_1q_2q_3'Q_4\bar{Q}_5\otimes\beta^{1/2}_{9},
			\notag\\
			\Psi_{A6}^{1/2} = q_1q_2q_3'Q_4\bar{Q}_5\otimes\beta^{1/2}_{11},
			\notag\\
			\Psi_{A7}^{1/2} = q_1q_2q_3'Q_4\bar{Q}_5\otimes\beta^{1/2}_{12},
			\notag\\
			\Psi_{A8}^{1/2} = q_1q_2q_3'Q_4\bar{Q}_5\otimes\beta^{1/2}_{14},
		\end{align}
		\item $J^{P}=3/2^{-}$:
		\begin{align}\label{eqn:wavefunc:total:A3/2}
			\Psi_{A1}^{3/2} = q_1q_2q_3'Q_4\bar{Q}_5\otimes\beta^{3/2}_{4},
			\notag\\
			\Psi_{A2}^{3/2} = q_1q_2q_3'Q_4\bar{Q}_5\otimes\beta^{3/2}_{5},
			\notag\\
			\Psi_{A3}^{3/2} = q_1q_2q_3'Q_4\bar{Q}_5\otimes\beta^{3/2}_{6},
			\notag\\
			\Psi_{A4}^{3/2} = q_1q_2q_3'Q_4\bar{Q}_5\otimes\beta^{3/2}_{7},
			\notag\\
			\Psi_{A5}^{3/2} = q_1q_2q_3'Q_4\bar{Q}_5\otimes\beta^{3/2}_{9},
			\notag\\
			\Psi_{A6}^{3/2} = q_1q_2q_3'Q_4\bar{Q}_5\otimes\beta^{3/2}_{10},
			\notag\\
			\Psi_{A7}^{3/2} = q_1q_2q_3'Q_4\bar{Q}_5\otimes\beta^{3/2}_{11},
		\end{align}
		\item $J^{P}=5/2^{-}$:
		\begin{align}\label{eqn:wavefunc:total:A5/2}
			\Psi_{A1}^{5/2} = q_1q_2q_3'Q_4\bar{Q}_5\otimes\beta^{5/2}_{2},
			\notag\\
			\Psi_{A2}^{5/2} = q_1q_2q_3'Q_4\bar{Q}_5\otimes\beta^{5/2}_{3},
		\end{align}
	\end{enumerate}
	\item Type B [$\text{Flavor}=(nnsQ\bar{Q})^{I=0}$]:
	\begin{enumerate}
		\item $J^{P}=1/2^{-}$:
		\begin{align}\label{eqn:wavefunc:total:B1/2}
			\Psi_{B1}^{1/2} = q_1q_2q_3'Q_4\bar{Q}_5\otimes\beta^{1/2}_{1},
			\notag\\
			\Psi_{B2}^{1/2} = q_1q_2q_3'Q_4\bar{Q}_5\otimes\beta^{1/2}_{2},
			\notag\\
			\Psi_{B3}^{1/2} = q_1q_2q_3'Q_4\bar{Q}_5\otimes\beta^{1/2}_{4},
			\notag\\
			\Psi_{B4}^{1/2} = q_1q_2q_3'Q_4\bar{Q}_5\otimes\beta^{1/2}_{8},
			\notag\\
			\Psi_{B5}^{1/2} = q_1q_2q_3'Q_4\bar{Q}_5\otimes\beta^{1/2}_{10},
			\notag\\
			\Psi_{B6}^{1/2} = q_1q_2q_3'Q_4\bar{Q}_5\otimes\beta^{1/2}_{13},
			\notag\\
			\Psi_{B7}^{1/2} = q_1q_2q_3'Q_4\bar{Q}_5\otimes\beta^{1/2}_{15},
		\end{align}
		\item $J^{P}=3/2^{-}$:
		\begin{align}\label{eqn:wavefunc:total:B3/2}
			\Psi_{B1}^{3/2} = q_1q_2q_3'Q_4\bar{Q}_5\otimes\beta^{3/2}_{1},
			\notag\\
			\Psi_{B2}^{3/2} = q_1q_2q_3'Q_4\bar{Q}_5\otimes\beta^{3/2}_{2},
			\notag\\
			\Psi_{B3}^{3/2} = q_1q_2q_3'Q_4\bar{Q}_5\otimes\beta^{3/2}_{3},
			\notag\\
			\Psi_{B4}^{3/2} = q_1q_2q_3'Q_4\bar{Q}_5\otimes\beta^{3/2}_{8},
			\notag\\
			\Psi_{B5}^{3/2} = q_1q_2q_3'Q_4\bar{Q}_5\otimes\beta^{3/2}_{12},
		\end{align}
		\item $J^{P}=5/2^{-}$:
		\begin{align}\label{eqn:wavefunc:total:B5/2}
			\Psi_{B1}^{5/2} = q_1q_2q_3'Q_4\bar{Q}_5\otimes\beta^{5/2}_{1},
		\end{align}
	\end{enumerate}
	\item Type C [$\text{Flavor}=\{(nnnQ\bar{Q})^{I=3/2},sssQ\bar{Q}\}$]:
	\begin{enumerate}
		\item $J^{P}=1/2^{-}$:
		\begin{align}\label{eqn:wavefunc:total:C1/2}
			&\Psi_{C1}^{1/2}= q_1q_2q_3Q_4\bar{Q}_5\otimes \frac{1}{\sqrt{2}}
			\left(\beta^{1/2}_3-\beta^{1/2}_7\right),
			\notag\\
			&\Psi_{C2}^{1/2}= q_1q_2q_3Q_4\bar{Q}_5\otimes \frac{1}{\sqrt{2}}
			\left(\beta^{1/2}_5-\beta^{1/2}_9\right),
			\notag\\
			&\Psi_{C3}^{1/2}= q_1q_2q_3Q_4\bar{Q}_5\otimes \beta^{1/2}_{11},
			%
		\end{align}
		\item $J^{P}=3/2^{-}$:
		\begin{align}\label{eqn:wavefunc:total:C3/2}
			&\Psi_{C1}^{3/2}= q_1q_2q_3Q_4\bar{Q}_5\otimes \frac{1}{\sqrt{2}}
			\left(\beta^{3/2}_4-\beta^{3/2}_7\right),
			\notag\\
			&\Psi_{C2}^{3/2}= q_1q_2q_3Q_4\bar{Q}_5\otimes \beta^{3/2}_9,
			\notag\\
			&\Psi_{C3}^{3/2}= q_1q_2q_3Q_4\bar{Q}_5\otimes \beta^{3/2}_{10},
		\end{align}
		\item $J^{P}=5/2^{-}$:
		\begin{equation}\label{eqn:wavefunc:total:C5/2}
			\Psi_{C1}^{5/2}= q_1q_2q_3Q_4\bar{Q}_5\otimes \beta^{5/2}_3,
		\end{equation}
	\end{enumerate}
	\item Type D [$\text{Flavor}=(nnnQ\bar{Q})^{I=1/2}$]:
	\begin{enumerate}
		\item $J^{P}=1/2^{-}$:
		\begin{align}\label{eqn:wavefunc:total:D1/2}
			\Psi_{D1}^{1/2} =& \frac{1}{\sqrt{2}} \Big(
			\{n_1n_2\}n_3Q_4\bar{Q}_5\otimes\beta_6^{1/2}
			\notag\\
			&\qquad- [n_1n_2]n_3Q_4\bar{Q}_5\otimes\beta_1^{1/2} \Big),
			\notag\\
			\Psi_{D2}^{1/2} =& \frac{1}{2} \Big[ \{n_1n_2\}n_3Q_4\bar{Q}_5
			\otimes \left(\beta_3^{1/2}+\beta_7^{1/2}\right)
			\notag\\
			&\qquad+ [n_1n_2]n_3Q_4\bar{Q}_5 \otimes
			\left(\beta_2^{1/2}-\beta_8^{1/2}\right) \Big],
			\notag\\
			\Psi_{D3}^{1/2} =& \frac{1}{2} \Big[ \{n_1n_2\}n_3Q_4\bar{Q}_5
			\otimes \left(\beta_5^{1/2}+\beta_9^{1/2}\right)
			\notag\\
			&\qquad+ [n_1n_2]n_3Q_4\bar{Q}_5 \otimes
			\left(\beta_4^{1/2}-\beta_{10}^{1/2}\right) \Big],
			\notag\\
			\Psi_{D4}^{1/2} =& \frac{1}{\sqrt{2}} \Big(
			\{n_1n_2\}n_3Q_4\bar{Q}_5\otimes\beta_{12}^{1/2}
			\notag\\
			&\qquad+ [n_1n_2]n_3Q_4\bar{Q}_5\otimes\beta_{13}^{1/2} \Big),
			\notag\\
			\Psi_{D5}^{1/2} =& \frac{1}{\sqrt{2}} \Big(
			\{n_1n_2\}n_3Q_4\bar{Q}_5\otimes\beta_{14}^{1/2}
			\notag\\
			&\qquad+ [n_1n_2]n_3Q_4\bar{Q}_5\otimes\beta_{15}^{1/2} \Big),
		\end{align}
		\item $J^{P}=3/2^{-}$:
		\begin{align}\label{eqn:wavefunc:total:D3/2}
			\Psi_{D1}^{3/2} =& \frac{1}{\sqrt{2}} \Big(
			\{n_1n_2\}n_3Q_4\bar{Q}_5\otimes\beta_5^{3/2}
			\notag\\
			&\qquad- [n_1n_2]n_3Q_4\bar{Q}_5\otimes\beta_1^{3/2} \Big),
			\notag\\
			\Psi_{D2}^{3/2} =& \frac{1}{\sqrt{2}} \Big(
			\{n_1n_2\}n_3Q_4\bar{Q}_5\otimes\beta_6^{3/2}
			\notag\\
			&\qquad- [n_1n_2]n_3Q_4\bar{Q}_5\otimes\beta_2^{3/2} \Big),
			\notag\\
			\Psi_{D3}^{3/2} =& \frac{1}{2} \Big[ \{n_1n_2\}n_3Q_4\bar{Q}_5
			\otimes \left(\beta_4^{3/2}+\beta_7^{3/2}\right)
			\notag\\
			&\qquad+ [n_1n_2]n_3Q_4\bar{Q}_5 \otimes
			\left(\beta_3^{3/2}-\beta_8^{3/2}\right) \Big],
			\notag\\
			\Psi_{D4}^{3/2} =& \frac{1}{\sqrt{2}} \Big(
			\{n_1n_2\}n_3Q_4\bar{Q}_5\otimes\beta_{11}^{3/2}
			\notag\\
			&\qquad+ [n_1n_2]n_3Q_4\bar{Q}_5\otimes\beta_{12}^{3/2} \Big),
		\end{align}
		\item $J^{P}=5/2^{-}$:
		\begin{align}\label{eqn:wavefunc:total:D5/2}
			\Psi_{D1}^{5/2} =& \frac{1}{\sqrt{2}} \Big(
			\{n_1n_2\}n_3Q_4\bar{Q}_5\otimes\beta_2^{5/2}
			\notag\\
			&\qquad- [n_1n_2]n_3Q_4\bar{Q}_5\otimes\beta_1^{5/2} \Big),
		\end{align}
	\end{enumerate}
	where $\{n_1n_2\}\equiv(n_1n_2)^{I=1}$ and
	$[n_1n_2]\equiv(n_1n_2)^{I=0}$.
\end{enumerate}
%

\section{The eigenvectors of the pentaquarks}
\label{App:eigenvec}

To obtain the relative widths of pentaquark decays into a light baryon and a charmonium, or into a charm baryon and an anticharm meson, one needs to transform the eigenvectors to the corresponding configuration.
We transform the eigenvectors of $qqqc\bar{c}$ pentaquark states into all possible configurations, as shown in Tables~\ref{table:wavefunc:nnnc.c}--\ref{table:wavefunc:sssc.c}.
Since we are only interested in the OZI-superallowed decays, we only present the color-singlet components.
%

\begin{table*}
	\centering \caption{The eigenvectors of the $nnnc\bar{c}$ pentaquark
		states. The masses are all in units of MeV.}
	\label{table:wavefunc:nnnc.c}
	\begin{tabular}{ccc|cccc|cccccc}
		\toprule[1pt] \toprule[1pt]
		&&&&\multicolumn{2}{c}{$nnn{\otimes}c\bar{c}$}&&&\multicolumn{4}{c}{$nnc{\otimes}n\bar{c}$} \\
		$I$ & $J^P$ & Mass & ${\Delta{J/\psi}}$ & ${\Delta\eta_{c}}$ &
		${N{J/\psi}}$ & ${N\eta_{c}}$ & ${\Sigma_{c}^{*}\bar{D}^{*}}$ &
		${\Sigma_{c}^{*}\bar{D}}$ & ${\Sigma_{c}\bar{D}^{*}}$ &
		${\Sigma_{c}\bar{D}}$ & ${\Lambda_{c}\bar{D}^{*}}$
		& ${\Lambda_{c}\bar{D}}$ \\
		\midrule[1pt] $\frac{3}{2}$ & $\frac{1}{2}^{-}$
		&  $4601.9$ & $-0.197$ &&&&  $0.146$ && $-0.563$ &  $0.304$ \\
		&& $4717.1$ &  $0.114$ &&&& $-0.621$ && $-0.218$ & $-0.081$ \\
		& $\frac{3}{2}^{-}$
		&  $4633.0$ & $-0.053$ & $-0.118$ &&& $-0.521$ &  $0.350$ & $-0.211$ & \\
		\midrule[1pt] $\frac{1}{2}$ & $\frac{1}{2}^{-}$
		&  $4327.0$ &&&  $0.084$ & $-0.134$ & $-0.075$ &&  $0.060$ &  $0.566$ & $-0.326$ &  $0.029$ \\
		&& $4372.4$ &&&  $0.093$ &  $0.077$ &  $0.322$ &&  $0.380$ &  $0.067$ &  $0.072$ & $-0.426$ \\
		&& $4480.9$ &&& $-0.069$ & $-0.035$ & $-0.403$ &&  $0.357$ &  $0.115$ &  $0.364$ &  $0.087$ \\
		& $\frac{3}{2}^{-}$
		&  $4367.4$ &&& $-0.072$ &&  $0.030$ & $-0.555$ & $-0.036$ && $-0.364$ & \\
		&& $4476.3$ &&& $-0.011$ &&  $0.124$ &  $0.119$ &  $0.602$ && $-0.230$ & \\
		&& $4524.5$ &&&  $0.056$ &&  $0.560$ &  $0.181$ & $-0.231$ && $-0.210$ & \\
		& $\frac{5}{2}^{-}$
		& $4546.0$ &&&&&  $0.667$ &&&&& \\
		\bottomrule[1pt] \bottomrule[1pt]
	\end{tabular}
\end{table*}
%

\begin{table*}
	\centering \caption{The eigenvectors for the $(nnsc\bar{c})^{I=1}$
		pentaquark states.
		The masses are all in units of MeV.}
	\label{table:wavefunc:nnsc.c:I=1}
	\begin{tabular}{cc|cccc|cccc|ccccccccccc}
		\toprule[1pt] \toprule[1pt]
		&& &\multicolumn{2}{c}{$nns{\otimes}c\bar{c}$}& & &\multicolumn{2}{c}{$nnc{\otimes}s\bar{c}$} & &\multicolumn{6}{c}{$nsc{\otimes}n\bar{c}$} & \\
		$J^{P}$ & Mass & ${\Sigma^{*}{J/\psi}}$ & ${\Sigma^{*}\eta_{c}}$ &
		${\Sigma{J/\psi}}$ & ${\Sigma\eta_{c}}$
		& ${\Sigma_{c}^{*}\bar{D}_{s}^{*}}$ & ${\Sigma_{c}^{*}\bar{D}_{s}}$
		& ${\Sigma_{c}\bar{D}_{s}^{*}}$ & ${\Sigma_{c}\bar{D}_{s}}$
		& ${\Xi_{c}^{*}\bar{D}^{*}}$ & ${\Xi_{c}^{*}\bar{D}}$ &
		${\Xi_{c}'\bar{D}^{*}}$ & ${\Xi_{c}'\bar{D}}$ &
		${\Xi_{c}\bar{D}^{*}}$
		& ${\Xi_{c}\bar{D}}$ \\
		\midrule[1pt] $\frac{1}{2}^{-}$
		&  $4442.8$ &$0.144$&&$0.134$&$-0.190$ &$-0.124$&&$0.206$&$0.856$ &$-0.089$&&$0.356$&$0.156$&$-0.050$&$0.269$ \\
		& $4522.2$ &$-0.008$&&$-0.105$&$-0.108$ &$-0.476$&&$-0.631$&$0.004$ &$-0.169$&&$0.038$&$-0.447$&$-0.254$&$0.221$ \\
		& $4612.6$ &$0.138$&&$-0.106$&$-0.034$ &$-0.664$&&$0.545$&$-0.004$ &$-0.195$&&$-0.360$&$-0.076$&$0.300$&$0.136$ \\
		& $4696.3$ &$0.229$&&$0.017$&$-0.005$ &$-0.017$&&$0.396$&$-0.238$ &$0.164$&&$-0.181$&$0.130$&$-0.624$&$0.259$ \\
		& $4808.1$ &$-0.140$&&$-0.012$&$-0.007$ &$0.525$&&$0.188$&$0.071$ &$-0.666$&&$-0.134$&$-0.098$&$-0.160$&$-0.041$ \\
		$\frac{3}{2}^{-}$
		& $4485.9$ &$0.789$&$-0.053$&$0.072$& &$0.030$&$0.688$&$0.283$& &$0.040$&$0.046$&$-0.265$&&$-0.003$& \\
		& $4488.4$ &$0.610$&$0.023$&$-0.091$& &$-0.007$&$-0.517$&$0.001$& &$-0.082$&$-0.446$&$0.138$&&$-0.344$& \\
		& $4584.9$ &$-0.022$&$-0.021$&$0.054$& &$-0.009$&$0.046$&$-0.907$& &$0.078$&$-0.022$&$-0.439$&&$-0.230$& \\
		& $4636.2$ &$-0.016$&$-0.048$&$-0.062$& &$-0.924$&$-0.098$&$0.068$& &$-0.257$&$-0.177$&$-0.245$&&$0.268$& \\
		& $4728.8$ &$-0.062$&$-0.132$&$0.004$& &$-0.302$&$0.301$&$-0.153$& &$0.602$&$-0.343$&$0.245$&&$0.116$& \\
		$\frac{5}{2}^{-}$
		& $4644.3$ &$-0.006$&&& &$0.940$&&& &$0.473$&&&&& \\
		\bottomrule[1pt] \bottomrule[1pt]
	\end{tabular}
\end{table*}
%
\begin{table*}
	\centering \caption{The eigenvectors for the $(nnsc\bar{c})^{I=0}$
		pentaquark states.
		The masses are all in units of MeV.}
	\label{table:wavefunc:nnsc.c:I=0}
	\begin{tabular}{cc|cc|cc|ccccccccccccccc}
		\toprule[1pt] \toprule[1pt]
		&& \multicolumn{2}{c}{$nns{\otimes}c\bar{c}$} &\multicolumn{2}{c}{$nnc{\otimes}s\bar{c}$}& \multicolumn{6}{c}{$nsc{\otimes}n\bar{c}$} & \\
		$J^{P}$ & Mass
		& ${\Lambda{J/\psi}}$ & ${\Lambda\eta_{c}}$
		& ${\Lambda_{c}\bar{D}_{s}^{*}}$ & ${\Lambda_{c}\bar{D}_{s}}$ &
		${\Xi_{c}^{*}\bar{D}^{*}}$ & ${\Xi_{c}^{*}\bar{D}}$ &
		${\Xi_{c}'\bar{D}^{*}}$ & ${\Xi_{c}'\bar{D}}$ &
		${\Xi_{c}\bar{D}^{*}}$
		& ${\Xi_{c}\bar{D}}$ \\
		\midrule[1pt] $\frac{1}{2}^{-}$
		& $4197.4$ &$0.652$&$-0.059$ &$-0.089$&$0.795$ &$0.028$&&$-0.028$&$0.340$&$-0.105$&$-0.061$ \\
		& $4208.6$ &$-0.735$&$-0.057$ &$-0.009$&$0.234$ &$0.104$&&$0.047$&$0.417$&$0.208$&$-0.371$ \\
		& $4386.6$ &$0.095$&$-0.078$ &$-0.894$&$-0.006$ &$-0.031$&&$-0.343$&$-0.184$&$0.267$&$-0.188$ \\
		& $4465.0$ &$-0.091$&$0.132$ &$0.103$&$-0.015$ &$-0.084$&&$-0.354$&$0.311$&$0.344$&$0.548$ \\
		& $4489.6$ &$-0.112$&$-0.096$ &$-0.052$&$0.474$ &$0.404$&&$0.318$&$-0.320$&$0.336$&$0.251$ \\
		& $4607.0$ &$0.076$&$0.041$ &$-0.365$&$-0.089$ &$-0.484$&&$0.537$&$0.148$&$0.191$&$0.104$ \\
		$\frac{3}{2}^{-}$
		& $4387.3$ &$-0.031$& &$-0.865$& &$-0.094$&$0.197$&$-0.446$&&$0.213$& \\
		& $4501.5$ &$-0.077$& &$-0.253$& &$-0.036$&$0.631$&$0.433$&&$-0.212$& \\
		& $4603.6$ &$-0.006$& &$-0.236$& &$-0.174$&$-0.189$&$-0.169$&&$-0.738$& \\
		& $4656.0$ &$-0.065$&  &$0.204$&  &$0.675$&$0.234$&$-0.319$&&$-0.178$& \\
		$\frac{5}{2}^{-}$
		& $4680.6$ && && &$0.817$&&&&& \\
		\bottomrule[1pt] \bottomrule[1pt]
	\end{tabular}
\end{table*}
%

\begin{table*}
	\centering \caption{The eigenvectors for the $ssnc\bar{c}$
		pentaquark states.
		The masses are all in units of MeV.}
	\label{table:wavefunc:ssnc.c}
	\begin{tabular}{cc|cccc|cccc|ccccccccccc}
		\toprule[1pt] \toprule[1pt]
		&& &\multicolumn{2}{c}{$ssn{\otimes}c\bar{c}$}& & &\multicolumn{2}{c}{$ssc{\otimes}n\bar{c}$} & &\multicolumn{6}{c}{$nsc{\otimes}s\bar{c}$} & \\
		$J^{P}$ & Mass & ${\Xi^{*}{J/\psi}}$ & ${\Xi^{*}\eta_{c}}$ &
		${\Xi{J/\psi}}$ & ${\Xi\eta_{c}}$ & ${\Omega_{c}^{*}\bar{D}^{*}}$ &
		${\Omega_{c}^{*}\bar{D}}$ & ${\Omega_{c}\bar{D}^{*}}$ &
		${\Omega_{c}\bar{D}}$ & ${\Xi_{c}^{*}\bar{D}_{s}^{*}}$ &
		${\Xi_{c}^{*}\bar{D}_{s}}$ & ${\Xi_{c}'\bar{D}_{s}^{*}}$ &
		${\Xi_{c}'\bar{D}_{s}}$ & ${\Xi_{c}\bar{D}_{s}^{*}}$
		& ${\Xi_{c}\bar{D}_{s}}$ \\
		\midrule[1pt] $\frac{1}{2}^{-}$
		& $4573.4$ &$0.068$&&$-0.171$&$0.199$ &$0.112$&&$-0.180$&$-0.725$ &$-0.038$&&$0.065$&$0.426$&$-0.397$&$0.017$ \\
		& $4621.7$ &$-0.035$&&$-0.136$&$-0.123$ &$-0.426$&&$-0.526$&$-0.078$ &$0.246$&&$0.312$&$0.003$&$0.135$&$-0.492$ \\
		& $4728.5$ &$-0.091$&&$-0.111$&$-0.046$ &$-0.542$&&$0.401$&$0.158$ &$0.335$&&$-0.298$&$-0.052$&$-0.447$&$-0.112$ \\
		& $4787.6$ &$-0.306$&&$0.010$&$-0.001$ &$0.164$&&$-0.653$&$0.260$ &$0.085$&&$-0.534$&$0.247$&$0.075$&$-0.007$ \\
		& $4902.2$ &$-0.174$&&$0.007$&$0.004$ &$0.673$&&$0.191$&$0.078$ &$0.602$&&$0.187$&$0.074$&$0.019$&$0.030$ \\
		$\frac{3}{2}^{-}$
		& $4614.5$ &$0.041$&$-0.047$&$-0.118$& &$-0.012$&$-0.751$&$-0.065$& &$-0.047$&$0.410$&$0.042$&&$0.449$& \\
		& $4715.2$ &$-0.015$&$-0.013$&$-0.023$& &$0.096$&$0.162$&$0.848$& &$-0.101$&$-0.053$&$-0.447$&&$0.276$& \\
		& $4769.1$ &$0.006$&$0.034$&$-0.079$& &$-0.702$&$-0.310$&$0.313$& &$0.486$&$0.070$&$-0.109$&&$-0.257$& \\
		& $4819.0$ &$0.077$&$0.162$&$0.007$& &$0.665$&$-0.323$&$0.208$& &$0.445$&$-0.330$&$0.200$&&$0.090$& \\
		$\frac{5}{2}^{-}$
		& $4790.0$ &$0.006$&&& &$0.945$&&& &$-0.469$&&&&& \\
		\bottomrule[1pt] \bottomrule[1pt]
	\end{tabular}
\end{table*}
%

\begin{table}
	\centering \caption{The eigenvectors for the $sssc\bar{c}$
		pentaquark states.
		The masses are all in units of MeV.}
	\label{table:wavefunc:sssc.c}
	\begin{tabular}{cc|cc|ccccccccc}
		\toprule[1pt] \toprule[1pt]
		&&\multicolumn{2}{c}{$sss{\otimes}c\bar{c}$}&\multicolumn{4}{c}{$ssc{\otimes}s\bar{c}$} \\
		$J^P$ & Mass & ${\Omega{J/\psi}}$ & ${\Omega\eta_{c}}$ &
		${\Omega_{c}^{*}\bar{D}_{s}^{*}}$ & ${\Omega_{c}^{*}\bar{D}_{s}}$ &
		${\Omega_{c}\bar{D}_{s}^{*}}$
		& ${\Omega_{c}\bar{D}_{s}}$ \\
		\midrule[1pt] $\frac{1}{2}^{-}$
		& $4894.4$ &$-0.367$& &$0.098$&&$-0.582$&$0.226$ \\
		& $5009.4$ &$0.203$& &$-0.628$&&$-0.176$&$-0.072$ \\
		$\frac{3}{2}^{-}$
		& $4924.1$ &$-0.087$&$-0.187$ &$-0.531$&$0.327$&$-0.203$& \\
		\bottomrule[1pt] \bottomrule[1pt]
	\end{tabular}
\end{table}
%

\end{appendix}
\bibliography{myreference}

\begin{thebibliography}{79}%
\makeatletter
\providecommand \@ifxundefined [1]{%
 \@ifx{#1\undefined}
}%
\providecommand \@ifnum [1]{%
 \ifnum #1\expandafter \@firstoftwo
 \else \expandafter \@secondoftwo
 \fi
}%
\providecommand \@ifx [1]{%
 \ifx #1\expandafter \@firstoftwo
 \else \expandafter \@secondoftwo
 \fi
}%
\providecommand \natexlab [1]{#1}%
\providecommand \enquote  [1]{``#1''}%
\providecommand \bibnamefont  [1]{#1}%
\providecommand \bibfnamefont [1]{#1}%
\providecommand \citenamefont [1]{#1}%
\providecommand \href@noop [0]{\@secondoftwo}%
\providecommand \href [0]{\begingroup \@sanitize@url \@href}%
\providecommand \@href[1]{\@@startlink{#1}\@@href}%
\providecommand \@@href[1]{\endgroup#1\@@endlink}%
\providecommand \@sanitize@url [0]{\catcode `\\12\catcode `\$12\catcode
  `\&12\catcode `\#12\catcode `\^12\catcode `\_12\catcode `\%12\relax}%
\providecommand \@@startlink[1]{}%
\providecommand \@@endlink[0]{}%
\providecommand \url  [0]{\begingroup\@sanitize@url \@url }%
\providecommand \@url [1]{\endgroup\@href {#1}{\urlprefix }}%
\providecommand \urlprefix  [0]{URL }%
\providecommand \Eprint [0]{\href }%
\providecommand \doibase [0]{http://dx.doi.org/}%
\providecommand \selectlanguage [0]{\@gobble}%
\providecommand \bibinfo  [0]{\@secondoftwo}%
\providecommand \bibfield  [0]{\@secondoftwo}%
\providecommand \translation [1]{[#1]}%
\providecommand \BibitemOpen [0]{}%
\providecommand \bibitemStop [0]{}%
\providecommand \bibitemNoStop [0]{.\EOS\space}%
\providecommand \EOS [0]{\spacefactor3000\relax}%
\providecommand \BibitemShut  [1]{\csname bibitem#1\endcsname}%
\let\auto@bib@innerbib\@empty
\bibitem [{\citenamefont {Gell-Mann}(1964)}]{GellMann:1964nj}%
  \BibitemOpen
  \bibfield  {author} {\bibinfo {author} {\bibfnamefont {M.}~\bibnamefont
  {Gell-Mann}},\ }\href {\doibase 10.1016/S0031-9163(64)92001-3} {\bibfield
  {journal} {\bibinfo  {journal} {Phys. Lett.}\ }\textbf {\bibinfo {volume}
  {8}},\ \bibinfo {pages} {214} (\bibinfo {year} {1964})}\BibitemShut {NoStop}%
\bibitem [{\citenamefont {Zweig}(1964)}]{Zweig:1964jf}%
  \BibitemOpen
  \bibfield  {author} {\bibinfo {author} {\bibfnamefont {G.}~\bibnamefont
  {Zweig}},\ }\bibfield  {booktitle} {\emph {\bibinfo {booktitle} {DEVELOPMENTS
  IN THE QUARK THEORY OF HADRONS. VOL. 1. 1964 - 1978}},\ }\href
  {http://cds.cern.ch/record/570209} {\bibfield  {journal} {\bibinfo  {journal}
  {Developments in the Quark Theory of Hadrons, Volume 1. Edited by D.
  Lichtenberg and S. Rosen. pp. 22-101}\ ,\ \bibinfo {pages} {22}} (\bibinfo
  {year} {1964})}\BibitemShut {NoStop}%
\bibitem [{\citenamefont {Jaffe}(1977{\natexlab{a}})}]{Jaffe:1976ig}%
  \BibitemOpen
  \bibfield  {author} {\bibinfo {author} {\bibfnamefont {R.~L.}\ \bibnamefont
  {Jaffe}},\ }\href {\doibase 10.1103/PhysRevD.15.267} {\bibfield  {journal}
  {\bibinfo  {journal} {Phys. Rev.}\ }\textbf {\bibinfo {volume} {D15}},\
  \bibinfo {pages} {267} (\bibinfo {year} {1977}{\natexlab{a}})}\BibitemShut
  {NoStop}%
\bibitem [{\citenamefont {Jaffe}(1977{\natexlab{b}})}]{Jaffe:1976ih}%
  \BibitemOpen
  \bibfield  {author} {\bibinfo {author} {\bibfnamefont {R.~L.}\ \bibnamefont
  {Jaffe}},\ }\href {\doibase 10.1103/PhysRevD.15.281} {\bibfield  {journal}
  {\bibinfo  {journal} {Phys. Rev.}\ }\textbf {\bibinfo {volume} {D15}},\
  \bibinfo {pages} {281} (\bibinfo {year} {1977}{\natexlab{b}})}\BibitemShut
  {NoStop}%
\bibitem [{\citenamefont {Chan}\ and\ \citenamefont
  {H{\o}gaasen}(1977)}]{Chan:1977st}%
  \BibitemOpen
  \bibfield  {author} {\bibinfo {author} {\bibfnamefont {H.-M.}\ \bibnamefont
  {Chan}}\ and\ \bibinfo {author} {\bibfnamefont {H.}~\bibnamefont
  {H{\o}gaasen}},\ }\href {\doibase 10.1016/0370-2693(77)90077-6} {\bibfield
  {journal} {\bibinfo  {journal} {Phys. Lett.}\ }\textbf {\bibinfo {volume}
  {72B}},\ \bibinfo {pages} {121} (\bibinfo {year} {1977})}\BibitemShut
  {NoStop}%
\bibitem [{\citenamefont {Chao}(1980)}]{Chao:1979mm}%
  \BibitemOpen
  \bibfield  {author} {\bibinfo {author} {\bibfnamefont {K.-T.}\ \bibnamefont
  {Chao}},\ }\href {\doibase 10.1016/0550-3213(80)90033-4} {\bibfield
  {journal} {\bibinfo  {journal} {Nucl. Phys.}\ }\textbf {\bibinfo {volume}
  {B169}},\ \bibinfo {pages} {281} (\bibinfo {year} {1980})}\BibitemShut
  {NoStop}%
\bibitem [{\citenamefont {Chao}(1981{\natexlab{a}})}]{Chao:1979tg}%
  \BibitemOpen
  \bibfield  {author} {\bibinfo {author} {\bibfnamefont {K.-T.}\ \bibnamefont
  {Chao}},\ }\href {\doibase 10.1016/0550-3213(81)90143-7} {\bibfield
  {journal} {\bibinfo  {journal} {Nucl. Phys.}\ }\textbf {\bibinfo {volume}
  {B183}},\ \bibinfo {pages} {435} (\bibinfo {year}
  {1981}{\natexlab{a}})}\BibitemShut {NoStop}%
\bibitem [{\citenamefont {Chao}(1981{\natexlab{b}})}]{Chao:1980dv}%
  \BibitemOpen
  \bibfield  {author} {\bibinfo {author} {\bibfnamefont {K.-T.}\ \bibnamefont
  {Chao}},\ }\href {\doibase 10.1007/BF01431564} {\bibfield  {journal}
  {\bibinfo  {journal} {Z. Phys.}\ }\textbf {\bibinfo {volume} {C7}},\ \bibinfo
  {pages} {317} (\bibinfo {year} {1981}{\natexlab{b}})}\BibitemShut {NoStop}%
\bibitem [{\citenamefont {Fukugita}\ \emph {et~al.}(1978)\citenamefont
  {Fukugita}, \citenamefont {Konishi},\ and\ \citenamefont
  {Hansson}}]{Fukugita:1978sn}%
  \BibitemOpen
  \bibfield  {author} {\bibinfo {author} {\bibfnamefont {M.}~\bibnamefont
  {Fukugita}}, \bibinfo {author} {\bibfnamefont {K.}~\bibnamefont {Konishi}}, \
  and\ \bibinfo {author} {\bibfnamefont {T.~H.}\ \bibnamefont {Hansson}},\
  }\href {\doibase 10.1016/0370-2693(78)90569-5} {\bibfield  {journal}
  {\bibinfo  {journal} {Phys. Lett.}\ }\textbf {\bibinfo {volume} {74B}},\
  \bibinfo {pages} {261} (\bibinfo {year} {1978})}\BibitemShut {NoStop}%
\bibitem [{\citenamefont {H{\"o}gaasen}\ and\ \citenamefont
  {Sorba}(1978)}]{Hogaasen:1978jw}%
  \BibitemOpen
  \bibfield  {author} {\bibinfo {author} {\bibfnamefont {H.}~\bibnamefont
  {H{\"o}gaasen}}\ and\ \bibinfo {author} {\bibfnamefont {P.}~\bibnamefont
  {Sorba}},\ }\href {\doibase 10.1016/0550-3213(78)90417-0} {\bibfield
  {journal} {\bibinfo  {journal} {Nucl. Phys.}\ }\textbf {\bibinfo {volume}
  {B145}},\ \bibinfo {pages} {119} (\bibinfo {year} {1978})}\BibitemShut
  {NoStop}%
\bibitem [{\citenamefont {Strottman}(1979)}]{Strottman:1979qu}%
  \BibitemOpen
  \bibfield  {author} {\bibinfo {author} {\bibfnamefont {D.}~\bibnamefont
  {Strottman}},\ }\href {\doibase 10.1103/PhysRevD.20.748} {\bibfield
  {journal} {\bibinfo  {journal} {Phys. Rev.}\ }\textbf {\bibinfo {volume}
  {D20}},\ \bibinfo {pages} {748} (\bibinfo {year} {1979})}\BibitemShut
  {NoStop}%
\bibitem [{\citenamefont {Choi}\ \emph {et~al.}(2003)\citenamefont {Choi} \emph
  {et~al.}}]{Choi:2003ue}%
  \BibitemOpen
  \bibfield  {author} {\bibinfo {author} {\bibfnamefont {S.~K.}\ \bibnamefont
  {Choi}} \emph {et~al.} (\bibinfo {collaboration} {Belle Collaboration}),\
  }\href {\doibase 10.1103/PhysRevLett.91.262001} {\bibfield  {journal}
  {\bibinfo  {journal} {Phys. Rev. Lett.}\ }\textbf {\bibinfo {volume} {91}},\
  \bibinfo {pages} {262001} (\bibinfo {year} {2003})},\ \Eprint
  {http://arxiv.org/abs/hep-ex/0309032} {arXiv:hep-ex/0309032 [hep-ex]}
  \BibitemShut {NoStop}%
\bibitem [{\citenamefont {Acosta}\ \emph {et~al.}(2004)\citenamefont {Acosta}
  \emph {et~al.}}]{Acosta:2003zx}%
  \BibitemOpen
  \bibfield  {author} {\bibinfo {author} {\bibfnamefont {D.}~\bibnamefont
  {Acosta}} \emph {et~al.} (\bibinfo {collaboration} {CDF Collaboration}),\
  }\href {\doibase 10.1103/PhysRevLett.93.072001} {\bibfield  {journal}
  {\bibinfo  {journal} {Phys. Rev. Lett.}\ }\textbf {\bibinfo {volume} {93}},\
  \bibinfo {pages} {072001} (\bibinfo {year} {2004})},\ \Eprint
  {http://arxiv.org/abs/hep-ex/0312021} {arXiv:hep-ex/0312021 [hep-ex]}
  \BibitemShut {NoStop}%
\bibitem [{\citenamefont {Abazov}\ \emph {et~al.}(2004)\citenamefont {Abazov}
  \emph {et~al.}}]{Abazov:2004kp}%
  \BibitemOpen
  \bibfield  {author} {\bibinfo {author} {\bibfnamefont {V.~M.}\ \bibnamefont
  {Abazov}} \emph {et~al.} (\bibinfo {collaboration} {D0 Collaboration}),\
  }\href {\doibase 10.1103/PhysRevLett.93.162002} {\bibfield  {journal}
  {\bibinfo  {journal} {Phys. Rev. Lett.}\ }\textbf {\bibinfo {volume} {93}},\
  \bibinfo {pages} {162002} (\bibinfo {year} {2004})},\ \Eprint
  {http://arxiv.org/abs/hep-ex/0405004} {arXiv:hep-ex/0405004 [hep-ex]}
  \BibitemShut {NoStop}%
\bibitem [{\citenamefont {Aubert}\ \emph
  {et~al.}(2005{\natexlab{a}})\citenamefont {Aubert} \emph
  {et~al.}}]{Aubert:2004ns}%
  \BibitemOpen
  \bibfield  {author} {\bibinfo {author} {\bibfnamefont {B.}~\bibnamefont
  {Aubert}} \emph {et~al.} (\bibinfo {collaboration} {BABAR Collaboration}),\
  }\href {\doibase 10.1103/PhysRevD.71.071103} {\bibfield  {journal} {\bibinfo
  {journal} {Phys. Rev.}\ }\textbf {\bibinfo {volume} {D71}},\ \bibinfo {pages}
  {071103} (\bibinfo {year} {2005}{\natexlab{a}})},\ \Eprint
  {http://arxiv.org/abs/hep-ex/0406022} {arXiv:hep-ex/0406022 [hep-ex]}
  \BibitemShut {NoStop}%
\bibitem [{\citenamefont {Aaij}\ \emph {et~al.}(2012)\citenamefont {Aaij} \emph
  {et~al.}}]{Aaij:2011sn}%
  \BibitemOpen
  \bibfield  {author} {\bibinfo {author} {\bibfnamefont {R.}~\bibnamefont
  {Aaij}} \emph {et~al.} (\bibinfo {collaboration} {LHCb Collaboration}),\
  }\href {\doibase 10.1140/epjc/s10052-012-1972-7} {\bibfield  {journal}
  {\bibinfo  {journal} {Eur. Phys. J.}\ }\textbf {\bibinfo {volume} {C72}},\
  \bibinfo {pages} {1972} (\bibinfo {year} {2012})},\ \Eprint
  {http://arxiv.org/abs/1112.5310} {arXiv:1112.5310 [hep-ex]} \BibitemShut
  {NoStop}%
\bibitem [{\citenamefont {Chatrchyan}\ \emph {et~al.}(2013)\citenamefont
  {Chatrchyan} \emph {et~al.}}]{Chatrchyan:2013cld}%
  \BibitemOpen
  \bibfield  {author} {\bibinfo {author} {\bibfnamefont {S.}~\bibnamefont
  {Chatrchyan}} \emph {et~al.} (\bibinfo {collaboration} {CMS Collaboration}),\
  }\href {\doibase 10.1007/JHEP04(2013)154} {\bibfield  {journal} {\bibinfo
  {journal} {JHEP}\ }\textbf {\bibinfo {volume} {04}},\ \bibinfo {pages} {154}
  (\bibinfo {year} {2013})},\ \Eprint {http://arxiv.org/abs/1302.3968}
  {arXiv:1302.3968 [hep-ex]} \BibitemShut {NoStop}%
\bibitem [{\citenamefont {Ablikim}\ \emph {et~al.}(2014)\citenamefont {Ablikim}
  \emph {et~al.}}]{Ablikim:2013dyn}%
  \BibitemOpen
  \bibfield  {author} {\bibinfo {author} {\bibfnamefont {M.}~\bibnamefont
  {Ablikim}} \emph {et~al.} (\bibinfo {collaboration} {BESIII Collaboration}),\
  }\href {\doibase 10.1103/PhysRevLett.112.092001} {\bibfield  {journal}
  {\bibinfo  {journal} {Phys. Rev. Lett.}\ }\textbf {\bibinfo {volume} {112}},\
  \bibinfo {pages} {092001} (\bibinfo {year} {2014})},\ \Eprint
  {http://arxiv.org/abs/1310.4101} {arXiv:1310.4101 [hep-ex]} \BibitemShut
  {NoStop}%
\bibitem [{\citenamefont {Abe}\ \emph {et~al.}(2005)\citenamefont {Abe} \emph
  {et~al.}}]{Abe:2004zs}%
  \BibitemOpen
  \bibfield  {author} {\bibinfo {author} {\bibfnamefont {K.}~\bibnamefont
  {Abe}} \emph {et~al.} (\bibinfo {collaboration} {Belle Collaboration}),\
  }\bibfield  {booktitle} {\emph {\bibinfo {booktitle} {{Proceedings, 32nd
  International Conference on High Energy Physics (ICHEP 2004): Beijing, China,
  August 16-22, 2004. Vol. 1+2}}},\ }\href {\doibase
  10.1103/PhysRevLett.94.182002} {\bibfield  {journal} {\bibinfo  {journal}
  {Phys. Rev. Lett.}\ }\textbf {\bibinfo {volume} {94}},\ \bibinfo {pages}
  {182002} (\bibinfo {year} {2005})},\ \Eprint
  {http://arxiv.org/abs/hep-ex/0408126} {arXiv:hep-ex/0408126 [hep-ex]}
  \BibitemShut {NoStop}%
\bibitem [{\citenamefont {Aaltonen}\ \emph {et~al.}(2009)\citenamefont
  {Aaltonen} \emph {et~al.}}]{Aaltonen:2009tz}%
  \BibitemOpen
  \bibfield  {author} {\bibinfo {author} {\bibfnamefont {T.}~\bibnamefont
  {Aaltonen}} \emph {et~al.} (\bibinfo {collaboration} {CDF Collaboration}),\
  }\href {\doibase 10.1103/PhysRevLett.102.242002} {\bibfield  {journal}
  {\bibinfo  {journal} {Phys. Rev. Lett.}\ }\textbf {\bibinfo {volume} {102}},\
  \bibinfo {pages} {242002} (\bibinfo {year} {2009})},\ \Eprint
  {http://arxiv.org/abs/0903.2229} {arXiv:0903.2229 [hep-ex]} \BibitemShut
  {NoStop}%
\bibitem [{\citenamefont {Aubert}\ \emph
  {et~al.}(2005{\natexlab{b}})\citenamefont {Aubert} \emph
  {et~al.}}]{Aubert:2005rm}%
  \BibitemOpen
  \bibfield  {author} {\bibinfo {author} {\bibfnamefont {B.}~\bibnamefont
  {Aubert}} \emph {et~al.} (\bibinfo {collaboration} {BaBar Collaboration}),\
  }\href {\doibase 10.1103/PhysRevLett.95.142001} {\bibfield  {journal}
  {\bibinfo  {journal} {Phys. Rev. Lett.}\ }\textbf {\bibinfo {volume} {95}},\
  \bibinfo {pages} {142001} (\bibinfo {year} {2005}{\natexlab{b}})},\ \Eprint
  {http://arxiv.org/abs/hep-ex/0506081} {arXiv:hep-ex/0506081 [hep-ex]}
  \BibitemShut {NoStop}%
\bibitem [{\citenamefont {Aubert}\ \emph {et~al.}(2007)\citenamefont {Aubert}
  \emph {et~al.}}]{Aubert:2007zz}%
  \BibitemOpen
  \bibfield  {author} {\bibinfo {author} {\bibfnamefont {B.}~\bibnamefont
  {Aubert}} \emph {et~al.} (\bibinfo {collaboration} {BaBar Collaboration}),\
  }\bibfield  {booktitle} {\emph {\bibinfo {booktitle} {{Proceedings of the
  33rd International Conference on High Energy Physics (ICHEP '06): Moscow,
  Russia, July 26-August 2, 2006}}},\ }\href {\doibase
  10.1103/PhysRevLett.98.212001} {\bibfield  {journal} {\bibinfo  {journal}
  {Phys. Rev. Lett.}\ }\textbf {\bibinfo {volume} {98}},\ \bibinfo {pages}
  {212001} (\bibinfo {year} {2007})},\ \Eprint
  {http://arxiv.org/abs/hep-ex/0610057} {arXiv:hep-ex/0610057 [hep-ex]}
  \BibitemShut {NoStop}%
\bibitem [{\citenamefont {Wang}\ \emph {et~al.}(2007)\citenamefont {Wang} \emph
  {et~al.}}]{Wang:2007ea}%
  \BibitemOpen
  \bibfield  {author} {\bibinfo {author} {\bibfnamefont {X.~L.}\ \bibnamefont
  {Wang}} \emph {et~al.} (\bibinfo {collaboration} {Belle Collaboration}),\
  }\href {\doibase 10.1103/PhysRevLett.99.142002} {\bibfield  {journal}
  {\bibinfo  {journal} {Phys. Rev. Lett.}\ }\textbf {\bibinfo {volume} {99}},\
  \bibinfo {pages} {142002} (\bibinfo {year} {2007})},\ \Eprint
  {http://arxiv.org/abs/0707.3699} {arXiv:0707.3699 [hep-ex]} \BibitemShut
  {NoStop}%
\bibitem [{\citenamefont {Swanson}(2004)}]{Swanson:2003tb}%
  \BibitemOpen
  \bibfield  {author} {\bibinfo {author} {\bibfnamefont {E.~S.}\ \bibnamefont
  {Swanson}},\ }\href {\doibase 10.1016/j.physletb.2004.03.033} {\bibfield
  {journal} {\bibinfo  {journal} {Phys. Lett.}\ }\textbf {\bibinfo {volume}
  {B588}},\ \bibinfo {pages} {189} (\bibinfo {year} {2004})},\ \Eprint
  {http://arxiv.org/abs/hep-ph/0311229} {arXiv:hep-ph/0311229 [hep-ph]}
  \BibitemShut {NoStop}%
\bibitem [{\citenamefont {Guo}\ \emph {et~al.}(2018)\citenamefont {Guo},
  \citenamefont {Hanhart}, \citenamefont {Meißner}, \citenamefont {Wang},
  \citenamefont {Zhao},\ and\ \citenamefont {Zou}}]{Guo:2017jvc}%
  \BibitemOpen
  \bibfield  {author} {\bibinfo {author} {\bibfnamefont {F.-K.}\ \bibnamefont
  {Guo}}, \bibinfo {author} {\bibfnamefont {C.}~\bibnamefont {Hanhart}},
  \bibinfo {author} {\bibfnamefont {U.-G.}\ \bibnamefont {Meißner}}, \bibinfo
  {author} {\bibfnamefont {Q.}~\bibnamefont {Wang}}, \bibinfo {author}
  {\bibfnamefont {Q.}~\bibnamefont {Zhao}}, \ and\ \bibinfo {author}
  {\bibfnamefont {B.-S.}\ \bibnamefont {Zou}},\ }\href {\doibase
  10.1103/RevModPhys.90.015004} {\bibfield  {journal} {\bibinfo  {journal}
  {Rev. Mod. Phys.}\ }\textbf {\bibinfo {volume} {90}},\ \bibinfo {pages}
  {015004} (\bibinfo {year} {2018})},\ \Eprint
  {http://arxiv.org/abs/1705.00141} {arXiv:1705.00141 [hep-ph]} \BibitemShut
  {NoStop}%
\bibitem [{\citenamefont {Zhu}(2005)}]{Zhu:2005hp}%
  \BibitemOpen
  \bibfield  {author} {\bibinfo {author} {\bibfnamefont {S.-L.}\ \bibnamefont
  {Zhu}},\ }\href {\doibase 10.1016/j.physletb.2005.08.068} {\bibfield
  {journal} {\bibinfo  {journal} {Phys. Lett.}\ }\textbf {\bibinfo {volume}
  {B625}},\ \bibinfo {pages} {212} (\bibinfo {year} {2005})},\ \Eprint
  {http://arxiv.org/abs/hep-ph/0507025} {arXiv:hep-ph/0507025 [hep-ph]}
  \BibitemShut {NoStop}%
\bibitem [{\citenamefont {Esposito}\ \emph {et~al.}(2016)\citenamefont
  {Esposito}, \citenamefont {Pilloni},\ and\ \citenamefont
  {Polosa}}]{Esposito:2016itg}%
  \BibitemOpen
  \bibfield  {author} {\bibinfo {author} {\bibfnamefont {A.}~\bibnamefont
  {Esposito}}, \bibinfo {author} {\bibfnamefont {A.}~\bibnamefont {Pilloni}}, \
  and\ \bibinfo {author} {\bibfnamefont {A.~D.}\ \bibnamefont {Polosa}},\
  }\href {\doibase 10.1016/j.physletb.2016.05.028} {\bibfield  {journal}
  {\bibinfo  {journal} {Phys. Lett.}\ }\textbf {\bibinfo {volume} {B758}},\
  \bibinfo {pages} {292} (\bibinfo {year} {2016})},\ \Eprint
  {http://arxiv.org/abs/1603.07667} {arXiv:1603.07667 [hep-ph]} \BibitemShut
  {NoStop}%
\bibitem [{\citenamefont {Cui}\ \emph {et~al.}(2007)\citenamefont {Cui},
  \citenamefont {Chen}, \citenamefont {Deng},\ and\ \citenamefont
  {Zhu}}]{Cui:2006mp}%
  \BibitemOpen
  \bibfield  {author} {\bibinfo {author} {\bibfnamefont {Y.}~\bibnamefont
  {Cui}}, \bibinfo {author} {\bibfnamefont {X.-L.}\ \bibnamefont {Chen}},
  \bibinfo {author} {\bibfnamefont {W.-Z.}\ \bibnamefont {Deng}}, \ and\
  \bibinfo {author} {\bibfnamefont {S.-L.}\ \bibnamefont {Zhu}},\ }\href@noop
  {} {\bibfield  {journal} {\bibinfo  {journal} {HEPNP}\ }\textbf {\bibinfo
  {volume} {31}},\ \bibinfo {pages} {7} (\bibinfo {year} {2007})},\ \Eprint
  {http://arxiv.org/abs/hep-ph/0607226} {arXiv:hep-ph/0607226 [hep-ph]}
  \BibitemShut {NoStop}%
\bibitem [{\citenamefont {Park}\ and\ \citenamefont
  {Lee}(2014)}]{Park:2013fda}%
  \BibitemOpen
  \bibfield  {author} {\bibinfo {author} {\bibfnamefont {W.}~\bibnamefont
  {Park}}\ and\ \bibinfo {author} {\bibfnamefont {S.~H.}\ \bibnamefont {Lee}},\
  }\href {\doibase 10.1016/j.nuclphysa.2014.02.008} {\bibfield  {journal}
  {\bibinfo  {journal} {Nucl. Phys.}\ }\textbf {\bibinfo {volume} {A925}},\
  \bibinfo {pages} {161} (\bibinfo {year} {2014})},\ \Eprint
  {http://arxiv.org/abs/1311.5330} {arXiv:1311.5330 [nucl-th]} \BibitemShut
  {NoStop}%
\bibitem [{\citenamefont {Lebed}\ \emph {et~al.}(2017)\citenamefont {Lebed},
  \citenamefont {Mitchell},\ and\ \citenamefont {Swanson}}]{Lebed:2016hpi}%
  \BibitemOpen
  \bibfield  {author} {\bibinfo {author} {\bibfnamefont {R.~F.}\ \bibnamefont
  {Lebed}}, \bibinfo {author} {\bibfnamefont {R.~E.}\ \bibnamefont {Mitchell}},
  \ and\ \bibinfo {author} {\bibfnamefont {E.~S.}\ \bibnamefont {Swanson}},\
  }\href {\doibase 10.1016/j.ppnp.2016.11.003} {\bibfield  {journal} {\bibinfo
  {journal} {Prog. Part. Nucl. Phys.}\ }\textbf {\bibinfo {volume} {93}},\
  \bibinfo {pages} {143} (\bibinfo {year} {2017})},\ \Eprint
  {http://arxiv.org/abs/1610.04528} {arXiv:1610.04528 [hep-ph]} \BibitemShut
  {NoStop}%
\bibitem [{\citenamefont {Esposito}\ \emph {et~al.}(2017)\citenamefont
  {Esposito}, \citenamefont {Pilloni},\ and\ \citenamefont
  {Polosa}}]{Esposito:2016noz}%
  \BibitemOpen
  \bibfield  {author} {\bibinfo {author} {\bibfnamefont {A.}~\bibnamefont
  {Esposito}}, \bibinfo {author} {\bibfnamefont {A.}~\bibnamefont {Pilloni}}, \
  and\ \bibinfo {author} {\bibfnamefont {A.~D.}\ \bibnamefont {Polosa}},\
  }\href {\doibase 10.1016/j.physrep.2016.11.002} {\bibfield  {journal}
  {\bibinfo  {journal} {Phys. Rept.}\ }\textbf {\bibinfo {volume} {668}},\
  \bibinfo {pages} {1} (\bibinfo {year} {2017})},\ \Eprint
  {http://arxiv.org/abs/1611.07920} {arXiv:1611.07920 [hep-ph]} \BibitemShut
  {NoStop}%
\bibitem [{\citenamefont {Chen}\ \emph
  {et~al.}(2016{\natexlab{a}})\citenamefont {Chen}, \citenamefont {Chen},
  \citenamefont {Liu},\ and\ \citenamefont {Zhu}}]{Chen:2016qju}%
  \BibitemOpen
  \bibfield  {author} {\bibinfo {author} {\bibfnamefont {H.-X.}\ \bibnamefont
  {Chen}}, \bibinfo {author} {\bibfnamefont {W.}~\bibnamefont {Chen}}, \bibinfo
  {author} {\bibfnamefont {X.}~\bibnamefont {Liu}}, \ and\ \bibinfo {author}
  {\bibfnamefont {S.-L.}\ \bibnamefont {Zhu}},\ }\href {\doibase
  10.1016/j.physrep.2016.05.004} {\bibfield  {journal} {\bibinfo  {journal}
  {Phys. Rept.}\ }\textbf {\bibinfo {volume} {639}},\ \bibinfo {pages} {1}
  (\bibinfo {year} {2016}{\natexlab{a}})},\ \Eprint
  {http://arxiv.org/abs/1601.02092} {arXiv:1601.02092 [hep-ph]} \BibitemShut
  {NoStop}%
\bibitem [{\citenamefont {Ali}\ \emph {et~al.}(2017)\citenamefont {Ali},
  \citenamefont {Lange},\ and\ \citenamefont {Stone}}]{Ali:2017jda}%
  \BibitemOpen
  \bibfield  {author} {\bibinfo {author} {\bibfnamefont {A.}~\bibnamefont
  {Ali}}, \bibinfo {author} {\bibfnamefont {J.~S.}\ \bibnamefont {Lange}}, \
  and\ \bibinfo {author} {\bibfnamefont {S.}~\bibnamefont {Stone}},\ }\href
  {\doibase 10.1016/j.ppnp.2017.08.003} {\bibfield  {journal} {\bibinfo
  {journal} {Prog. Part. Nucl. Phys.}\ }\textbf {\bibinfo {volume} {97}},\
  \bibinfo {pages} {123} (\bibinfo {year} {2017})},\ \Eprint
  {http://arxiv.org/abs/1706.00610} {arXiv:1706.00610 [hep-ph]} \BibitemShut
  {NoStop}%
\bibitem [{\citenamefont {Liu}\ \emph {et~al.}(2019{\natexlab{a}})\citenamefont
  {Liu}, \citenamefont {Chen}, \citenamefont {Chen}, \citenamefont {Liu},\ and\
  \citenamefont {Zhu}}]{Liu:2019zoy}%
  \BibitemOpen
  \bibfield  {author} {\bibinfo {author} {\bibfnamefont {Y.-R.}\ \bibnamefont
  {Liu}}, \bibinfo {author} {\bibfnamefont {H.-X.}\ \bibnamefont {Chen}},
  \bibinfo {author} {\bibfnamefont {W.}~\bibnamefont {Chen}}, \bibinfo {author}
  {\bibfnamefont {X.}~\bibnamefont {Liu}}, \ and\ \bibinfo {author}
  {\bibfnamefont {S.-L.}\ \bibnamefont {Zhu}},\ }\href {\doibase
  10.1016/j.ppnp.2019.04.003} {\bibfield  {journal} {\bibinfo  {journal} {Prog.
  Part. Nucl. Phys.}\ }\textbf {\bibinfo {volume} {107}},\ \bibinfo {pages}
  {237} (\bibinfo {year} {2019}{\natexlab{a}})},\ \Eprint
  {http://arxiv.org/abs/1903.11976} {arXiv:1903.11976 [hep-ph]} \BibitemShut
  {NoStop}%
\bibitem [{\citenamefont {Aaij}\ \emph {et~al.}(2015)\citenamefont {Aaij} \emph
  {et~al.}}]{Aaij:2015tga}%
  \BibitemOpen
  \bibfield  {author} {\bibinfo {author} {\bibfnamefont {R.}~\bibnamefont
  {Aaij}} \emph {et~al.} (\bibinfo {collaboration} {LHCb Collaboration}),\
  }\href {\doibase 10.1103/PhysRevLett.115.072001} {\bibfield  {journal}
  {\bibinfo  {journal} {Phys. Rev. Lett.}\ }\textbf {\bibinfo {volume} {115}},\
  \bibinfo {pages} {072001} (\bibinfo {year} {2015})},\ \Eprint
  {http://arxiv.org/abs/1507.03414} {arXiv:1507.03414 [hep-ex]} \BibitemShut
  {NoStop}%
\bibitem [{\citenamefont {Aaij}\ \emph {et~al.}(2019)\citenamefont {Aaij} \emph
  {et~al.}}]{Aaij:2019vzc}%
  \BibitemOpen
  \bibfield  {author} {\bibinfo {author} {\bibfnamefont {R.}~\bibnamefont
  {Aaij}} \emph {et~al.} (\bibinfo {collaboration} {LHCb Collaboration}),\
  }\href {\doibase 10.1103/PhysRevLett.122.222001} {\bibfield  {journal}
  {\bibinfo  {journal} {Phys. Rev. Lett.}\ }\textbf {\bibinfo {volume} {122}},\
  \bibinfo {pages} {222001} (\bibinfo {year} {2019})},\ \Eprint
  {http://arxiv.org/abs/1904.03947} {arXiv:1904.03947 [hep-ex]} \BibitemShut
  {NoStop}%
\bibitem [{\citenamefont {Chen}\ \emph
  {et~al.}(2015{\natexlab{a}})\citenamefont {Chen}, \citenamefont {Liu},
  \citenamefont {Li},\ and\ \citenamefont {Zhu}}]{Chen:2015loa}%
  \BibitemOpen
  \bibfield  {author} {\bibinfo {author} {\bibfnamefont {R.}~\bibnamefont
  {Chen}}, \bibinfo {author} {\bibfnamefont {X.}~\bibnamefont {Liu}}, \bibinfo
  {author} {\bibfnamefont {X.-Q.}\ \bibnamefont {Li}}, \ and\ \bibinfo {author}
  {\bibfnamefont {S.-L.}\ \bibnamefont {Zhu}},\ }\href {\doibase
  10.1103/PhysRevLett.115.132002} {\bibfield  {journal} {\bibinfo  {journal}
  {Phys. Rev. Lett.}\ }\textbf {\bibinfo {volume} {115}},\ \bibinfo {pages}
  {132002} (\bibinfo {year} {2015}{\natexlab{a}})},\ \Eprint
  {http://arxiv.org/abs/1507.03704} {arXiv:1507.03704 [hep-ph]} \BibitemShut
  {NoStop}%
\bibitem [{\citenamefont {Chen}\ \emph
  {et~al.}(2015{\natexlab{b}})\citenamefont {Chen}, \citenamefont {Chen},
  \citenamefont {Liu}, \citenamefont {Steele},\ and\ \citenamefont
  {Zhu}}]{Chen:2015moa}%
  \BibitemOpen
  \bibfield  {author} {\bibinfo {author} {\bibfnamefont {H.-X.}\ \bibnamefont
  {Chen}}, \bibinfo {author} {\bibfnamefont {W.}~\bibnamefont {Chen}}, \bibinfo
  {author} {\bibfnamefont {X.}~\bibnamefont {Liu}}, \bibinfo {author}
  {\bibfnamefont {T.~G.}\ \bibnamefont {Steele}}, \ and\ \bibinfo {author}
  {\bibfnamefont {S.-L.}\ \bibnamefont {Zhu}},\ }\href {\doibase
  10.1103/PhysRevLett.115.172001} {\bibfield  {journal} {\bibinfo  {journal}
  {Phys. Rev. Lett.}\ }\textbf {\bibinfo {volume} {115}},\ \bibinfo {pages}
  {172001} (\bibinfo {year} {2015}{\natexlab{b}})},\ \Eprint
  {http://arxiv.org/abs/1507.03717} {arXiv:1507.03717 [hep-ph]} \BibitemShut
  {NoStop}%
\bibitem [{\citenamefont {Huang}\ \emph {et~al.}(2016)\citenamefont {Huang},
  \citenamefont {Deng}, \citenamefont {Ping},\ and\ \citenamefont
  {Wang}}]{Huang:2015uda}%
  \BibitemOpen
  \bibfield  {author} {\bibinfo {author} {\bibfnamefont {H.}~\bibnamefont
  {Huang}}, \bibinfo {author} {\bibfnamefont {C.}~\bibnamefont {Deng}},
  \bibinfo {author} {\bibfnamefont {J.}~\bibnamefont {Ping}}, \ and\ \bibinfo
  {author} {\bibfnamefont {F.}~\bibnamefont {Wang}},\ }\href {\doibase
  10.1140/epjc/s10052-016-4476-z} {\bibfield  {journal} {\bibinfo  {journal}
  {Eur. Phys. J.}\ }\textbf {\bibinfo {volume} {C76}},\ \bibinfo {pages} {624}
  (\bibinfo {year} {2016})},\ \Eprint {http://arxiv.org/abs/1510.04648}
  {arXiv:1510.04648 [hep-ph]} \BibitemShut {NoStop}%
\bibitem [{\citenamefont {Mei{\ss}ner}\ and\ \citenamefont
  {Oller}(2015)}]{Meissner:2015mza}%
  \BibitemOpen
  \bibfield  {author} {\bibinfo {author} {\bibfnamefont {U.-G.}\ \bibnamefont
  {Mei{\ss}ner}}\ and\ \bibinfo {author} {\bibfnamefont {J.~A.}\ \bibnamefont
  {Oller}},\ }\href {\doibase 10.1016/j.physletb.2015.10.015} {\bibfield
  {journal} {\bibinfo  {journal} {Phys. Lett.}\ }\textbf {\bibinfo {volume}
  {B751}},\ \bibinfo {pages} {59} (\bibinfo {year} {2015})},\ \Eprint
  {http://arxiv.org/abs/1507.07478} {arXiv:1507.07478 [hep-ph]} \BibitemShut
  {NoStop}%
\bibitem [{\citenamefont {Roca}\ \emph {et~al.}(2015)\citenamefont {Roca},
  \citenamefont {Nieves},\ and\ \citenamefont {Oset}}]{Roca:2015dva}%
  \BibitemOpen
  \bibfield  {author} {\bibinfo {author} {\bibfnamefont {L.}~\bibnamefont
  {Roca}}, \bibinfo {author} {\bibfnamefont {J.}~\bibnamefont {Nieves}}, \ and\
  \bibinfo {author} {\bibfnamefont {E.}~\bibnamefont {Oset}},\ }\href {\doibase
  10.1103/PhysRevD.92.094003} {\bibfield  {journal} {\bibinfo  {journal} {Phys.
  Rev.}\ }\textbf {\bibinfo {volume} {D92}},\ \bibinfo {pages} {094003}
  (\bibinfo {year} {2015})},\ \Eprint {http://arxiv.org/abs/1507.04249}
  {arXiv:1507.04249 [hep-ph]} \BibitemShut {NoStop}%
\bibitem [{\citenamefont {Azizi}\ \emph {et~al.}(2017)\citenamefont {Azizi},
  \citenamefont {Sarac},\ and\ \citenamefont {Sundu}}]{Azizi:2016dhy}%
  \BibitemOpen
  \bibfield  {author} {\bibinfo {author} {\bibfnamefont {K.}~\bibnamefont
  {Azizi}}, \bibinfo {author} {\bibfnamefont {Y.}~\bibnamefont {Sarac}}, \ and\
  \bibinfo {author} {\bibfnamefont {H.}~\bibnamefont {Sundu}},\ }\href
  {\doibase 10.1103/PhysRevD.95.094016} {\bibfield  {journal} {\bibinfo
  {journal} {Phys. Rev.}\ }\textbf {\bibinfo {volume} {D95}},\ \bibinfo {pages}
  {094016} (\bibinfo {year} {2017})},\ \Eprint
  {http://arxiv.org/abs/1612.07479} {arXiv:1612.07479 [hep-ph]} \BibitemShut
  {NoStop}%
\bibitem [{\citenamefont {Chen}\ \emph
  {et~al.}(2016{\natexlab{b}})\citenamefont {Chen}, \citenamefont {Liu},\ and\
  \citenamefont {Zhu}}]{Chen:2016heh}%
  \BibitemOpen
  \bibfield  {author} {\bibinfo {author} {\bibfnamefont {R.}~\bibnamefont
  {Chen}}, \bibinfo {author} {\bibfnamefont {X.}~\bibnamefont {Liu}}, \ and\
  \bibinfo {author} {\bibfnamefont {S.-L.}\ \bibnamefont {Zhu}},\ }\href
  {\doibase 10.1016/j.nuclphysa.2016.04.012} {\bibfield  {journal} {\bibinfo
  {journal} {Nucl. Phys.}\ }\textbf {\bibinfo {volume} {A954}},\ \bibinfo
  {pages} {406} (\bibinfo {year} {2016}{\natexlab{b}})},\ \Eprint
  {http://arxiv.org/abs/1601.03233} {arXiv:1601.03233 [hep-ph]} \BibitemShut
  {NoStop}%
\bibitem [{\citenamefont {Chen}\ \emph
  {et~al.}(2016{\natexlab{c}})\citenamefont {Chen}, \citenamefont {Cui},
  \citenamefont {Chen}, \citenamefont {Liu}, \citenamefont {Steele},\ and\
  \citenamefont {Zhu}}]{Chen:2016otp}%
  \BibitemOpen
  \bibfield  {author} {\bibinfo {author} {\bibfnamefont {H.-X.}\ \bibnamefont
  {Chen}}, \bibinfo {author} {\bibfnamefont {E.-L.}\ \bibnamefont {Cui}},
  \bibinfo {author} {\bibfnamefont {W.}~\bibnamefont {Chen}}, \bibinfo {author}
  {\bibfnamefont {X.}~\bibnamefont {Liu}}, \bibinfo {author} {\bibfnamefont
  {T.~G.}\ \bibnamefont {Steele}}, \ and\ \bibinfo {author} {\bibfnamefont
  {S.-L.}\ \bibnamefont {Zhu}},\ }\href {\doibase
  10.1140/epjc/s10052-016-4438-5} {\bibfield  {journal} {\bibinfo  {journal}
  {Eur. Phys. J.}\ }\textbf {\bibinfo {volume} {C76}},\ \bibinfo {pages} {572}
  (\bibinfo {year} {2016}{\natexlab{c}})},\ \Eprint
  {http://arxiv.org/abs/1602.02433} {arXiv:1602.02433 [hep-ph]} \BibitemShut
  {NoStop}%
\bibitem [{\citenamefont {Chen}\ \emph
  {et~al.}(2019{\natexlab{a}})\citenamefont {Chen}, \citenamefont {Sun},
  \citenamefont {Liu},\ and\ \citenamefont {Zhu}}]{Chen:2019asm}%
  \BibitemOpen
  \bibfield  {author} {\bibinfo {author} {\bibfnamefont {R.}~\bibnamefont
  {Chen}}, \bibinfo {author} {\bibfnamefont {Z.-F.}\ \bibnamefont {Sun}},
  \bibinfo {author} {\bibfnamefont {X.}~\bibnamefont {Liu}}, \ and\ \bibinfo
  {author} {\bibfnamefont {S.-L.}\ \bibnamefont {Zhu}},\ }\href {\doibase
  10.1103/PhysRevD.100.011502} {\bibfield  {journal} {\bibinfo  {journal}
  {Phys. Rev.}\ }\textbf {\bibinfo {volume} {D100}},\ \bibinfo {pages} {011502}
  (\bibinfo {year} {2019}{\natexlab{a}})},\ \Eprint
  {http://arxiv.org/abs/1903.11013} {arXiv:1903.11013 [hep-ph]} \BibitemShut
  {NoStop}%
\bibitem [{\citenamefont {Chen}\ \emph
  {et~al.}(2019{\natexlab{b}})\citenamefont {Chen}, \citenamefont {Chen},\ and\
  \citenamefont {Zhu}}]{Chen:2019bip}%
  \BibitemOpen
  \bibfield  {author} {\bibinfo {author} {\bibfnamefont {H.-X.}\ \bibnamefont
  {Chen}}, \bibinfo {author} {\bibfnamefont {W.}~\bibnamefont {Chen}}, \ and\
  \bibinfo {author} {\bibfnamefont {S.-L.}\ \bibnamefont {Zhu}},\ }\href@noop
  {} {\  (\bibinfo {year} {2019}{\natexlab{b}})},\ \Eprint
  {http://arxiv.org/abs/1903.11001} {arXiv:1903.11001 [hep-ph]} \BibitemShut
  {NoStop}%
\bibitem [{\citenamefont {Guo}\ \emph {et~al.}(2019)\citenamefont {Guo},
  \citenamefont {Jing}, \citenamefont {Meißner},\ and\ \citenamefont
  {Sakai}}]{Guo:2019fdo}%
  \BibitemOpen
  \bibfield  {author} {\bibinfo {author} {\bibfnamefont {F.-K.}\ \bibnamefont
  {Guo}}, \bibinfo {author} {\bibfnamefont {H.-J.}\ \bibnamefont {Jing}},
  \bibinfo {author} {\bibfnamefont {U.-G.}\ \bibnamefont {Meißner}}, \ and\
  \bibinfo {author} {\bibfnamefont {S.}~\bibnamefont {Sakai}},\ }\href
  {\doibase 10.1103/PhysRevD.99.091501} {\bibfield  {journal} {\bibinfo
  {journal} {Phys. Rev.}\ }\textbf {\bibinfo {volume} {D99}},\ \bibinfo {pages}
  {091501} (\bibinfo {year} {2019})},\ \Eprint
  {http://arxiv.org/abs/1903.11503} {arXiv:1903.11503 [hep-ph]} \BibitemShut
  {NoStop}%
\bibitem [{\citenamefont {He}(2019)}]{He:2019ify}%
  \BibitemOpen
  \bibfield  {author} {\bibinfo {author} {\bibfnamefont {J.}~\bibnamefont
  {He}},\ }\href {\doibase 10.1140/epjc/s10052-019-6906-1} {\bibfield
  {journal} {\bibinfo  {journal} {Eur. Phys. J.}\ }\textbf {\bibinfo {volume}
  {C79}},\ \bibinfo {pages} {393} (\bibinfo {year} {2019})},\ \Eprint
  {http://arxiv.org/abs/1903.11872} {arXiv:1903.11872 [hep-ph]} \BibitemShut
  {NoStop}%
\bibitem [{\citenamefont {Liu}\ \emph {et~al.}(2019{\natexlab{b}})\citenamefont
  {Liu}, \citenamefont {Pan}, \citenamefont {Peng}, \citenamefont
  {S\'anchez~S\'anchez}, \citenamefont {Geng}, \citenamefont {Hosaka},\ and\
  \citenamefont {Pavon~Valderrama}}]{Liu:2019tjn}%
  \BibitemOpen
  \bibfield  {author} {\bibinfo {author} {\bibfnamefont {M.-Z.}\ \bibnamefont
  {Liu}}, \bibinfo {author} {\bibfnamefont {Y.-W.}\ \bibnamefont {Pan}},
  \bibinfo {author} {\bibfnamefont {F.-Z.}\ \bibnamefont {Peng}}, \bibinfo
  {author} {\bibfnamefont {M.}~\bibnamefont {S\'anchez~S\'anchez}}, \bibinfo
  {author} {\bibfnamefont {L.-S.}\ \bibnamefont {Geng}}, \bibinfo {author}
  {\bibfnamefont {A.}~\bibnamefont {Hosaka}}, \ and\ \bibinfo {author}
  {\bibfnamefont {M.}~\bibnamefont {Pavon~Valderrama}},\ }\href {\doibase
  10.1103/PhysRevLett.122.242001} {\bibfield  {journal} {\bibinfo  {journal}
  {Phys. Rev. Lett.}\ }\textbf {\bibinfo {volume} {122}},\ \bibinfo {pages}
  {242001} (\bibinfo {year} {2019}{\natexlab{b}})},\ \Eprint
  {http://arxiv.org/abs/1903.11560} {arXiv:1903.11560 [hep-ph]} \BibitemShut
  {NoStop}%
\bibitem [{\citenamefont {Lebed}(2015)}]{Lebed:2015tna}%
  \BibitemOpen
  \bibfield  {author} {\bibinfo {author} {\bibfnamefont {R.~F.}\ \bibnamefont
  {Lebed}},\ }\href {\doibase 10.1016/j.physletb.2015.08.032} {\bibfield
  {journal} {\bibinfo  {journal} {Phys. Lett.}\ }\textbf {\bibinfo {volume}
  {B749}},\ \bibinfo {pages} {454} (\bibinfo {year} {2015})},\ \Eprint
  {http://arxiv.org/abs/1507.05867} {arXiv:1507.05867 [hep-ph]} \BibitemShut
  {NoStop}%
\bibitem [{\citenamefont {Maiani}\ \emph {et~al.}(2015)\citenamefont {Maiani},
  \citenamefont {Polosa},\ and\ \citenamefont {Riquer}}]{Maiani:2015vwa}%
  \BibitemOpen
  \bibfield  {author} {\bibinfo {author} {\bibfnamefont {L.}~\bibnamefont
  {Maiani}}, \bibinfo {author} {\bibfnamefont {A.~D.}\ \bibnamefont {Polosa}},
  \ and\ \bibinfo {author} {\bibfnamefont {V.}~\bibnamefont {Riquer}},\ }\href
  {\doibase 10.1016/j.physletb.2015.08.008} {\bibfield  {journal} {\bibinfo
  {journal} {Phys. Lett.}\ }\textbf {\bibinfo {volume} {B749}},\ \bibinfo
  {pages} {289} (\bibinfo {year} {2015})},\ \Eprint
  {http://arxiv.org/abs/1507.04980} {arXiv:1507.04980 [hep-ph]} \BibitemShut
  {NoStop}%
\bibitem [{\citenamefont {Mironov}\ and\ \citenamefont
  {Morozov}(2015)}]{Mironov:2015ica}%
  \BibitemOpen
  \bibfield  {author} {\bibinfo {author} {\bibfnamefont {A.}~\bibnamefont
  {Mironov}}\ and\ \bibinfo {author} {\bibfnamefont {A.}~\bibnamefont
  {Morozov}},\ }\href {\doibase 10.1134/S0021364015170099} {\bibfield
  {journal} {\bibinfo  {journal} {JETP Lett.}\ }\textbf {\bibinfo {volume}
  {102}},\ \bibinfo {pages} {271} (\bibinfo {year} {2015})},\ \bibinfo {note}
  {[Pisma Zh. Eksp. Teor. Fiz.102,no.5,302(2015)]},\ \Eprint
  {http://arxiv.org/abs/1507.04694} {arXiv:1507.04694 [hep-ph]} \BibitemShut
  {NoStop}%
\bibitem [{\citenamefont {Wang}(2016)}]{Wang:2015epa}%
  \BibitemOpen
  \bibfield  {author} {\bibinfo {author} {\bibfnamefont {Z.-G.}\ \bibnamefont
  {Wang}},\ }\href {\doibase 10.1140/epjc/s10052-016-3920-4} {\bibfield
  {journal} {\bibinfo  {journal} {Eur. Phys. J.}\ }\textbf {\bibinfo {volume}
  {C76}},\ \bibinfo {pages} {70} (\bibinfo {year} {2016})},\ \Eprint
  {http://arxiv.org/abs/1508.01468} {arXiv:1508.01468 [hep-ph]} \BibitemShut
  {NoStop}%
\bibitem [{\citenamefont {Zhu}\ and\ \citenamefont {Qiao}(2016)}]{Zhu:2015bba}%
  \BibitemOpen
  \bibfield  {author} {\bibinfo {author} {\bibfnamefont {R.}~\bibnamefont
  {Zhu}}\ and\ \bibinfo {author} {\bibfnamefont {C.-F.}\ \bibnamefont {Qiao}},\
  }\href {\doibase 10.1016/j.physletb.2016.03.022} {\bibfield  {journal}
  {\bibinfo  {journal} {Phys. Lett.}\ }\textbf {\bibinfo {volume} {B756}},\
  \bibinfo {pages} {259} (\bibinfo {year} {2016})},\ \Eprint
  {http://arxiv.org/abs/1510.08693} {arXiv:1510.08693 [hep-ph]} \BibitemShut
  {NoStop}%
\bibitem [{\citenamefont {Santopinto}\ and\ \citenamefont
  {Giachino}(2017)}]{Santopinto:2016pkp}%
  \BibitemOpen
  \bibfield  {author} {\bibinfo {author} {\bibfnamefont {E.}~\bibnamefont
  {Santopinto}}\ and\ \bibinfo {author} {\bibfnamefont {A.}~\bibnamefont
  {Giachino}},\ }\href {\doibase 10.1103/PhysRevD.96.014014} {\bibfield
  {journal} {\bibinfo  {journal} {Phys. Rev.}\ }\textbf {\bibinfo {volume}
  {D96}},\ \bibinfo {pages} {014014} (\bibinfo {year} {2017})},\ \Eprint
  {http://arxiv.org/abs/1604.03769} {arXiv:1604.03769 [hep-ph]} \BibitemShut
  {NoStop}%
\bibitem [{\citenamefont {Richard}\ \emph {et~al.}(2017)\citenamefont
  {Richard}, \citenamefont {Valcarce},\ and\ \citenamefont
  {Vijande}}]{Richard:2017una}%
  \BibitemOpen
  \bibfield  {author} {\bibinfo {author} {\bibfnamefont {J.~M.}\ \bibnamefont
  {Richard}}, \bibinfo {author} {\bibfnamefont {A.}~\bibnamefont {Valcarce}}, \
  and\ \bibinfo {author} {\bibfnamefont {J.}~\bibnamefont {Vijande}},\ }\href
  {\doibase 10.1016/j.physletb.2017.10.036} {\bibfield  {journal} {\bibinfo
  {journal} {Phys. Lett.}\ }\textbf {\bibinfo {volume} {B774}},\ \bibinfo
  {pages} {710} (\bibinfo {year} {2017})},\ \Eprint
  {http://arxiv.org/abs/1710.08239} {arXiv:1710.08239 [hep-ph]} \BibitemShut
  {NoStop}%
\bibitem [{\citenamefont {Ali}\ and\ \citenamefont
  {Parkhomenko}(2019)}]{Ali:2019npk}%
  \BibitemOpen
  \bibfield  {author} {\bibinfo {author} {\bibfnamefont {A.}~\bibnamefont
  {Ali}}\ and\ \bibinfo {author} {\bibfnamefont {A.~{\relax Ya}.}\ \bibnamefont
  {Parkhomenko}},\ }\href {\doibase 10.1016/j.physletb.2019.05.002} {\bibfield
  {journal} {\bibinfo  {journal} {Phys. Lett.}\ }\textbf {\bibinfo {volume}
  {B793}},\ \bibinfo {pages} {365} (\bibinfo {year} {2019})},\ \Eprint
  {http://arxiv.org/abs/1904.00446} {arXiv:1904.00446 [hep-ph]} \BibitemShut
  {NoStop}%
\bibitem [{\citenamefont {Ne'eman}(1961)}]{Neeman:1961jhl}%
  \BibitemOpen
  \bibfield  {author} {\bibinfo {author} {\bibfnamefont {Y.}~\bibnamefont
  {Ne'eman}},\ }\href {\doibase 10.1016/0029-5582(61)90134-1} {\bibfield
  {journal} {\bibinfo  {journal} {Nucl. Phys.}\ }\textbf {\bibinfo {volume}
  {26}},\ \bibinfo {pages} {222} (\bibinfo {year} {1961})},\ \bibinfo {note}
  {[,34(1961)]}\BibitemShut {NoStop}%
\bibitem [{\citenamefont {Gell-Mann}(1962)}]{GellMann:1962xb}%
  \BibitemOpen
  \bibfield  {author} {\bibinfo {author} {\bibfnamefont {M.}~\bibnamefont
  {Gell-Mann}},\ }\href {\doibase 10.1103/PhysRev.125.1067} {\bibfield
  {journal} {\bibinfo  {journal} {Phys. Rev.}\ }\textbf {\bibinfo {volume}
  {125}},\ \bibinfo {pages} {1067} (\bibinfo {year} {1962})}\BibitemShut
  {NoStop}%
\bibitem [{\citenamefont {Eichten}\ \emph {et~al.}(1978)\citenamefont
  {Eichten}, \citenamefont {Gottfried}, \citenamefont {Kinoshita},
  \citenamefont {Lane},\ and\ \citenamefont {Yan}}]{Eichten:1978tg}%
  \BibitemOpen
  \bibfield  {author} {\bibinfo {author} {\bibfnamefont {E.}~\bibnamefont
  {Eichten}}, \bibinfo {author} {\bibfnamefont {K.}~\bibnamefont {Gottfried}},
  \bibinfo {author} {\bibfnamefont {T.}~\bibnamefont {Kinoshita}}, \bibinfo
  {author} {\bibfnamefont {K.~D.}\ \bibnamefont {Lane}}, \ and\ \bibinfo
  {author} {\bibfnamefont {T.-M.}\ \bibnamefont {Yan}},\ }\href {\doibase
  10.1103/PhysRevD.17.3090} {\bibfield  {journal} {\bibinfo  {journal} {Phys.
  Rev.}\ }\textbf {\bibinfo {volume} {D17}},\ \bibinfo {pages} {3090} (\bibinfo
  {year} {1978})},\ \bibinfo {note} {[Erratum: Phys. Rev.D 21, 313
  (1980)]}\BibitemShut {NoStop}%
\bibitem [{\citenamefont {De~R{\'u}jula}\ \emph {et~al.}(1975)\citenamefont
  {De~R{\'u}jula}, \citenamefont {Georgi},\ and\ \citenamefont
  {Glashow}}]{DeRujula:1975qlm}%
  \BibitemOpen
  \bibfield  {author} {\bibinfo {author} {\bibfnamefont {A.}~\bibnamefont
  {De~R{\'u}jula}}, \bibinfo {author} {\bibfnamefont {H.}~\bibnamefont
  {Georgi}}, \ and\ \bibinfo {author} {\bibfnamefont {S.~L.}\ \bibnamefont
  {Glashow}},\ }\href {\doibase 10.1103/PhysRevD.12.147} {\bibfield  {journal}
  {\bibinfo  {journal} {Phys. Rev.}\ }\textbf {\bibinfo {volume} {D12}},\
  \bibinfo {pages} {147} (\bibinfo {year} {1975})}\BibitemShut {NoStop}%
\bibitem [{\citenamefont {Isgur}\ and\ \citenamefont
  {Karl}(1977)}]{Isgur:1977ef}%
  \BibitemOpen
  \bibfield  {author} {\bibinfo {author} {\bibfnamefont {N.}~\bibnamefont
  {Isgur}}\ and\ \bibinfo {author} {\bibfnamefont {G.}~\bibnamefont {Karl}},\
  }\href {\doibase 10.1016/0370-2693(77)90074-0} {\bibfield  {journal}
  {\bibinfo  {journal} {Phys. Lett.}\ }\textbf {\bibinfo {volume} {72B}},\
  \bibinfo {pages} {109} (\bibinfo {year} {1977})}\BibitemShut {NoStop}%
\bibitem [{\citenamefont {Basdevant}\ and\ \citenamefont
  {Boukraa}(1985)}]{Basdevant:1984rk}%
  \BibitemOpen
  \bibfield  {author} {\bibinfo {author} {\bibfnamefont {J.~L.}\ \bibnamefont
  {Basdevant}}\ and\ \bibinfo {author} {\bibfnamefont {S.}~\bibnamefont
  {Boukraa}},\ }\href {\doibase 10.1007/BF01413604} {\bibfield  {journal}
  {\bibinfo  {journal} {Z. Phys.}\ }\textbf {\bibinfo {volume} {C28}},\
  \bibinfo {pages} {413} (\bibinfo {year} {1985})}\BibitemShut {NoStop}%
\bibitem [{\citenamefont {Godfrey}\ and\ \citenamefont
  {Isgur}(1985)}]{Godfrey:1985xj}%
  \BibitemOpen
  \bibfield  {author} {\bibinfo {author} {\bibfnamefont {S.}~\bibnamefont
  {Godfrey}}\ and\ \bibinfo {author} {\bibfnamefont {N.}~\bibnamefont
  {Isgur}},\ }\href {\doibase 10.1103/PhysRevD.32.189} {\bibfield  {journal}
  {\bibinfo  {journal} {Phys. Rev. D}\ }\textbf {\bibinfo {volume} {32}},\
  \bibinfo {pages} {189} (\bibinfo {year} {1985})}\BibitemShut {NoStop}%
\bibitem [{\citenamefont {Capstick}\ and\ \citenamefont
  {Isgur}(1986)}]{Capstick:1986bm}%
  \BibitemOpen
  \bibfield  {author} {\bibinfo {author} {\bibfnamefont {S.}~\bibnamefont
  {Capstick}}\ and\ \bibinfo {author} {\bibfnamefont {N.}~\bibnamefont
  {Isgur}},\ }\bibfield  {booktitle} {\emph {\bibinfo {booktitle}
  {{Proceedings, International Conference on Hadron Spectroscopy: College Park,
  Maryland, April 20-22, 1985}}},\ }\href {\doibase 10.1103/PhysRevD.34.2809}
  {\bibfield  {journal} {\bibinfo  {journal} {Phys. Rev.}\ }\textbf {\bibinfo
  {volume} {D34}},\ \bibinfo {pages} {2809} (\bibinfo {year} {1986})},\
  \bibinfo {note} {[AIP Conf. Proc.132,267(1985)]}\BibitemShut {NoStop}%
\bibitem [{\citenamefont {Zeldovich}\ and\ \citenamefont
  {Sakharov}(1966)}]{Sakharov:1966tua}%
  \BibitemOpen
  \bibfield  {author} {\bibinfo {author} {\bibfnamefont {Y.~B.}\ \bibnamefont
  {Zeldovich}}\ and\ \bibinfo {author} {\bibfnamefont {A.~D.}\ \bibnamefont
  {Sakharov}},\ }\href
  {https://www.osti.gov/biblio/4529076-quark-structure-masses-strongly-interacting-particles}
  {\bibfield  {journal} {\bibinfo  {journal} {Yad. Fiz.}\ }\textbf {\bibinfo
  {volume} {4}},\ \bibinfo {pages} {395} (\bibinfo {year} {1966})}\BibitemShut
  {NoStop}%
\bibitem [{\citenamefont {Zeldovich}\ and\ \citenamefont
  {Sakharov}(1967)}]{Sakharov:1967}%
  \BibitemOpen
  \bibfield  {author} {\bibinfo {author} {\bibfnamefont {Y.~B.}\ \bibnamefont
  {Zeldovich}}\ and\ \bibinfo {author} {\bibfnamefont {A.~D.}\ \bibnamefont
  {Sakharov}},\ }\href@noop {} {\bibfield  {journal} {\bibinfo  {journal} {Sov.
  J. Nucl. Phys.}\ }\textbf {\bibinfo {volume} {4}},\ \bibinfo {pages} {283}
  (\bibinfo {year} {1967})}\BibitemShut {NoStop}%
\bibitem [{\citenamefont {DeGrand}\ \emph {et~al.}(1975)\citenamefont
  {DeGrand}, \citenamefont {Jaffe}, \citenamefont {Johnson},\ and\
  \citenamefont {Kiskis}}]{DeGrand:1975cf}%
  \BibitemOpen
  \bibfield  {author} {\bibinfo {author} {\bibfnamefont {T.~A.}\ \bibnamefont
  {DeGrand}}, \bibinfo {author} {\bibfnamefont {R.~L.}\ \bibnamefont {Jaffe}},
  \bibinfo {author} {\bibfnamefont {K.}~\bibnamefont {Johnson}}, \ and\
  \bibinfo {author} {\bibfnamefont {J.~E.}\ \bibnamefont {Kiskis}},\ }\href
  {\doibase 10.1103/PhysRevD.12.2060} {\bibfield  {journal} {\bibinfo
  {journal} {Phys. Rev.}\ }\textbf {\bibinfo {volume} {D12}},\ \bibinfo {pages}
  {2060} (\bibinfo {year} {1975})}\BibitemShut {NoStop}%
\bibitem [{\citenamefont {Cui}\ \emph {et~al.}(2006)\citenamefont {Cui},
  \citenamefont {Chen}, \citenamefont {Deng},\ and\ \citenamefont
  {Zhu}}]{Cui:2005az}%
  \BibitemOpen
  \bibfield  {author} {\bibinfo {author} {\bibfnamefont {Y.}~\bibnamefont
  {Cui}}, \bibinfo {author} {\bibfnamefont {X.-L.}\ \bibnamefont {Chen}},
  \bibinfo {author} {\bibfnamefont {W.-Z.}\ \bibnamefont {Deng}}, \ and\
  \bibinfo {author} {\bibfnamefont {S.-L.}\ \bibnamefont {Zhu}},\ }\href
  {\doibase 10.1103/PhysRevD.73.014018} {\bibfield  {journal} {\bibinfo
  {journal} {Phys. Rev.}\ }\textbf {\bibinfo {volume} {D73}},\ \bibinfo {pages}
  {014018} (\bibinfo {year} {2006})},\ \Eprint
  {http://arxiv.org/abs/hep-ph/0511150} {arXiv:hep-ph/0511150 [hep-ph]}
  \BibitemShut {NoStop}%
\bibitem [{\citenamefont {Buccella}\ \emph {et~al.}(2007)\citenamefont
  {Buccella}, \citenamefont {Hogaasen}, \citenamefont {Richard},\ and\
  \citenamefont {Sorba}}]{Buccella:2006fn}%
  \BibitemOpen
  \bibfield  {author} {\bibinfo {author} {\bibfnamefont {F.}~\bibnamefont
  {Buccella}}, \bibinfo {author} {\bibfnamefont {H.}~\bibnamefont {Hogaasen}},
  \bibinfo {author} {\bibfnamefont {J.-M.}\ \bibnamefont {Richard}}, \ and\
  \bibinfo {author} {\bibfnamefont {P.}~\bibnamefont {Sorba}},\ }\href
  {\doibase 10.1140/epjc/s10052-006-0142-1} {\bibfield  {journal} {\bibinfo
  {journal} {Eur. Phys. J.}\ }\textbf {\bibinfo {volume} {C49}},\ \bibinfo
  {pages} {743} (\bibinfo {year} {2007})},\ \Eprint
  {http://arxiv.org/abs/hep-ph/0608001} {arXiv:hep-ph/0608001 [hep-ph]}
  \BibitemShut {NoStop}%
\bibitem [{\citenamefont {Wu}\ \emph {et~al.}(2017)\citenamefont {Wu},
  \citenamefont {Liu}, \citenamefont {Chen}, \citenamefont {Liu},\ and\
  \citenamefont {Zhu}}]{Wu:2017weo}%
  \BibitemOpen
  \bibfield  {author} {\bibinfo {author} {\bibfnamefont {J.}~\bibnamefont
  {Wu}}, \bibinfo {author} {\bibfnamefont {Y.-R.}\ \bibnamefont {Liu}},
  \bibinfo {author} {\bibfnamefont {K.}~\bibnamefont {Chen}}, \bibinfo {author}
  {\bibfnamefont {X.}~\bibnamefont {Liu}}, \ and\ \bibinfo {author}
  {\bibfnamefont {S.-L.}\ \bibnamefont {Zhu}},\ }\href {\doibase
  10.1103/PhysRevD.95.034002} {\bibfield  {journal} {\bibinfo  {journal} {Phys.
  Rev.}\ }\textbf {\bibinfo {volume} {D95}},\ \bibinfo {pages} {034002}
  (\bibinfo {year} {2017})},\ \Eprint {http://arxiv.org/abs/1701.03873}
  {arXiv:1701.03873 [hep-ph]} \BibitemShut {NoStop}%
\bibitem [{\citenamefont {Karliner}\ \emph {et~al.}(2017)\citenamefont
  {Karliner}, \citenamefont {Nussinov},\ and\ \citenamefont
  {Rosner}}]{Karliner:2016zzc}%
  \BibitemOpen
  \bibfield  {author} {\bibinfo {author} {\bibfnamefont {M.}~\bibnamefont
  {Karliner}}, \bibinfo {author} {\bibfnamefont {S.}~\bibnamefont {Nussinov}},
  \ and\ \bibinfo {author} {\bibfnamefont {J.~L.}\ \bibnamefont {Rosner}},\
  }\href {\doibase 10.1103/PhysRevD.95.034011} {\bibfield  {journal} {\bibinfo
  {journal} {Phys. Rev.}\ }\textbf {\bibinfo {volume} {D95}},\ \bibinfo {pages}
  {034011} (\bibinfo {year} {2017})},\ \Eprint
  {http://arxiv.org/abs/1611.00348} {arXiv:1611.00348 [hep-ph]} \BibitemShut
  {NoStop}%
\bibitem [{\citenamefont {H{\o}gaasen}\ \emph {et~al.}(2014)\citenamefont
  {H{\o}gaasen}, \citenamefont {Kou}, \citenamefont {Richard},\ and\
  \citenamefont {Sorba}}]{Hogaasen:2013nca}%
  \BibitemOpen
  \bibfield  {author} {\bibinfo {author} {\bibfnamefont {H.}~\bibnamefont
  {H{\o}gaasen}}, \bibinfo {author} {\bibfnamefont {E.}~\bibnamefont {Kou}},
  \bibinfo {author} {\bibfnamefont {J.-M.}\ \bibnamefont {Richard}}, \ and\
  \bibinfo {author} {\bibfnamefont {P.}~\bibnamefont {Sorba}},\ }\href
  {\doibase 10.1016/j.physletb.2014.03.027} {\bibfield  {journal} {\bibinfo
  {journal} {Phys. Lett.}\ }\textbf {\bibinfo {volume} {B732}},\ \bibinfo
  {pages} {97} (\bibinfo {year} {2014})},\ \Eprint
  {http://arxiv.org/abs/1309.2049} {arXiv:1309.2049 [hep-ph]} \BibitemShut
  {NoStop}%
\bibitem [{\citenamefont {Chan}\ \emph {et~al.}(1978)\citenamefont {Chan},
  \citenamefont {Fukugita}, \citenamefont {Hansson}, \citenamefont {Hoffman},
  \citenamefont {Konishi}, \citenamefont {Hogaasen},\ and\ \citenamefont
  {Tsou}}]{Chan:1978nk}%
  \BibitemOpen
  \bibfield  {author} {\bibinfo {author} {\bibfnamefont {H.-M.}\ \bibnamefont
  {Chan}}, \bibinfo {author} {\bibfnamefont {M.}~\bibnamefont {Fukugita}},
  \bibinfo {author} {\bibfnamefont {T.~H.}\ \bibnamefont {Hansson}}, \bibinfo
  {author} {\bibfnamefont {H.~J.}\ \bibnamefont {Hoffman}}, \bibinfo {author}
  {\bibfnamefont {K.}~\bibnamefont {Konishi}}, \bibinfo {author} {\bibfnamefont
  {H.}~\bibnamefont {Hogaasen}}, \ and\ \bibinfo {author} {\bibfnamefont
  {S.~T.}\ \bibnamefont {Tsou}},\ }\href {\doibase
  10.1016/0370-2693(78)90872-9} {\bibfield  {journal} {\bibinfo  {journal}
  {Phys. Lett.}\ }\textbf {\bibinfo {volume} {76B}},\ \bibinfo {pages} {634}
  (\bibinfo {year} {1978})}\BibitemShut {NoStop}%
\bibitem [{\citenamefont {Weng}\ \emph {et~al.}(2018)\citenamefont {Weng},
  \citenamefont {Chen},\ and\ \citenamefont {Deng}}]{Weng:2018mmf}%
  \BibitemOpen
  \bibfield  {author} {\bibinfo {author} {\bibfnamefont {X.-Z.}\ \bibnamefont
  {Weng}}, \bibinfo {author} {\bibfnamefont {X.-L.}\ \bibnamefont {Chen}}, \
  and\ \bibinfo {author} {\bibfnamefont {W.-Z.}\ \bibnamefont {Deng}},\ }\href
  {\doibase 10.1103/PhysRevD.97.054008} {\bibfield  {journal} {\bibinfo
  {journal} {Phys. Rev.}\ }\textbf {\bibinfo {volume} {D97}},\ \bibinfo {pages}
  {054008} (\bibinfo {year} {2018})},\ \Eprint
  {http://arxiv.org/abs/1801.08644} {arXiv:1801.08644 [hep-ph]} \BibitemShut
  {NoStop}%
\bibitem [{\citenamefont {Aaij}\ \emph {et~al.}(2017)\citenamefont {Aaij} \emph
  {et~al.}}]{Aaij:2017ueg}%
  \BibitemOpen
  \bibfield  {author} {\bibinfo {author} {\bibfnamefont {R.}~\bibnamefont
  {Aaij}} \emph {et~al.} (\bibinfo {collaboration} {LHCb Collaboration}),\
  }\href {\doibase 10.1103/PhysRevLett.119.112001} {\bibfield  {journal}
  {\bibinfo  {journal} {Phys. Rev. Lett.}\ }\textbf {\bibinfo {volume} {119}},\
  \bibinfo {pages} {112001} (\bibinfo {year} {2017})},\ \Eprint
  {http://arxiv.org/abs/1707.01621} {arXiv:1707.01621 [hep-ex]} \BibitemShut
  {NoStop}%
\bibitem [{\citenamefont {H{\o}gaasen}\ \emph {et~al.}(2006)\citenamefont
  {H{\o}gaasen}, \citenamefont {Richard},\ and\ \citenamefont
  {Sorba}}]{Hogaasen:2005jv}%
  \BibitemOpen
  \bibfield  {author} {\bibinfo {author} {\bibfnamefont {H.}~\bibnamefont
  {H{\o}gaasen}}, \bibinfo {author} {\bibfnamefont {J.~M.}\ \bibnamefont
  {Richard}}, \ and\ \bibinfo {author} {\bibfnamefont {P.}~\bibnamefont
  {Sorba}},\ }\href {\doibase 10.1103/PhysRevD.73.054013} {\bibfield  {journal}
  {\bibinfo  {journal} {Phys. Rev.}\ }\textbf {\bibinfo {volume} {D73}},\
  \bibinfo {pages} {054013} (\bibinfo {year} {2006})},\ \Eprint
  {http://arxiv.org/abs/hep-ph/0511039} {arXiv:hep-ph/0511039 [hep-ph]}
  \BibitemShut {NoStop}%
\bibitem [{\citenamefont {Zhao}\ \emph {et~al.}(2014)\citenamefont {Zhao},
  \citenamefont {Deng},\ and\ \citenamefont {Zhu}}]{Zhao:2014qva}%
  \BibitemOpen
  \bibfield  {author} {\bibinfo {author} {\bibfnamefont {L.}~\bibnamefont
  {Zhao}}, \bibinfo {author} {\bibfnamefont {W.-Z.}\ \bibnamefont {Deng}}, \
  and\ \bibinfo {author} {\bibfnamefont {S.-L.}\ \bibnamefont {Zhu}},\ }\href
  {\doibase 10.1103/PhysRevD.90.094031} {\bibfield  {journal} {\bibinfo
  {journal} {Phys. Rev.}\ }\textbf {\bibinfo {volume} {D90}},\ \bibinfo {pages}
  {094031} (\bibinfo {year} {2014})},\ \Eprint {http://arxiv.org/abs/1408.3924}
  {arXiv:1408.3924 [hep-ph]} \BibitemShut {NoStop}%
\bibitem [{\citenamefont {Gao}(1992)}]{Gao-1992-Group}%
  \BibitemOpen
  \bibfield  {author} {\bibinfo {author} {\bibfnamefont {C.}~\bibnamefont
  {Gao}},\ }\href@noop {} {\emph {\bibinfo {title} {Group Theory and its
  Application in Particle Physics (in Chinese)}}}\ (\bibinfo  {publisher}
  {Higher Education Press},\ \bibinfo {year} {1992})\BibitemShut {NoStop}%
\end{thebibliography}%
\end{document}